\documentclass[useAMS,usenatbib]{mn2e}

\usepackage{graphicx}

\newcommand{\apj}{ApJ}
\newcommand{\apjs}{ApJS}

\newcommand{\aap}{A\&A}
\newcommand{\aaps}{A\&AS}
\newcommand{\apss}{Ap\&SS}
\newcommand{\mnras}{MNRAS}

\title[Dynamical masses, absolute radii and 3D orbits of HD~181068]{Dynamical masses, absolute radii and 3D orbits of the triply eclipsing star HD~181068 from {\it Kepler} photometry}
\author[T. Borkovits et al.]{T. Borkovits$^{1,2,3}$\thanks{E-mail:
borko@electra.bajaobs.hu (TB)}, A. Derekas$^{2,4}$, L. L. Kiss$^{2,3,4}$, A. Kir\'aly$^{2,5}$, E. Forg\'acs-Dajka$^{5,6}$ \\
\newauthor I. B. B\'\i r\'o$^{1}$, T. R. Bedding$^4$, S.T. Bryson$^{7}$, D. Huber$^{7,4}$, R. Szab\'o$^{2}$\\
$^{1}$Baja Astronomical Observatory, H-6500 Baja, Szegedi \'ut, Kt. 766, Hungary\\
$^{2}$Konkoly Observatory, MTA CSFK, H-1121 Budapest, Konkoly Thege M. \'ut 15-17, Hungary\\
$^{3}$ELTE Gothard-Lend\"ulet Research Group, H-9700 Szombathely, Szent Imre herceg \'ut 112, Hungary\\
$^{4}$Sydney Institute for Astronomy, School of Physics, University of Sydney, NSW 2006, Australia\\
$^{5}$Astronomical Department of E\"otv\"os University, H-1118 P\'azm\'any P\'eter stny. 1/A, Budapest, Hungary\\
$^{6}$University of Vienna, T\"urkenschanzstrasse 17, 1180 Vienna, Austria \\
$^{7}$NASA Ames Research Center, Moffett Field, CA 94035, USA}
\begin{document}

\date{Accepted ??? Received ???; in original form ???}

\pagerange{\pageref{firstpage}--\pageref{lastpage}} \pubyear{2012}

\maketitle

\label{firstpage}

\begin{abstract}
HD~181068 is the brighter of the two known triply eclipsing hierarchical triple stars in the {\it Kepler} field. It has been continuously observed for more than 2 years with the {\it Kepler} space telescope. Of the nine quarters of the data, three have been obtained in short-cadence mode, that is one point per 58.9~s. Here we analyse this unique dataset to determine absolute physical parameters (most importantly the masses and radii) and full orbital configuration using a sophisticated novel approach. We measure eclipse timing variations (ETVs), which are then combined with the single-lined radial velocity measurements to yield masses in a manner equivalent to double-lined spectroscopic binaries. We have also developed a new light curve synthesis code that is used to model the triple, mutual eclipses and the effects of the changing tidal field on the stellar surface and the relativistic Doppler-beaming. By combining the stellar masses from the ETV study with the simultaneous light curve analysis we determine the absolute radii of the three stars. Our results indicate that the close and the wide subsystems revolve in almost exactly coplanar and prograde orbits. The newly determined parameters draw a consistent picture of the system with such details that have been beyond reach before. 
\end{abstract}

\begin{keywords}
stars: multiple -- stars: eclipsing -- stars: individual: HD 181068
\end{keywords}

\section{Introduction}
\label{Intro}

The {\it Kepler} space telescope, in addition to its primary science aims, has led to a new era in the investigation of multiple star systems. Among the highlights we find the discoveries of the first triply eclipsing triple systems \citep{carteretal11,der11} and some interesting studies of multiple star systems \citep{steffenetal11,fei11,gie12,leh12}.

Binary and multiple systems have an important role in astrophysics. The most accurate way to measure stellar parameters is through eclipsing binaries, and their distance determination is also very accurate. Their light curves provide essential information on the internal structure of the components, their atmospheres and their magnetic activity. In the case of noncircular orbits and multiple systems, the orbital elements can change significantly, allowing detailed insight into the time variation of these parameters. The special geometry of the very rare and new category of eclipsing systems, namely the triply (or mutually) eclipsing triple systems, enables us fast and easy determination of further characteristics that otherwise could only be studied with great effort on a long time-scale. 

As an example, we refer to the spatial configuration of such hierarchical triple systems, which is a key-parameter in understanding their origin and evolution \citep[see e.~g.][and references therein]{fabryckytremaine07}. In the absence of mutual eclipses, the two ways to determine the mutual (or relative) inclination in a hierarchical system are $(a)$ astrometric (or, more rarely, polarimetric) measurements of the spatial orientations of the two orbits individually, or $(b)$ indirect dynamical calculation from the measured mutual gravitational perturbations of the bodies. The first method requires long baseline optical (or very-long baseline radio) interferometric measurements for the most interesting close binaries, which typically have milli-arcsecond angular separations. It is therefore not suprising that, starting with the pioneering work by \citet{lestradeetal93} on Algol, this method has only been applied to about a dozen binaries  \citep[see also][for more recent results]{baronetal12,sanbornzavala12,petersonetal11,obrienetal11}. The applicability of polarimetric measurements (although does not require high-category instruments) in this field is even more restricted \citep[see e.~g.][]{piirola10}. The second method, the detection of gravitational perturbations, requires accurate, frequent and continuous photometric eclipse time determination. This method will be described in detail in the next section. 

The situation is much easier in the case of mutual eclipses, where the shape of the light curve (especially around the ingress and egress phases) contains direct and unique information about the system geometry. This is discussed in detail by \citet{ragozzineholman10} and \citet{pal12}. The former authors list several other values of multi-transiting systems, mainly in the context of multiple planetary systems. Their model has been succesfully applied to analysing complex light curves and determining the corresponding geometrical and physical parameters (both for the orbits and the individual bodies) for different multiple-transiting planetary (\citealp[][Kepler-11]{lissaueretal11}, \citealp[][Kepler-16]{doyleetal11}, \citealp[][Kepler-34b-35b]{welshetal12}, \citealp[][Kepler-36]{carteretal12}) and stellar systems \citep[][KOI-126]{carteretal11}. 

KOI-126 and HD~181068 are the first representatives of the new category of the triply eclipsing triple systems. Both are also members of a very small group of compact hierarchical triple stellar systems. They contain a close binary, with orbital periods $P_1^{\rmn{KOI}}=1.77$ and $P_1^{\rmn{HD}}=0.91$ days, and a more distant component forming a wider binary with the centre of mass of the close pair with periods $P_2^{\rmn{KOI}}=33.92$, and $P_2^{\rmn{HD}}=45.47$ days, respectively. The main speciality of the two systems is their triply eclipsing nature, which means that both the inner and the outer binaries show eclipses. They have other, very peculiar characteristics. Both belong to the most compact triple stellar systems, and there is only one known hierarchical triple system with a shorter outer period, namely $\lambda$~Tau, with $P_2=33.03$ days. Furthermore, these two systems are unusual even amongst the very few similarly compact triples, in having reversed outer mass-ratio. In other words, in these two objects the wide, single component is the more massive star, and also the largest and brightest. Before {\it Kepler}, the highest known outer mass-ratio did not reach 1.5, and for 97\% of known hierarchical triplets it remained under 1, i.~e. almost in all the catalogized systems, the total mass of the close binary exceeded the mass of the tertiary component \citep[see][]{tokovinin08}. (The question of whether this comes from observational bias is not discussed here.) In contrast, the outer mass ratios of these two new systems are$q_\rmn{AB}^{\rmn{KOI}}\sim3.0$, and $q_\rmn{AB}^{\rmn{HD}}\sim1.9$, respectively. 

Despite the similarities of KOI-126 and HD~181068 to each other, there are remarkable differences between the two systems. On one hand, KOI-126 consists of three nearly spherical main sequence stars, where the members of the close binary have such a low surface brightnesses that their light curve modelling is largely equivalent to those of the multiple planetary systems. This is not true for HD~181068, where all the three stars are tidally distorted, have almost equal surface brightnesses and show evidence of intrinsic light variations, all of which make light curve modelling of HD~181068 more difficult than for KOI-126. On the other hand, dynamical analysis of HD~181068 is much less complex than for KOI-126, because of the much simpler and apparently constant orbital configurations. As a consequence, our method of light-curve analysis is much closer to the traditional eclipsing binary star light curve modeling methods \citep[see][for a review]{kallrathmilone09} than the procedures applied for systems like KOI-126.

In this paper, we analyse more than 2 years of {\it Kepler} observations of HD~181068. We mainly concentrate on determining the fundamental astrophysical parameters of the three stars and orbital elements of the close and wide orbits. These quantities by themselves carry very important information already about the system and their members' origin and evolution and, furthermore, give the necessary input parameters for other forthcoming studies, for example for a comprehensive study of pulsations of the red giant component. Nevertheless, due to the uniqueness of the studied system, our aim is not simply to give a case study. The specifics of HD~181068 allow us to present methods never used before. For example, in our period study (Section 3), which depends on the analysis of the eclipse timing variations (ETV) for both the close and the wide systems, we determine the (inclination-dependent) masses of the wide binary members in a new manner. While the radial velocity curve of the most massive $A$ component is known, the missing second radial velocity curve of the spectroscopically unseen $B$ component (i.~e. the close binary itself) is replaced by the light-time orbit of the $B$ component deduced from the ETV analysis of the shallow eclipses. This method is fundamentally different from the one followed by \citet{steffenetal11} for KOI-928, for example, because it does not use the dynamical part of the ETV, only the simple geometrical light-time contribution. More details are given in Section 3. In Section\ \ref{sect:lcanalysis}, the light curve analysis procedure is described in detail, while Section 5 contains the discussion of the results. Finally, the details of our light curve synthesis and analysis code, and some additional examples of calculations of certain quantities purely in a photometrical and geometrical way from the mutual eclipses, are given in the appendices.

It is important to establish a clear notation for this system. In \citet{der11} the three components were labelled A, B, and C (in order of decreasing masses and luminosities). Here, we use the more clarified and expressive denotations, $A$, $Ba$, $Bb$. As before, $A$ denotes the most massive and luminous component (the main component of the wider $A-B$ binary), while $Ba$ and $Bb$ refer to the members of the close  binary formed by the two red dwarfs, formerly denoted by $B$ and $C$. When referring to any physical quantities of the individual stars, we use subscripts. For example, $m_\rmn{A}$ and $m_\rmn{Ba}$ denote the masses of the $A$ and $Ba$ components, respectively, but $m_\rmn{B}$ refers to the total mass of the close binary, i.e. ($m_\rmn{Ba}+m_\rmn{Bb}$), and $m_\rmn{AB}$ stands for the total mass of the hierarchical triple. With this notation we can avoid the confusion with the indices of the orbital parameters of different orbits used for the period study. Namely, following the common usage, the elements of relative orbit of the $Bb$ component around its companion, $Ba$ is subscripted with $1$, whereas the relative orbit of the ternary component $A$, around the center of mass of the $Ba-Bb$ subsystem (symbolically represented with $B$) is associated with the subscript $2$. However, in terms of light-time and the radial velocity, the absolute orbit (i.e. the orbit of some star around the center of mass) is to be considered, rather than the relative orbits. In these cases, those absolute orbital elements, which numerically differ from the corresponding relative orbital element, were naturally denoted by the alphabetic sign of the given star, or subsystem. 

\section[]{Observation and data reduction}
\label{Sect2}

The analysis is based on photometry from the {\it Kepler} space telescope \citep{bor10,gil10,koc10,jen10a,jen10b}. The dataset is 775 days long, observed in 6 quarters (Q1-Q6) at long cadence (time resolution of 29.4 min) and 3 quarters (Q7-Q9) at short cadence (time resolution of 58.9 sec). Since HD~181068 is a $\sim$~7 magnitude star, it is heavily saturated, resulting in charge bleeding. Therefore, the short-cadence observations were obtained using a Custom Made Aperture Mask. This was uploaded directly to the spacecraft lookup table and shaped precisely to match the shape of the target on the detector including the bleeding area.

\subsection{Measuring the times of minima}
\label{measuremin}

\begin{figure}
\includegraphics[width=84mm]{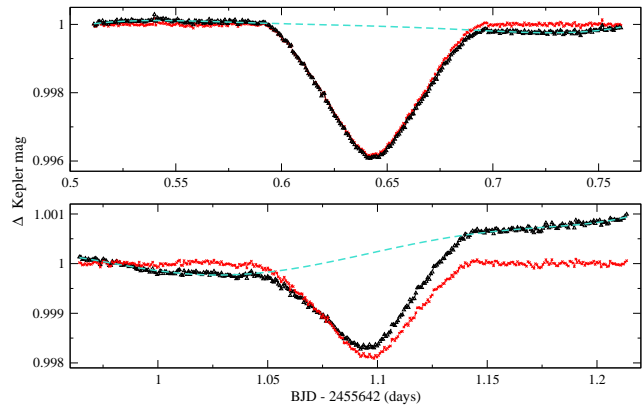}
 \caption{\label{minima} Example of a primary ({\it upper panel}) and a secondary ({\it lower panel}) shallow minimum to illustrate the states before (black triangles) and after (red crosses) the detrending of the minima. The dashed line is the fit used for the detrending (see Section\ \ref{measuremin}).}
\end{figure}

The 2.1 year-long observations cover $\sim$~885 orbital cycles of the close pair and 17 revolutions of the wide system. Approximately 10\% of the eclipses of the close binary (hereafter we refer to them as shallow minima) occur during the eclipse events of the wide system (hereafter deep minima), and cannot be observed. Additionally, a few hundred events escaped observation due to data gaps. In all, 1177 of the 1770 shallow minima were analysed. The analysis of these minima was quite a complex task. As shown by \citet{der11}, the red giant component shows oscillations on a time scale similar to the half of the orbital period of the short period binary. In addition, there are long term variations, discussed in Sect.\ \ref{sect:lcanalysis}, which slightly distort the shape of the shallow minima, as shown in  Fig.\ \ref{minima}. This distortion has a significant effect on the measurement of the exact times of minima. 

To correct for these distortions, we applied the following method in determining the times of minima. We took the $\pm$~0.225 days interval around each minimum and fitted low-order (4-6) polynomials outside the eclipses. Then we corrected each subset, which resulted in a detrended light curve. Finally, to determine the times of minima, we fitted low order (5-6) polynomials to the lowest parts of the minima. 

We also analysed the available deep minima. Out of the 34 events we were able to determine times of minima in 28 cases. (One of these events was omitted from the final analysis, due to its large deviation from the general trend of the data, which might be caused by its incomplete sampling.) To determine these times of minima, first we removed the effects of the intrinsic brightness variations from the light curves, and then fitted each outer transit and occultation event individually with our newly developed simultaneous light curve solution code. Both the code and the complete light curve analysis are described in Sect.\ \ref{sect:lcanalysis}.

The determined times of minima are listed in Tables~\ref{tab:ToM} and \ref{tab:greatToM} for the close and the wide pairs, respectively.

\begin{table*}
 \caption{Times of minima for the close pair}
 \label{tab:ToM}
 \begin{tabular}{@{}lcclcclcclcc}
  \hline
BJD & $\sigma$ & Type & BJD & $\sigma$ & Type & BJD & $\sigma$ & Type & BJD & $\sigma$ & Type \\
\hline
2454963.8399 & 0.0010 & II & 2454994.6312 & 0.0010 & II & 2455101.5010 & 0.0010 & II & 2455132.2946 & 0.0010 & II \\ 
2454964.2926 & 0.0010 &  I & 2454995.0838 & 0.0010 &  I & 2455101.9551 & 0.0010 &  I & 2455132.7470 & 0.0010 &  I \\ 
2454965.1967 & 0.0010 &  I & 2454995.5359 & 0.0010 & II & 2455102.4099 & 0.0010 & II & 2455133.1999 & 0.0010 & II \\ 
2454965.6478 & 0.0010 & II & 2454995.9891 & 0.0010 &  I & 2455102.8605 & 0.0010 &  I & 2455133.6511 & 0.0010 &  I \\ 
2454966.1021 & 0.0010 &  I & 2454996.4429 & 0.0010 & II & 2455103.3130 & 0.0010 & II & 2455134.1046 & 0.0010 & II \\ 
2454966.5546 & 0.0010 & II & 2454996.8929 & 0.0010 &  I & 2455103.7657 & 0.0010 &  I & 2455134.5572 & 0.0010 &  I \\ 
2454967.0071 & 0.0010 &  I & 2454997.3488 & 0.0010 & II & 2455104.2174 & 0.0010 & II & 2455135.0101 & 0.0010 & II \\ 
2454967.4605 & 0.0010 & II & 2454997.8017 & 0.0010 &  I & 2455104.6722 & 0.0010 &  I & 2455137.2785 & 0.0010 &  I \\ 
2454967.9144 & 0.0010 &  I & 2454998.2542 & 0.0010 & II & 2455105.1258 & 0.0010 & II & 2455137.7282 & 0.0010 & II \\ 
2454968.3676 & 0.0010 & II & 2454998.7059 & 0.0010 &  I & 2455105.5781 & 0.0010 &  I & 2455138.1807 & 0.0010 &  I \\ 
2454968.8194 & 0.0010 &  I & 2454999.1610 & 0.0010 & II & 2455106.0310 & 0.0010 & II & 2455138.6355 & 0.0010 & II \\ 
2454969.2732 & 0.0010 & II & 2454999.6116 & 0.0010 &  I & 2455106.4840 & 0.0010 &  I & 2455139.0857 & 0.0010 &  I \\ 
2454969.7251 & 0.0010 &  I & 2455003.2343 & 0.0010 &  I & 2455106.9385 & 0.0010 & II & 2455139.5388 & 0.0010 & II \\ 
2454970.1781 & 0.0010 & II & 2455003.6883 & 0.0010 & II & 2455107.3897 & 0.0010 &  I & 2455139.9900 & 0.0010 &  I \\ 
2454970.6311 & 0.0010 &  I & 2455004.1399 & 0.0010 &  I & 2455107.8435 & 0.0010 & II & 2455140.4443 & 0.0010 & II \\ 
2454971.0852 & 0.0010 & II & 2455004.5945 & 0.0010 & II & 2455108.2951 & 0.0010 &  I & 2455140.8985 & 0.0010 &  I \\ 
2454971.5369 & 0.0010 &  I & 2455005.0453 & 0.0010 &  I & 2455108.7484 & 0.0010 & II & 2455141.3563 & 0.0010 & II \\ 
2454971.9897 & 0.0010 & II & 2455005.4987 & 0.0010 & II & 2455109.2005 & 0.0010 &  I & 2455141.8030 & 0.0010 &  I \\ 
2454972.4439 & 0.0010 &  I & 2455005.9514 & 0.0010 &  I & 2455109.6543 & 0.0010 & II & 2455142.2567 & 0.0010 & II \\ 
2454972.8964 & 0.0010 & II & 2455006.4069 & 0.0010 & II & 2455110.1077 & 0.0010 &  I & 2455142.7091 & 0.0010 &  I \\ 
2454973.3487 & 0.0010 &  I & 2455006.8569 & 0.0010 &  I & 2455110.5593 & 0.0010 & II & 2455143.1628 & 0.0010 & II \\ 
2454973.8012 & 0.0010 & II & 2455007.3125 & 0.0010 & II & 2455111.0126 & 0.0010 &  I & 2455143.6156 & 0.0010 &  I \\ 
2454974.2544 & 0.0010 &  I & 2455007.7626 & 0.0010 &  I & 2455111.9171 & 0.0010 &  I & 2455144.0683 & 0.0010 & II \\ 
2454974.7077 & 0.0010 & II & 2455008.2172 & 0.0010 & II & 2455114.6330 & 0.0010 &  I & 2455144.5208 & 0.0010 &  I \\ 
2454975.1600 & 0.0010 &  I & 2455008.6681 & 0.0010 &  I & 2455115.0886 & 0.0010 & II & 2455144.9733 & 0.0010 & II \\ 
2454975.6110 & 0.0010 & II & 2455009.1237 & 0.0010 & II & 2455115.9942 & 0.0010 & II & 2455145.4263 & 0.0010 &  I \\ 
2454976.0664 & 0.0010 &  I & 2455010.0269 & 0.0010 & II & 2455116.4469 & 0.0010 &  I & 2455145.8797 & 0.0010 & II \\ 
2454976.5207 & 0.0010 & II & 2455010.4798 & 0.0010 &  I & 2455116.8992 & 0.0010 & II & 2455146.3337 & 0.0010 &  I \\ 
2454976.9719 & 0.0010 &  I & 2455010.9336 & 0.0010 & II & 2455117.3522 & 0.0010 &  I & 2455146.7850 & 0.0010 & II \\ 
2454981.0479 & 0.0010 & II & 2455011.8403 & 0.0010 & II & 2455117.8044 & 0.0010 & II & 2455147.2386 & 0.0010 &  I \\ 
2454981.5003 & 0.0010 &  I & 2455012.2914 & 0.0010 &  I & 2455118.2563 & 0.0010 &  I & 2455147.6920 & 0.0010 & II \\ 
2454981.9544 & 0.0010 & II & 2455012.7449 & 0.0010 & II & 2455118.7105 & 0.0010 & II & 2455148.1429 & 0.0010 &  I \\ 
2454982.4048 & 0.0010 &  I & 2455013.6507 & 0.0010 & II & 2455119.1635 & 0.0010 &  I & 2455148.5975 & 0.0010 & II \\ 
2454982.8580 & 0.0010 & II & 2455014.1035 & 0.0010 &  I & 2455119.6164 & 0.0010 & II & 2455149.0511 & 0.0010 &  I \\ 
2454983.3116 & 0.0010 &  I & 2455014.5572 & 0.0010 & II & 2455120.0681 & 0.0010 &  I & 2455149.9553 & 0.0010 &  I \\ 
2454983.7671 & 0.0010 & II & 2455015.0081 & 0.0010 &  I & 2455120.5200 & 0.0010 & II & 2455150.4070 & 0.0010 & II \\ 
2454984.2167 & 0.0010 &  I & 2455016.3699 & 0.0010 & II & 2455120.9754 & 0.0010 &  I & 2455150.8602 & 0.0010 &  I \\ 
2454984.6695 & 0.0010 & II & 2455016.8215 & 0.0010 &  I & 2455121.4274 & 0.0010 & II & 2455151.3110 & 0.0010 & II \\ 
2454985.1226 & 0.0010 &  I & 2455019.5391 & 0.0010 &  I & 2455121.8792 & 0.0010 &  I & 2455151.7693 & 0.0010 &  I \\ 
2454985.5771 & 0.0010 & II & 2455019.9919 & 0.0010 & II & 2455122.3325 & 0.0010 & II & 2455152.2221 & 0.0010 & II \\ 
2454986.0275 & 0.0010 &  I & 2455020.4449 & 0.0010 &  I & 2455122.7838 & 0.0010 &  I & 2455152.6729 & 0.0010 &  I \\ 
2454986.4819 & 0.0010 & II & 2455020.8964 & 0.0010 & II & 2455123.2389 & 0.0010 & II & 2455153.1263 & 0.0010 & II \\ 
2454986.9334 & 0.0010 &  I & 2455093.3498 & 0.0010 & II & 2455124.5962 & 0.0010 &  I & 2455153.5789 & 0.0010 &  I \\ 
2454987.3857 & 0.0010 & II & 2455093.8030 & 0.0010 &  I & 2455125.0489 & 0.0010 & II & 2455154.0318 & 0.0010 & II \\ 
2454987.8395 & 0.0010 &  I & 2455094.2549 & 0.0010 & II & 2455125.5019 & 0.0010 &  I & 2455156.7506 & 0.0010 & II \\ 
2454988.2928 & 0.0010 & II & 2455094.7076 & 0.0010 &  I & 2455125.9538 & 0.0010 & II & 2455157.2025 & 0.0010 &  I \\ 
2454988.7449 & 0.0010 &  I & 2455095.1606 & 0.0010 & II & 2455126.4076 & 0.0010 &  I & 2455157.6564 & 0.0010 & II \\ 
2454989.1967 & 0.0010 & II & 2455095.6140 & 0.0010 &  I & 2455126.8605 & 0.0010 & II & 2455160.3727 & 0.0010 & II \\ 
2454989.6504 & 0.0010 &  I & 2455096.0667 & 0.0010 & II & 2455127.3131 & 0.0010 &  I & 2455160.8254 & 0.0010 &  I \\ 
2454990.1049 & 0.0010 & II & 2455096.5190 & 0.0010 &  I & 2455127.7654 & 0.0010 & II & 2455161.2782 & 0.0010 & II \\ 
2454990.5561 & 0.0010 &  I & 2455096.9724 & 0.0010 & II & 2455128.2180 & 0.0010 &  I & 2455161.7294 & 0.0010 &  I \\ 
2454991.0091 & 0.0010 & II & 2455097.4240 & 0.0010 &  I & 2455128.6723 & 0.0010 & II & 2455162.1839 & 0.0010 & II \\ 
2454991.4619 & 0.0010 &  I & 2455097.8796 & 0.0010 & II & 2455129.1244 & 0.0010 &  I & 2455162.6355 & 0.0010 &  I \\ 
2454991.9159 & 0.0010 & II & 2455098.3312 & 0.0010 &  I & 2455129.5755 & 0.0010 & II & 2455163.0890 & 0.0010 & II \\ 
2454992.3675 & 0.0010 &  I & 2455098.7829 & 0.0010 & II & 2455130.0306 & 0.0010 &  I & 2455163.5420 & 0.0010 &  I \\ 
2454992.8209 & 0.0010 & II & 2455099.2375 & 0.0010 &  I & 2455130.4822 & 0.0010 & II & 2455163.9946 & 0.0010 & II \\ 
2454993.2726 & 0.0010 &  I & 2455099.6907 & 0.0010 & II & 2455130.9330 & 0.0010 &  I & 2455164.4462 & 0.0010 &  I \\ 
2454993.7237 & 0.0010 & II & 2455100.1415 & 0.0010 &  I & 2455131.3871 & 0.0010 & II & 2455164.8980 & 0.0010 & II \\ 
2454994.1785 & 0.0010 &  I & 2455101.0502 & 0.0010 &  I & 2455131.8408 & 0.0010 &  I & 2455165.3521 & 0.0010 &  I \\ 
\hline
\end{tabular}
\end{table*}
\setcounter{table}{0}
\begin{table*}
 \caption{Times of minima for the close pair (continued)}
 \begin{tabular}{@{}lcclcclcclcc}
  \hline
BJD & $\sigma$ & Type & BJD & $\sigma$ & Type & BJD & $\sigma$ & Type & BJD & $\sigma$ & Type \\
\hline
2455165.8057 & 0.0010 & II & 2455196.5973 & 0.0010 & II & 2455229.2014 & 0.0010 & II & 2455262.2589 & 0.0010 &  I \\ 
2455166.2571 & 0.0010 &  I & 2455197.0506 & 0.0010 &  I & 2455229.6530 & 0.0010 &  I & 2455262.7143 & 0.0010 & II \\ 
2455166.7118 & 0.0010 & II & 2455197.5016 & 0.0010 & II & 2455234.1820 & 0.0010 &  I & 2455263.1643 & 0.0010 &  I \\ 
2455167.1625 & 0.0010 &  I & 2455197.9563 & 0.0010 &  I & 2455234.6354 & 0.0010 & II & 2455263.6172 & 0.0010 & II \\ 
2455167.6143 & 0.0010 & II & 2455198.8642 & 0.0010 &  I & 2455235.0871 & 0.0010 &  I & 2455264.0693 & 0.0010 &  I \\ 
2455168.0689 & 0.0010 &  I & 2455199.3154 & 0.0010 & II & 2455235.5403 & 0.0010 & II & 2455264.5227 & 0.0010 & II \\ 
2455168.5204 & 0.0010 & II & 2455199.7682 & 0.0010 &  I & 2455235.9946 & 0.0010 &  I & 2455264.9751 & 0.0010 &  I \\ 
2455168.9745 & 0.0010 &  I & 2455200.2232 & 0.0010 & II & 2455236.4474 & 0.0010 & II & 2455265.4305 & 0.0010 & II \\ 
2455169.4273 & 0.0010 & II & 2455200.6731 & 0.0010 &  I & 2455236.8990 & 0.0010 &  I & 2455265.8805 & 0.0010 &  I \\ 
2455169.8793 & 0.0010 &  I & 2455201.1280 & 0.0010 & II & 2455237.3518 & 0.0010 & II & 2455266.3335 & 0.0010 & II \\ 
2455170.3314 & 0.0010 & II & 2455201.5798 & 0.0010 &  I & 2455237.8052 & 0.0010 &  I & 2455266.7867 & 0.0010 &  I \\ 
2455170.7846 & 0.0010 &  I & 2455202.0321 & 0.0010 & II & 2455238.7127 & 0.0010 &  I & 2455267.2403 & 0.0010 & II \\ 
2455171.2377 & 0.0010 & II & 2455202.4852 & 0.0010 &  I & 2455239.1626 & 0.0010 & II & 2455267.6921 & 0.0010 &  I \\ 
2455171.6922 & 0.0010 &  I & 2455202.9387 & 0.0010 & II & 2455239.6170 & 0.0010 &  I & 2455268.1446 & 0.0010 & II \\ 
2455172.1444 & 0.0010 & II & 2455205.6558 & 0.0010 & II & 2455240.0709 & 0.0010 & II & 2455268.5972 & 0.0010 &  I \\ 
2455172.5969 & 0.0010 &  I & 2455206.1085 & 0.0010 &  I & 2455240.5226 & 0.0010 &  I & 2455269.0530 & 0.0010 & II \\ 
2455173.0483 & 0.0010 & II & 2455206.5624 & 0.0010 & II & 2455240.9766 & 0.0010 & II & 2455269.9571 & 0.0010 & II \\ 
2455173.5019 & 0.0010 &  I & 2455207.0141 & 0.0010 &  I & 2455241.4291 & 0.0010 &  I & 2455270.4094 & 0.0010 &  I \\ 
2455173.9555 & 0.0010 & II & 2455207.4666 & 0.0010 & II & 2455241.8823 & 0.0010 & II & 2455270.8620 & 0.0010 & II \\ 
2455174.4083 & 0.0010 &  I & 2455207.9207 & 0.0010 &  I & 2455242.3343 & 0.0010 &  I & 2455271.3125 & 0.0010 &  I \\ 
2455174.8618 & 0.0010 & II & 2455208.3747 & 0.0010 & II & 2455242.7872 & 0.0010 & II & 2455274.0311 & 0.0010 &  I \\ 
2455175.3120 & 0.0010 &  I & 2455208.8256 & 0.0010 &  I & 2455243.2405 & 0.0010 &  I & 2455274.4849 & 0.0010 & II \\ 
2455176.2170 & 0.0010 &  I & 2455209.2773 & 0.0010 & II & 2455243.6902 & 0.0010 & II & 2455274.9361 & 0.0010 &  I \\ 
2455176.6701 & 0.0010 & II & 2455209.7301 & 0.0010 &  I & 2455244.1458 & 0.0010 &  I & 2455277.2007 & 0.0010 & II \\ 
2455177.1247 & 0.0010 &  I & 2455210.1846 & 0.0010 & II & 2455244.5992 & 0.0010 & II & 2455278.1082 & 0.0010 & II \\ 
2455177.5766 & 0.0010 & II & 2455210.6363 & 0.0010 &  I & 2455245.0515 & 0.0010 &  I & 2455279.9181 & 0.0010 & II \\ 
2455178.0299 & 0.0010 &  I & 2455211.0904 & 0.0010 & II & 2455245.5051 & 0.0010 & II & 2455280.8252 & 0.0010 & II \\ 
2455178.4846 & 0.0010 & II & 2455211.5409 & 0.0010 &  I & 2455245.9575 & 0.0010 &  I & 2455281.7304 & 0.0010 & II \\ 
2455178.9346 & 0.0010 &  I & 2455211.9941 & 0.0010 & II & 2455246.4102 & 0.0010 & II & 2455282.6374 & 0.0010 & II \\ 
2455179.3877 & 0.0010 & II & 2455212.4471 & 0.0010 &  I & 2455246.8628 & 0.0010 &  I & 2455283.5419 & 0.0010 & II \\ 
2455179.8418 & 0.0010 &  I & 2455212.8994 & 0.0010 & II & 2455247.3158 & 0.0010 & II & 2455284.4490 & 0.0010 & II \\ 
2455180.2948 & 0.0010 & II & 2455213.3535 & 0.0010 &  I & 2455247.7698 & 0.0010 &  I & 2455285.3552 & 0.0010 & II \\ 
2455180.7447 & 0.0010 &  I & 2455213.8046 & 0.0010 & II & 2455248.2229 & 0.0010 & II & 2455286.2593 & 0.0010 & II \\ 
2455184.8249 & 0.0010 & II & 2455214.2586 & 0.0010 &  I & 2455248.6748 & 0.0010 &  I & 2455287.1656 & 0.0010 & II \\ 
2455185.2776 & 0.0010 &  I & 2455214.7105 & 0.0010 & II & 2455250.9482 & 0.0010 & II & 2455288.0726 & 0.0010 & II \\ 
2455185.7273 & 0.0010 & II & 2455215.1632 & 0.0010 &  I & 2455251.3928 & 0.0010 &  I & 2455288.9782 & 0.0010 & II \\ 
2455186.1799 & 0.0010 &  I & 2455215.6139 & 0.0010 & II & 2455251.8454 & 0.0010 & II & 2455289.8833 & 0.0010 & II \\ 
2455186.6346 & 0.0010 & II & 2455216.0699 & 0.0010 &  I & 2455252.2979 & 0.0010 &  I & 2455290.7894 & 0.0010 & II \\ 
2455187.0869 & 0.0010 &  I & 2455217.4272 & 0.0010 & II & 2455252.7506 & 0.0010 & II & 2455291.6945 & 0.0010 & II \\ 
2455187.5412 & 0.0010 & II & 2455217.8792 & 0.0010 &  I & 2455253.2041 & 0.0010 &  I & 2455292.5995 & 0.0010 & II \\ 
2455187.9934 & 0.0010 &  I & 2455218.3336 & 0.0010 & II & 2455253.6582 & 0.0010 & II & 2455293.5069 & 0.0010 & II \\ 
2455188.4450 & 0.0010 & II & 2455218.7867 & 0.0010 &  I & 2455254.1080 & 0.0010 &  I & 2455297.1281 & 0.0010 & II \\ 
2455188.8981 & 0.0010 &  I & 2455219.2412 & 0.0010 & II & 2455254.5616 & 0.0010 & II & 2455298.0349 & 0.0010 & II \\ 
2455189.3515 & 0.0010 & II & 2455219.6907 & 0.0010 &  I & 2455255.0146 & 0.0010 &  I & 2455298.9417 & 0.0010 & II \\ 
2455189.8027 & 0.0010 &  I & 2455220.1447 & 0.0010 & II & 2455255.4681 & 0.0010 & II & 2455299.8474 & 0.0010 & II \\ 
2455190.2577 & 0.0010 & II & 2455220.5967 & 0.0010 &  I & 2455255.9214 & 0.0010 &  I & 2455300.7532 & 0.0010 & II \\ 
2455190.7087 & 0.0010 &  I & 2455221.0499 & 0.0010 & II & 2455256.3741 & 0.0010 & II & 2455301.6557 & 0.0010 & II \\ 
2455191.1655 & 0.0010 & II & 2455221.5019 & 0.0010 &  I & 2455256.8255 & 0.0010 &  I & 2455302.5648 & 0.0010 & II \\ 
2455191.6160 & 0.0010 &  I & 2455221.9583 & 0.0010 & II & 2455257.2747 & 0.0010 & II & 2455303.4691 & 0.0010 & II \\ 
2455192.0681 & 0.0010 & II & 2455222.4097 & 0.0010 &  I & 2455257.7299 & 0.0010 &  I & 2455305.2800 & 0.0010 & II \\ 
2455192.5204 & 0.0010 &  I & 2455222.8609 & 0.0010 & II & 2455258.1856 & 0.0010 & II & 2455306.1858 & 0.0010 & II \\ 
2455192.9764 & 0.0010 & II & 2455223.3136 & 0.0010 &  I & 2455258.6384 & 0.0010 &  I & 2455307.0881 & 0.0010 & II \\ 
2455193.4266 & 0.0010 &  I & 2455223.7642 & 0.0010 & II & 2455259.0912 & 0.0010 & II & 2455309.8044 & 0.0010 & II \\ 
2455193.8803 & 0.0010 & II & 2455224.2188 & 0.0010 &  I & 2455259.5424 & 0.0010 &  I & 2455310.7133 & 0.0010 & II \\ 
2455194.3331 & 0.0010 &  I & 2455224.6721 & 0.0010 & II & 2455259.9961 & 0.0010 & II & 2455311.6198 & 0.0010 & II \\ 
2455194.7853 & 0.0010 & II & 2455225.1257 & 0.0010 &  I & 2455260.4492 & 0.0010 &  I & 2455313.4275 & 0.0010 & II \\ 
2455195.2387 & 0.0010 &  I & 2455226.0314 & 0.0010 &  I & 2455260.8994 & 0.0010 & II & 2455314.3352 & 0.0010 & II \\ 
2455195.6946 & 0.0010 & II & 2455228.2937 & 0.0010 & II & 2455261.3547 & 0.0010 &  I & 2455315.2384 & 0.0010 & II \\ 
2455196.1452 & 0.0010 &  I & 2455228.7472 & 0.0010 &  I & 2455261.8072 & 0.0010 & II & 2455316.1422 & 0.0010 & II \\ 
\hline
\end{tabular}
\end{table*}
\setcounter{table}{0}
\begin{table*}
 \caption{Times of minima for the close pair (continued)}
 \begin{tabular}{@{}lcclcclcclcc}
  \hline
BJD & $\sigma$ & Type & BJD & $\sigma$ & Type & BJD & $\sigma$ & Type & BJD & $\sigma$ & Type \\
\hline
2455317.0475 & 0.0010 & II & 2455377.7335 & 0.0010 & II & 2455408.0720 & 0.0010 &  I & 2455439.7717 & 0.0010 &  I \\ 
2455319.7697 & 0.0010 & II & 2455378.1862 & 0.0010 &  I & 2455408.5241 & 0.0010 & II & 2455440.2272 & 0.0010 & II \\ 
2455320.6756 & 0.0010 & II & 2455378.6391 & 0.0010 & II & 2455411.2440 & 0.0010 & II & 2455440.6775 & 0.0010 &  I \\ 
2455321.5787 & 0.0010 & II & 2455379.0916 & 0.0010 &  I & 2455411.6947 & 0.0010 &  I & 2455441.1326 & 0.0010 & II \\ 
2455322.4873 & 0.0010 & II & 2455379.5443 & 0.0010 & II & 2455412.1468 & 0.0010 & II & 2455441.5837 & 0.0010 &  I \\ 
2455323.3916 & 0.0010 & II & 2455379.9967 & 0.0010 &  I & 2455412.6000 & 0.0010 &  I & 2455442.0395 & 0.0010 & II \\ 
2455324.2975 & 0.0010 & II & 2455380.4486 & 0.0010 & II & 2455413.0556 & 0.0010 & II & 2455442.4897 & 0.0010 &  I \\ 
2455325.2031 & 0.0010 & II & 2455380.9024 & 0.0010 &  I & 2455413.5075 & 0.0010 &  I & 2455442.9436 & 0.0010 & II \\ 
2455326.1094 & 0.0010 & II & 2455381.3567 & 0.0010 & II & 2455413.9629 & 0.0010 & II & 2455443.3940 & 0.0010 &  I \\ 
2455327.0152 & 0.0010 & II & 2455381.8100 & 0.0010 &  I & 2455414.4120 & 0.0010 &  I & 2455443.8493 & 0.0010 & II \\ 
2455327.9208 & 0.0010 & II & 2455382.2629 & 0.0010 & II & 2455414.8646 & 0.0010 & II & 2455444.3003 & 0.0010 &  I \\ 
2455328.8275 & 0.0010 & II & 2455382.7152 & 0.0010 &  I & 2455415.3159 & 0.0010 &  I & 2455444.7546 & 0.0010 & II \\ 
2455329.7327 & 0.0010 & II & 2455383.1670 & 0.0010 & II & 2455415.7715 & 0.0010 & II & 2455445.2046 & 0.0010 &  I \\ 
2455330.6358 & 0.0010 & II & 2455383.6215 & 0.0010 &  I & 2455416.2234 & 0.0010 &  I & 2455445.6593 & 0.0010 & II \\ 
2455331.5422 & 0.0010 & II & 2455384.0749 & 0.0010 & II & 2455416.6769 & 0.0010 & II & 2455446.1110 & 0.0010 &  I \\ 
2455332.4492 & 0.0010 & II & 2455384.5267 & 0.0010 &  I & 2455417.1288 & 0.0010 &  I & 2455446.5625 & 0.0010 & II \\ 
2455333.3539 & 0.0010 & II & 2455384.9805 & 0.0010 & II & 2455417.5823 & 0.0010 & II & 2455447.0161 & 0.0010 &  I \\ 
2455334.2614 & 0.0010 & II & 2455385.4313 & 0.0010 &  I & 2455418.0353 & 0.0010 &  I & 2455447.4660 & 0.0010 & II \\ 
2455335.1677 & 0.0010 & II & 2455385.8849 & 0.0010 & II & 2455418.4868 & 0.0010 & II & 2455447.9214 & 0.0010 &  I \\ 
2455336.0729 & 0.0010 & II & 2455388.6028 & 0.0010 & II & 2455418.9404 & 0.0010 &  I & 2455448.3760 & 0.0010 & II \\ 
2455337.8848 & 0.0010 & II & 2455389.0558 & 0.0010 &  I & 2455419.3938 & 0.0010 & II & 2455448.8280 & 0.0010 &  I \\ 
2455338.7889 & 0.0010 & II & 2455389.5085 & 0.0010 & II & 2455419.8471 & 0.0010 &  I & 2455449.2839 & 0.0010 & II \\ 
2455339.6974 & 0.0010 & II & 2455389.9616 & 0.0010 &  I & 2455420.3012 & 0.0010 & II & 2455449.7337 & 0.0010 &  I \\ 
2455342.4122 & 0.0010 & II & 2455390.4144 & 0.0010 & II & 2455420.7525 & 0.0010 &  I & 2455450.1861 & 0.0010 & II \\ 
2455343.3198 & 0.0010 & II & 2455390.8672 & 0.0010 &  I & 2455421.2057 & 0.0010 & II & 2455450.6387 & 0.0010 &  I \\ 
2455344.2237 & 0.0010 & II & 2455391.3201 & 0.0010 & II & 2455421.6579 & 0.0010 &  I & 2455451.0936 & 0.0010 & II \\ 
2455345.1306 & 0.0010 & II & 2455391.7716 & 0.0010 &  I & 2455422.1103 & 0.0010 & II & 2455451.5456 & 0.0010 &  I \\ 
2455346.0355 & 0.0010 & II & 2455392.2254 & 0.0010 & II & 2455422.5636 & 0.0010 &  I & 2455451.9975 & 0.0010 & II \\ 
2455346.9414 & 0.0010 & II & 2455392.6779 & 0.0010 &  I & 2455423.0166 & 0.0010 & II & 2455452.4485 & 0.0010 &  I \\ 
2455347.8465 & 0.0010 & II & 2455393.1300 & 0.0010 & II & 2455423.4695 & 0.0010 &  I & 2455452.9030 & 0.0010 & II \\ 
2455348.7514 & 0.0010 & II & 2455393.5834 & 0.0010 &  I & 2455423.9224 & 0.0010 & II & 2455453.3558 & 0.0010 &  I \\ 
2455349.6589 & 0.0010 & II & 2455394.0365 & 0.0010 & II & 2455424.3746 & 0.0010 &  I & 2455453.8064 & 0.0010 & II \\ 
2455350.5622 & 0.0010 & II & 2455394.4891 & 0.0010 &  I & 2455424.8276 & 0.0010 & II & 2455454.2636 & 0.0010 &  I \\ 
2455351.4699 & 0.0010 & II & 2455394.9404 & 0.0010 & II & 2455425.2819 & 0.0010 &  I & 2455456.5287 & 0.0010 & II \\ 
2455352.3738 & 0.0010 & II & 2455395.3944 & 0.0010 &  I & 2455425.7354 & 0.0010 & II & 2455456.9785 & 0.0010 &  I \\ 
2455353.2807 & 0.0010 & II & 2455395.8469 & 0.0010 & II & 2455426.6409 & 0.0010 & II & 2455457.4319 & 0.0010 & II \\ 
2455354.1831 & 0.0010 & II & 2455396.2995 & 0.0010 &  I & 2455427.0935 & 0.0010 &  I & 2455457.8844 & 0.0010 &  I \\ 
2455355.0909 & 0.0010 & II & 2455396.7526 & 0.0010 & II & 2455427.5467 & 0.0010 & II & 2455458.3384 & 0.0010 & II \\ 
2455355.9974 & 0.0010 & II & 2455397.2049 & 0.0010 &  I & 2455427.9997 & 0.0010 &  I & 2455458.7885 & 0.0010 &  I \\ 
2455356.9013 & 0.0010 & II & 2455397.6620 & 0.0010 & II & 2455428.4520 & 0.0010 & II & 2455459.2414 & 0.0010 & II \\ 
2455357.8074 & 0.0010 & II & 2455398.1130 & 0.0010 &  I & 2455428.9047 & 0.0010 &  I & 2455459.6944 & 0.0010 &  I \\ 
2455358.7129 & 0.0010 & II & 2455398.5619 & 0.0010 & II & 2455429.3569 & 0.0010 & II & 2455460.1483 & 0.0010 & II \\ 
2455359.6172 & 0.0010 & II & 2455399.0156 & 0.0010 &  I & 2455429.8087 & 0.0010 &  I & 2455460.6005 & 0.0010 &  I \\ 
2455360.5266 & 0.0010 & II & 2455399.4689 & 0.0010 & II & 2455430.2654 & 0.0010 & II & 2455461.0525 & 0.0010 & II \\ 
2455361.4287 & 0.0010 & II & 2455399.9218 & 0.0010 &  I & 2455430.7160 & 0.0010 &  I & 2455461.5072 & 0.0010 &  I \\ 
2455362.3342 & 0.0010 & II & 2455401.7331 & 0.0010 &  I & 2455431.1699 & 0.0010 & II & 2455461.9624 & 0.0010 & II \\ 
2455365.0530 & 0.0010 & II & 2455402.1855 & 0.0010 & II & 2455433.8849 & 0.0010 & II & 2455462.4128 & 0.0010 &  I \\ 
2455365.9592 & 0.0010 & II & 2455402.6390 & 0.0010 &  I & 2455434.3388 & 0.0010 &  I & 2455462.8674 & 0.0010 & II \\ 
2455366.8637 & 0.0010 & II & 2455403.0901 & 0.0010 & II & 2455434.7910 & 0.0010 & II & 2455463.7717 & 0.0005 & II \\ 
2455371.8469 & 0.0010 &  I & 2455403.5441 & 0.0010 &  I & 2455435.2442 & 0.0010 &  I & 2455464.2261 & 0.0005 &  I \\ 
2455372.2998 & 0.0010 & II & 2455403.9981 & 0.0010 & II & 2455435.6987 & 0.0010 & II & 2455464.6770 & 0.0005 & II \\ 
2455374.1114 & 0.0010 & II & 2455404.4514 & 0.0010 &  I & 2455436.1498 & 0.0010 &  I & 2455465.1315 & 0.0005 &  I \\ 
2455374.5610 & 0.0010 &  I & 2455404.9002 & 0.0010 & II & 2455436.6029 & 0.0010 & II & 2455465.5842 & 0.0005 & II \\ 
2455375.0148 & 0.0010 & II & 2455405.3552 & 0.0010 &  I & 2455437.0560 & 0.0010 &  I & 2455466.0369 & 0.0005 &  I \\ 
2455375.4687 & 0.0010 &  I & 2455405.8120 & 0.0010 & II & 2455437.5097 & 0.0010 & II & 2455466.4878 & 0.0005 & II \\ 
2455375.9223 & 0.0010 & II & 2455406.2622 & 0.0010 &  I & 2455437.9610 & 0.0010 &  I & 2455466.9428 & 0.0005 &  I \\ 
2455376.3742 & 0.0010 &  I & 2455406.7133 & 0.0010 & II & 2455438.4152 & 0.0010 & II & 2455467.3976 & 0.0005 & II \\ 
2455376.8286 & 0.0010 & II & 2455407.1677 & 0.0010 &  I & 2455438.8682 & 0.0010 &  I & 2455467.8483 & 0.0005 &  I \\ 
2455377.2789 & 0.0010 &  I & 2455407.6199 & 0.0010 & II & 2455439.3209 & 0.0010 & II & 2455468.3019 & 0.0005 & II \\ 
\hline
\end{tabular}
\end{table*}
\setcounter{table}{0}
\begin{table*}
 \caption{Times of minima for the close pair (continued)}
 \begin{tabular}{@{}lcclcclcclcc}
  \hline
BJD & $\sigma$ & Type & BJD & $\sigma$ & Type & BJD & $\sigma$ & Type & BJD & $\sigma$ & Type \\
\hline
2455468.7542 & 0.0005 &  I & 2455501.3575 & 0.0005 &  I & 2455531.2464 & 0.0005 &  I & 2455578.3420 & 0.0005 &  I \\ 
2455469.2073 & 0.0005 & II & 2455501.8090 & 0.0005 & II & 2455531.6999 & 0.0005 & II & 2455578.7932 & 0.0005 & II \\ 
2455469.6603 & 0.0005 &  I & 2455502.2632 & 0.0005 &  I & 2455532.1515 & 0.0005 &  I & 2455579.6986 & 0.0005 & II \\ 
2455470.1139 & 0.0005 & II & 2455502.7161 & 0.0005 & II & 2455532.6062 & 0.0005 & II & 2455580.1521 & 0.0005 &  I \\ 
2455470.5665 & 0.0005 &  I & 2455503.1681 & 0.0005 &  I & 2455533.0574 & 0.0005 &  I & 2455580.6049 & 0.0005 & II \\ 
2455471.0186 & 0.0005 & II & 2455503.6213 & 0.0005 & II & 2455533.5100 & 0.0005 & II & 2455581.0574 & 0.0005 &  I \\ 
2455471.4718 & 0.0005 &  I & 2455504.0741 & 0.0005 &  I & 2455533.9625 & 0.0005 &  I & 2455581.5105 & 0.0005 & II \\ 
2455471.9251 & 0.0005 & II & 2455504.5265 & 0.0005 & II & 2455534.4158 & 0.0005 & II & 2455581.9643 & 0.0005 &  I \\ 
2455472.3781 & 0.0005 &  I & 2455504.9801 & 0.0005 &  I & 2455534.8681 & 0.0005 &  I & 2455582.4163 & 0.0005 & II \\ 
2455472.8314 & 0.0005 & II & 2455505.4323 & 0.0005 & II & 2455535.3218 & 0.0005 & II & 2455582.8691 & 0.0005 &  I \\ 
2455473.2831 & 0.0005 &  I & 2455505.8865 & 0.0005 &  I & 2455535.7741 & 0.0005 &  I & 2455583.3218 & 0.0005 & II \\ 
2455473.7377 & 0.0005 & II & 2455506.3385 & 0.0005 & II & 2455536.2268 & 0.0005 & II & 2455583.7738 & 0.0005 &  I \\ 
2455474.6430 & 0.0005 & II & 2455506.7915 & 0.0005 &  I & 2455536.6788 & 0.0005 &  I & 2455584.2276 & 0.0005 & II \\ 
2455475.0950 & 0.0005 &  I & 2455507.2443 & 0.0005 & II & 2455537.1329 & 0.0005 & II & 2455584.6789 & 0.0005 &  I \\ 
2455475.5475 & 0.0005 & II & 2455507.6974 & 0.0005 &  I & 2455537.5842 & 0.0005 &  I & 2455585.1325 & 0.0005 & II \\ 
2455476.0015 & 0.0005 &  I & 2455508.6032 & 0.0005 &  I & 2455538.0368 & 0.0005 & II & 2455585.5851 & 0.0005 &  I \\ 
2455478.7173 & 0.0005 &  I & 2455509.0573 & 0.0005 & II & 2455538.4896 & 0.0005 &  I & 2455586.4906 & 0.0005 &  I \\ 
2455479.1718 & 0.0005 & II & 2455509.5096 & 0.0005 &  I & 2455538.9423 & 0.0005 & II & 2455586.9434 & 0.0005 & II \\ 
2455479.6237 & 0.0005 &  I & 2455509.9625 & 0.0005 & II & 2455539.3957 & 0.0005 &  I & 2455587.3954 & 0.0005 &  I \\ 
2455480.0770 & 0.0005 & II & 2455510.4144 & 0.0005 &  I & 2455539.8489 & 0.0005 & II & 2455587.8488 & 0.0005 & II \\ 
2455480.5302 & 0.0005 &  I & 2455510.8680 & 0.0005 & II & 2455540.3005 & 0.0005 &  I & 2455588.3015 & 0.0005 &  I \\ 
2455480.9821 & 0.0005 & II & 2455511.3213 & 0.0005 &  I & 2455540.7539 & 0.0005 & II & 2455588.7542 & 0.0005 & II \\ 
2455481.4349 & 0.0005 &  I & 2455511.7738 & 0.0005 & II & 2455541.2068 & 0.0005 &  I & 2455589.2072 & 0.0005 &  I \\ 
2455481.8883 & 0.0005 & II & 2455512.2263 & 0.0005 &  I & 2455541.6590 & 0.0005 & II & 2455589.6597 & 0.0005 & II \\ 
2455482.3405 & 0.0005 &  I & 2455512.6809 & 0.0005 & II & 2455542.1120 & 0.0005 &  I & 2455592.3773 & 0.0005 & II \\ 
2455482.7922 & 0.0005 & II & 2455513.1336 & 0.0005 &  I & 2455542.5656 & 0.0005 & II & 2455592.8297 & 0.0005 &  I \\ 
2455483.2465 & 0.0005 &  I & 2455513.5859 & 0.0005 & II & 2455543.0184 & 0.0005 &  I & 2455593.2826 & 0.0005 & II \\ 
2455483.7020 & 0.0005 & II & 2455514.0381 & 0.0005 &  I & 2455543.4702 & 0.0005 & II & 2455593.7360 & 0.0005 &  I \\ 
2455484.1517 & 0.0005 &  I & 2455514.4902 & 0.0005 & II & 2455547.0937 & 0.0005 & II & 2455596.9066 & 0.0005 & II \\ 
2455484.6053 & 0.0005 & II & 2455514.9437 & 0.0005 &  I & 2455547.5458 & 0.0005 &  I & 2455597.3587 & 0.0005 &  I \\ 
2455485.0578 & 0.0005 &  I & 2455515.3971 & 0.0005 & II & 2455547.9990 & 0.0005 & II & 2455597.8120 & 0.0005 & II \\ 
2455485.5097 & 0.0005 & II & 2455515.8504 & 0.0005 &  I & 2455548.4525 & 0.0005 &  I & 2455598.2641 & 0.0005 &  I \\ 
2455485.9627 & 0.0005 &  I & 2455516.3035 & 0.0005 & II & 2455548.9056 & 0.0005 & II & 2455598.7184 & 0.0005 & II \\ 
2455486.4147 & 0.0005 & II & 2455516.7557 & 0.0005 &  I & 2455549.3570 & 0.0005 &  I & 2455599.1709 & 0.0005 &  I \\ 
2455486.8686 & 0.0005 &  I & 2455517.2081 & 0.0005 & II & 2455549.8100 & 0.0005 & II & 2455599.6232 & 0.0005 & II \\ 
2455487.3220 & 0.0005 & II & 2455517.6619 & 0.0005 &  I & 2455550.2637 & 0.0005 &  I & 2455600.0759 & 0.0005 &  I \\ 
2455487.7737 & 0.0005 &  I & 2455518.1156 & 0.0005 & II & 2455550.7169 & 0.0005 & II & 2455600.5298 & 0.0005 & II \\ 
2455488.2261 & 0.0005 & II & 2455518.5680 & 0.0005 &  I & 2455551.1691 & 0.0005 &  I & 2455600.9820 & 0.0005 &  I \\ 
2455488.6789 & 0.0005 &  I & 2455519.0206 & 0.0005 & II & 2455551.6238 & 0.0005 & II & 2455601.4349 & 0.0005 & II \\ 
2455489.1323 & 0.0005 & II & 2455519.4744 & 0.0005 &  I & 2455552.0754 & 0.0005 &  I & 2455601.8876 & 0.0005 &  I \\ 
2455489.5846 & 0.0005 &  I & 2455519.9260 & 0.0005 & II & 2455569.7379 & 0.0005 & II & 2455602.3424 & 0.0005 & II \\ 
2455490.0371 & 0.0005 & II & 2455520.3791 & 0.0005 &  I & 2455570.1922 & 0.0005 &  I & 2455602.7933 & 0.0005 &  I \\ 
2455490.4899 & 0.0005 &  I & 2455520.8313 & 0.0005 & II & 2455570.6430 & 0.0005 & II & 2455603.2462 & 0.0005 & II \\ 
2455490.9432 & 0.0005 & II & 2455521.2845 & 0.0005 &  I & 2455571.0980 & 0.0005 &  I & 2455603.6989 & 0.0005 &  I \\ 
2455491.3953 & 0.0005 &  I & 2455524.4545 & 0.0005 & II & 2455571.5494 & 0.0005 & II & 2455604.1543 & 0.0005 & II \\ 
2455491.8476 & 0.0005 & II & 2455524.9077 & 0.0005 &  I & 2455572.0033 & 0.0005 &  I & 2455604.6052 & 0.0005 &  I \\ 
2455492.3012 & 0.0005 &  I & 2455525.3594 & 0.0005 & II & 2455572.4559 & 0.0005 & II & 2455605.0599 & 0.0005 & II \\ 
2455492.7532 & 0.0005 & II & 2455525.8125 & 0.0005 &  I & 2455572.9093 & 0.0005 &  I & 2455605.5115 & 0.0005 &  I \\ 
2455493.2069 & 0.0005 &  I & 2455526.2669 & 0.0005 & II & 2455573.3596 & 0.0005 & II & 2455605.9647 & 0.0005 & II \\ 
2455494.5648 & 0.0005 & II & 2455526.7179 & 0.0005 &  I & 2455573.8147 & 0.0005 &  I & 2455606.4173 & 0.0005 &  I \\ 
2455495.0172 & 0.0005 &  I & 2455527.1722 & 0.0005 & II & 2455574.2668 & 0.0005 & II & 2455606.8715 & 0.0005 & II \\ 
2455495.4689 & 0.0005 & II & 2455527.6246 & 0.0005 &  I & 2455574.7207 & 0.0005 &  I & 2455607.3236 & 0.0005 &  I \\ 
2455495.9226 & 0.0005 &  I & 2455528.0772 & 0.0005 & II & 2455575.1715 & 0.0005 & II & 2455607.7765 & 0.0005 & II \\ 
2455496.3746 & 0.0005 & II & 2455528.5289 & 0.0005 &  I & 2455575.6251 & 0.0005 &  I & 2455608.2291 & 0.0005 &  I \\ 
2455496.8289 & 0.0005 &  I & 2455528.9816 & 0.0005 & II & 2455576.0767 & 0.0005 & II & 2455608.6823 & 0.0005 & II \\ 
2455497.2808 & 0.0005 & II & 2455529.4345 & 0.0005 &  I & 2455576.5308 & 0.0005 &  I & 2455609.1350 & 0.0005 &  I \\ 
2455497.7340 & 0.0005 &  I & 2455529.8870 & 0.0005 & II & 2455576.9824 & 0.0005 & II & 2455609.5880 & 0.0005 & II \\ 
2455498.1876 & 0.0005 & II & 2455530.3409 & 0.0005 &  I & 2455577.4365 & 0.0005 &  I & 2455610.0405 & 0.0005 &  I \\ 
2455498.6395 & 0.0005 &  I & 2455530.7942 & 0.0005 & II & 2455577.8889 & 0.0005 & II & 2455610.4941 & 0.0005 & II \\ 
\hline
\end{tabular}
\end{table*}
\setcounter{table}{0}
\begin{table*}
 \caption{Times of minima for the close pair (continued)}
 \begin{tabular}{@{}lcclcclcclcc}
  \hline
BJD & $\sigma$ & Type & BJD & $\sigma$ & Type & BJD & $\sigma$ & Type & BJD & $\sigma$ & Type \\
\hline
2455610.9466 & 0.0005 &  I & 2455647.1719 & 0.0005 &  I & 2455677.0593 & 0.0005 &  I & 2455710.5712 & 0.0005 &  I \\ 
2455611.3995 & 0.0005 & II & 2455647.6256 & 0.0005 & II & 2455677.5115 & 0.0005 & II & 2455711.0220 & 0.0005 & II \\ 
2455611.8529 & 0.0005 &  I & 2455648.0790 & 0.0005 &  I & 2455678.8702 & 0.0005 &  I & 2455711.4773 & 0.0005 &  I \\ 
2455612.3055 & 0.0005 & II & 2455648.5305 & 0.0005 & II & 2455679.3227 & 0.0005 & II & 2455711.9281 & 0.0005 & II \\ 
2455615.0236 & 0.0005 & II & 2455648.9834 & 0.0005 &  I & 2455679.7759 & 0.0005 &  I & 2455712.3825 & 0.0005 &  I \\ 
2455615.4750 & 0.0005 &  I & 2455649.4363 & 0.0005 & II & 2455680.2286 & 0.0005 & II & 2455712.8348 & 0.0005 & II \\ 
2455615.9271 & 0.0005 & II & 2455649.8897 & 0.0005 &  I & 2455680.6813 & 0.0005 &  I & 2455713.2877 & 0.0005 &  I \\ 
2455616.3809 & 0.0005 &  I & 2455650.3425 & 0.0005 & II & 2455683.3990 & 0.0005 &  I & 2455713.7399 & 0.0005 & II \\ 
2455617.2863 & 0.0005 &  I & 2455650.7951 & 0.0005 &  I & 2455683.8511 & 0.0005 & II & 2455714.1931 & 0.0005 &  I \\ 
2455617.7385 & 0.0005 & II & 2455651.2478 & 0.0005 & II & 2455684.3045 & 0.0005 &  I & 2455714.6455 & 0.0005 & II \\ 
2455618.1924 & 0.0005 &  I & 2455651.7015 & 0.0005 &  I & 2455684.7566 & 0.0005 & II & 2455715.0986 & 0.0005 &  I \\ 
2455618.6455 & 0.0005 & II & 2455652.1542 & 0.0005 & II & 2455685.2096 & 0.0005 &  I & 2455715.5518 & 0.0005 & II \\ 
2455619.0977 & 0.0005 &  I & 2455653.0602 & 0.0005 & II & 2455685.6613 & 0.0005 & II & 2455716.4582 & 0.0005 & II \\ 
2455619.5523 & 0.0005 & II & 2455653.5132 & 0.0005 &  I & 2455686.1165 & 0.0005 &  I & 2455716.9101 & 0.0005 &  I \\ 
2455620.0035 & 0.0005 &  I & 2455653.9665 & 0.0005 & II & 2455686.5683 & 0.0005 & II & 2455717.3626 & 0.0005 & II \\ 
2455620.4562 & 0.0005 & II & 2455654.4189 & 0.0005 &  I & 2455687.0217 & 0.0005 &  I & 2455717.8152 & 0.0005 &  I \\ 
2455620.9085 & 0.0005 &  I & 2455654.8719 & 0.0005 & II & 2455687.4731 & 0.0005 & II & 2455718.2668 & 0.0005 & II \\ 
2455621.3623 & 0.0005 & II & 2455655.3245 & 0.0005 &  I & 2455687.9278 & 0.0005 &  I & 2455718.7203 & 0.0005 &  I \\ 
2455621.8147 & 0.0005 &  I & 2455655.7771 & 0.0005 & II & 2455688.3801 & 0.0005 & II & 2455719.1725 & 0.0005 & II \\ 
2455622.2670 & 0.0005 & II & 2455656.2302 & 0.0005 &  I & 2455688.8330 & 0.0005 &  I & 2455719.6261 & 0.0005 &  I \\ 
2455622.7196 & 0.0005 &  I & 2455656.6842 & 0.0005 & II & 2455689.2853 & 0.0005 & II & 2455720.0793 & 0.0005 & II \\ 
2455623.1728 & 0.0005 & II & 2455657.1370 & 0.0005 &  I & 2455689.7390 & 0.0005 &  I & 2455720.5319 & 0.0005 &  I \\ 
2455623.6254 & 0.0005 &  I & 2455657.5885 & 0.0005 & II & 2455690.1923 & 0.0005 & II & 2455720.9837 & 0.0005 & II \\ 
2455624.0789 & 0.0005 & II & 2455660.7584 & 0.0005 &  I & 2455690.6444 & 0.0005 &  I & 2455721.4378 & 0.0005 &  I \\ 
2455624.5310 & 0.0005 &  I & 2455661.2120 & 0.0005 & II & 2455691.0981 & 0.0005 & II & 2455721.8891 & 0.0005 & II \\ 
2455624.9824 & 0.0005 & II & 2455661.6649 & 0.0005 &  I & 2455691.5507 & 0.0005 &  I & 2455722.3431 & 0.0005 &  I \\ 
2455625.4362 & 0.0005 &  I & 2455662.1179 & 0.0005 & II & 2455692.0032 & 0.0005 & II & 2455722.7959 & 0.0005 & II \\ 
2455625.8874 & 0.0005 & II & 2455662.5705 & 0.0005 &  I & 2455692.4558 & 0.0005 &  I & 2455723.2483 & 0.0005 &  I \\ 
2455626.3419 & 0.0005 &  I & 2455663.0229 & 0.0005 & II & 2455692.9090 & 0.0005 & II & 2455723.7019 & 0.0005 & II \\ 
2455626.7942 & 0.0005 & II & 2455663.4762 & 0.0005 &  I & 2455693.3622 & 0.0005 &  I & 2455724.1541 & 0.0005 &  I \\ 
2455627.2478 & 0.0005 &  I & 2455663.9279 & 0.0005 & II & 2455693.8154 & 0.0005 & II & 2455724.6072 & 0.0005 & II \\ 
2455627.6989 & 0.0005 & II & 2455664.3806 & 0.0005 &  I & 2455694.2680 & 0.0005 &  I & 2455725.0599 & 0.0005 &  I \\ 
2455628.1529 & 0.0005 &  I & 2455664.8344 & 0.0005 & II & 2455694.7206 & 0.0005 & II & 2455725.5131 & 0.0005 & II \\ 
2455628.6048 & 0.0005 & II & 2455665.2877 & 0.0005 &  I & 2455695.1733 & 0.0005 &  I & 2455725.9651 & 0.0005 &  I \\ 
2455629.0582 & 0.0005 &  I & 2455665.7389 & 0.0005 & II & 2455695.6255 & 0.0005 & II & 2455728.6829 & 0.0005 &  I \\ 
2455629.5099 & 0.0005 & II & 2455666.1933 & 0.0005 &  I & 2455696.0795 & 0.0005 &  I & 2455729.1358 & 0.0005 & II \\ 
2455629.9642 & 0.0005 &  I & 2455666.6453 & 0.0005 & II & 2455696.5309 & 0.0005 & II & 2455729.5884 & 0.0005 &  I \\ 
2455630.4169 & 0.0005 & II & 2455667.0988 & 0.0005 &  I & 2455696.9853 & 0.0005 &  I & 2455730.0397 & 0.0005 & II \\ 
2455630.8700 & 0.0005 &  I & 2455667.5509 & 0.0005 & II & 2455697.4391 & 0.0005 & II & 2455730.4938 & 0.0005 &  I \\ 
2455631.3218 & 0.0005 & II & 2455668.0040 & 0.0005 &  I & 2455697.8911 & 0.0005 &  I & 2455730.9468 & 0.0005 & II \\ 
2455631.7755 & 0.0005 &  I & 2455668.4563 & 0.0005 & II & 2455698.3443 & 0.0005 & II & 2455731.3993 & 0.0005 &  I \\ 
2455632.2287 & 0.0005 & II & 2455668.9087 & 0.0005 &  I & 2455698.7968 & 0.0005 &  I & 2455731.8516 & 0.0005 & II \\ 
2455632.6800 & 0.0005 &  I & 2455669.3622 & 0.0005 & II & 2455699.2494 & 0.0005 & II & 2455732.3060 & 0.0005 &  I \\ 
2455633.1334 & 0.0005 & II & 2455669.8148 & 0.0005 &  I & 2455699.7030 & 0.0005 &  I & 2455732.7564 & 0.0005 & II \\ 
2455633.5857 & 0.0005 &  I & 2455670.2678 & 0.0005 & II & 2455700.1542 & 0.0005 & II & 2455733.2114 & 0.0005 &  I \\ 
2455634.0386 & 0.0005 & II & 2455670.7202 & 0.0005 &  I & 2455700.6085 & 0.0005 &  I & 2455733.6647 & 0.0005 & II \\ 
2455634.4916 & 0.0005 &  I & 2455671.1723 & 0.0005 & II & 2455701.0623 & 0.0005 & II & 2455734.1170 & 0.0005 &  I \\ 
2455641.7369 & 0.0005 &  I & 2455671.6268 & 0.0005 &  I & 2455701.5145 & 0.0005 &  I & 2455734.5686 & 0.0005 & II \\ 
2455642.1894 & 0.0005 & II & 2455672.0785 & 0.0005 & II & 2455701.9651 & 0.0005 & II & 2455735.0218 & 0.0005 &  I \\ 
2455642.6431 & 0.0005 &  I & 2455672.5312 & 0.0005 &  I & 2455702.4201 & 0.0005 &  I & 2455735.4758 & 0.0005 & II \\ 
2455643.0961 & 0.0005 & II & 2455672.9824 & 0.0005 & II & 2455702.8724 & 0.0005 & II & 2455735.9284 & 0.0005 &  I \\ 
2455643.5483 & 0.0005 &  I & 2455673.4376 & 0.0005 &  I & 2455703.3256 & 0.0005 &  I & 2455736.3806 & 0.0005 & II \\ 
2455644.0018 & 0.0005 & II & 2455673.8893 & 0.0005 & II & 2455706.0422 & 0.0005 &  I & 2455737.2865 & 0.0005 & II \\ 
2455644.4544 & 0.0005 &  I & 2455674.3431 & 0.0005 &  I & 2455706.4951 & 0.0005 & II & 2455737.7402 & 0.0005 &  I \\ 
2455644.9086 & 0.0005 & II & 2455674.7954 & 0.0005 & II & 2455706.9482 & 0.0005 &  I & 2455738.1930 & 0.0005 & II \\ 
2455645.3605 & 0.0005 &  I & 2455675.2484 & 0.0005 &  I & 2455708.3071 & 0.0005 & II & 2455738.6457 & 0.0005 &  I \\ 
2455645.8139 & 0.0005 & II & 2455675.6992 & 0.0005 & II & 2455708.7601 & 0.0005 &  I &  &  & \\ 
2455646.2670 & 0.0005 &  I & 2455676.1537 & 0.0005 &  I & 2455709.2132 & 0.0005 & II &  &  & \\ 
2455646.7209 & 0.0005 & II & 2455676.6071 & 0.0005 & II & 2455709.6667 & 0.0005 &  I &  &  & \\ 
\hline 
\end{tabular}
\end{table*}

\section[]{Analysis of the Eclipse Timing Variations (ETV)}

\subsection[]{The close binary}

\begin{figure*}
\includegraphics[width=168mm]{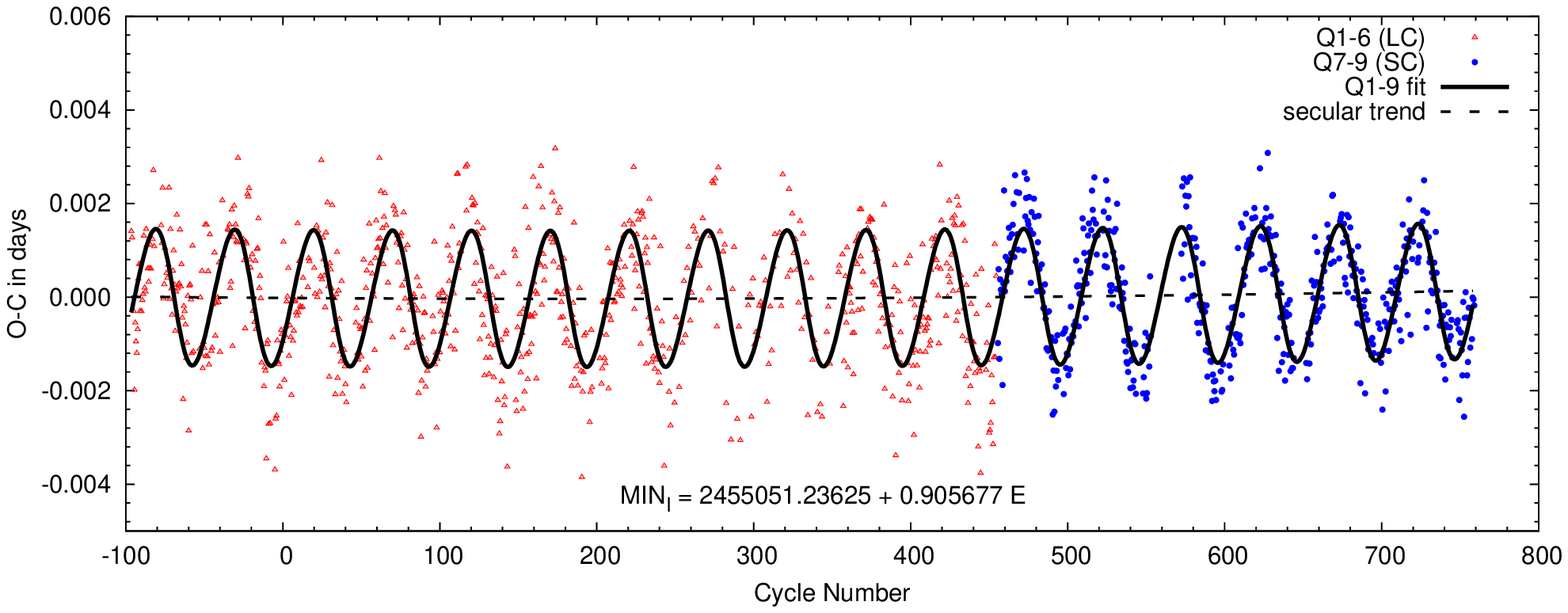}
 \caption{Eclipse timing variations in the shallow minima. Triangles and circles mark the LC and SC data, respectively. The solid line stands for the $Q1-Q9$ fit, while the dashed line represents the weak secular (parabolic) trend.}
 \label{Fig:O-Clinear}
\end{figure*}

In order to study the eclipse timing variations (ETV), the following linear ephemeris
was calculated for the shallow minima:
\begin{equation}
   MIN_\rmn{I-shallow}~\rmn{[BJD]}=2\,455\,051.23625+0\fd905677\times E ,
\end{equation}
where $E$ is the cycle number. The corresponding ETV diagram is plotted in Fig.~\ref{Fig:O-Clinear}.

We see a sinusoidal variation with a period identical to the eclipsing period of the wide system. There is also a smaller, long-term variation, that might either be part of a longer period variation, or represent a secular trend, as is the case with several  close binary systems. First we analyse the periodic behaviour of the ETV, and then the possible secular (parabolic) term will also be discussed.

\subsubsection[]{Short-period variations: General remarks}

\begin{figure}
\includegraphics[width=84mm]{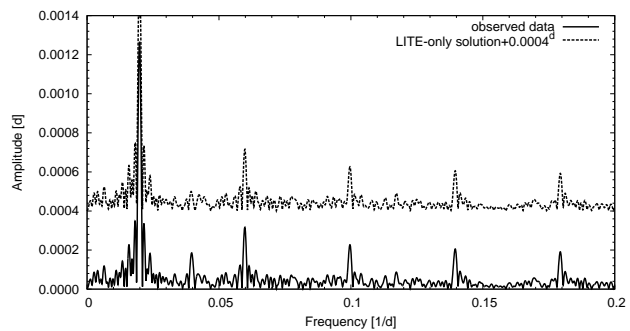}
 \caption{The DFT amplitude spectrum of the ETV curve (lower solid line). 
In order to illustrate the possible data-sampling origin of the odd harmonics
the spectrum of a similarly sampled sine function with $f_0$-frequency is also plotted (upper dashed line).}
 \label{Fig:DFTobs}
\end{figure}

To detect further periodicities, a discrete Fourier transform was calculated for the ETV curve. The resulting amplitude spectrum shows that the odd harmonics of the  fundamental frequency are also present (see Fig.~\ref{Fig:DFTobs}), while only the first even harmonic (i.e. $2f_0$) exists, and its amplitude is smaller than that of the $3f_0$ and $5f_0$ components. To check whether this structure is a consequence of the non-uniform sampling (i.e., the missing data during the deep eclipses, when the eclipse-events of the close pair cannot be observed, see Fig.~\ref{fig:ETVphase} below), we calculated a simple circular light-time orbit solution (i.e. we first fitted a sine curve with the fundamental frequency of the DFT spectrum). Sampling this solution at the locations (i.e. cycle numbers) of the observed data, and calculating the DFT spectrum of this dataset, we found that the two spectra have very similar structure (see Fig.~\ref{Fig:DFTobs}), confirming our conjecture that the odd peaks are a data-sampling effect. Consequently, we restrict our analysis on the main peak ($f_0$) and its second harmonic ($2f_0$).


\begin{table}
 \caption{Times of minima for the wide system}
 \label{tab:greatToM}
 \begin{tabular}{@{}lclc}
  \hline
  BJD & Cycle number$^a$ &  BJD & Cycle number$^a$ \\
  \hline
2454977.0831 & $-11.5$ & 2455363.5693 &  $-3.0$ \\
2455022.5375 & $-10.5$ & 2455386.3163 &  $-2.5$ \\
2455045.2970 & $-10.0$ & 2455409.0662 &  $-2.0$ \\
2455068.0335 &  $-9.5$ & 2455431.7818 &  $-1.5$ \\
2455113.5169 &  $-8.5$ & 2455454.5345 &  $-1.0$ \\
2455136.2170 &  $-8.0$ & 2455477.2681 &  $-0.5$ \\
2455158.9550 &  $-7.5$ & 2455499.9950 &   $0.0$ \\
2455204.4405 &  $-6.5$ & 2455545.4559 &   $1.0$ \\
2455227.1669 &  $-6.0$ & 2455590.9390 &   $2.0$ \\
2455249.9048 &  $-5.5$ & 2455613.6734 &   $2.5$ \\
2455272.6355 &  $-5.0$ & 2455659.1425 &   $3.5$ \\
2455295.3893 &  $-4.5$ & 2455681.8955 &   $4.0$ \\
2455318.1113 &  $-4.0$ & 2455704.6063 &   $4.5$ \\
2455340.8384 &  $-3.5$ & 2455727.3559 &   $5.0$ \\
\hline
 \end{tabular}

$^a$: half-integer values refer to secondary minima

\end{table}

%

Considering the fundamental term, it is clear that its main source should be the gravitational interaction between the inner, close binary, and the wider, more massive giant star. This interaction has at least two consequences: $(i)$ the geometrical light-time effect (LITE), and $(ii)$ a dynamical effect, due to the gravitational perturbations of the third body on the close, inner binary. In the case of LITE, the amplitude of the effect increases with the separation, as seen in dozens of systems \citep[see e.~g.][for most recent examples]{qianetal12,popvamos12}. Conversely, the amplitudes of the dynamical terms scale with ($P_1^2/P_2$) which, due to various observational biases, makes this phenomenon difficult to detect with traditional ground-based observations. A detailed analysis of this topic can be found in \citet{borkovitsetal03,borkovitsetal11}. To our knowledge, the only system in which the dynamical effect was clearly detected by classical ground-based, small-aperture photometric observations, is IU~Aurigae \citep{mayer90,ozdemiretal03}. Nevertheless, for compact systems like the recently discovered KOI-126 \citep{carteretal11}, KOI-928 \citep{steffenetal11}, the amplitude ratio may be reversed, as it was clearly shown for KOI-928 by \citet{steffenetal11}.

For HD~181068, we first consider the LITE contribution. Its shape and amplitude are:
\begin{equation}
ETV_\rmn{LITE}=\frac{a_\rmn{B}\sin i_2}{c}\frac{\left(1-e_2^2\right)\sin u_\rmn{B} }{1+e_2\cos v_2}, 
\end{equation}
\begin{equation}
A_\rmn{LITE}\approx1\fd1\times10^{-4}\frac{m_\rmn{A}}{m_\rmn{AB}^{2/3}}\sin i_2P_2^{2/3}\left(1-e^2_2\cos^2\omega_\rmn{B}\right)^{1/2},
\label{Eq:A_LITE}
\end{equation}
where $a_\rmn{B}$, $i_2$, $e_2$, $\omega_\rmn{B}$, $P_2$ are the semi-major axis, inclination, eccentricity, argument of periastron, and period of the binary's orbit around the common centre of mass of the triple system. Furthermore, $v_2$ is the true anomaly of the eclipsing pair in this orbit, $u_\rmn{B}=v_2+\omega_\rmn{B}$ is its true longitude measured from the intersection of the orbital plane and the plane of the sky, and $c$ is the speed of light. (Inclination, eccentricity, period and true anomaly are simply given subscript $2$, because their values are identical to those of the relative wider orbit, traditionally centered on the inner binary.) Note also that in Eq.~(\ref{Eq:A_LITE}) masses should be given in solar masses, while period in days. Substituting the values found by \citet{der11} (i.e., $m_\rmn{A}\approx3\rmn{M}_\odot$, $m_{AB}\approx4.6\rmn{M}_\odot$,
$i_2\approx87\fdg6$, $P_2\approx45\fd5$, $e_2=0$), we get
\begin{equation}
A_\rmn{LITE}\approx1.5\times10^{-3}\mathrm{~d},
\end{equation}
or $\sim2.4$ minutes. 

Now, considering the dynamical perturbation term, whose amplitude should be proportional to 
\begin{equation}
A_\rmn{dyn}\sim\frac{1}{2\pi}\frac{m_\rmn{A}}{m_\rmn{AB}}\frac{P_1^2}{P_2}\frac{\left(1-e_1^2\right)^{1/2}}{\left(1-e_2^2\right)^{3/2}},
\end{equation}
\citep{borkovitsetal11}. For the present system this results in
\begin{equation}
A_\rmn{dyn}\sim1.9\times10^{-3}\mathrm{~d},
\end{equation}
which is similar to the LITE. However, as we now point out, a more detailed analysis shows that the ETV curve should be LITE-dominated. Although the harmonics of the fundamental frequency could arise from the eccentricity of one of the orbits, there is strong evidence from the radial velocity solution of \citep{der11} that both orbits are circular, which is further supported by the locations and shapes of the secondary minima with respect to the primary minima in both the close and wide orbits (see next Section). 

Accepting that both orbits are nearly (or exactly) circular, the LITE contribution is restricted to the fundamental term, and there is no dynamical addition to this term. In this situation, the only dynamical terms that can give non-vanishing contributions are as follows:
\begin{eqnarray}
ETV_\rmn{dyn}&=&\frac{3}{8\pi}\frac{m_\rmn{A}}{m_\rmn{AB}}\frac{P_1^2}{P_2}\left\{\sin^2i_\rmn{m}\sin2(u_2-u_{\rmn{m}2})\right. \nonumber \\
&&+\frac{1}{2}\cot i_1\sin i_\rmn{m}\left[\sin u_{\rmn{m}1}\cos2(u_2-u_{\rmn{m}2})\right. \nonumber \\
&&\left.+\cos i_\mathrm{m}\cos u_{\rmn{m}1}\sin2(u_2-u_{\rmn{m}2})\right]\big\}
\end{eqnarray}
(see Eq.~(46)\footnote{We corrected here the erroneous negative sign in the nodal term (i.e. in front of $\frac{1}{2}\cot i_1$).} \citealt{borkovitsetal03}). As before, indices 1 and 2 refer to the elements of the close and wide relative orbits, respectively. Furthermore, $i_\rmn{m}$ denotes the mutual inclination of the two orbital planes, while $u_\rmn{m1}$ and $u_\rmn{m2}$ stand for the angular distances of the intersection of the two orbits from the plane of the sky, measured on the respective planes (see Fig.~\ref{fig:krsz} in Appendix~\ref{AppA}). We see that in the case of coplanarity, all these terms vanish due to $\sin i_\rmn{m}=0$. For the present situation, the second and third terms, arising from nodal regression (the precession of the orbital plane of the close pair) can also be simply omitted independently from the mutual inclination, due to the almost edge-on view of the orbital plane, as $\frac{1}{2}\cot i_1=\frac{1}{2}\cot 87\fdg6\approx0.02$.

As a consequence, we are in a very fortunate situation. Provided we accept that the $45.5$-day-period sinusoidal ETV is caused by the above described geometrical and dynamical effects, the signals of the two phenomena could very easily be disentangled. Firstly, the amplitude of the $P_2$-period component gives information about the physical dimensions of the close binary's orbit around the centre of mass of the triple system. Combining this result with radial velocity measurements of the giant companion makes it possible to determine the masses $m_\rmn{A}$ and $m_\rmn{AB}$ (as a function of the photometrically known $\sin i_2$), in a similar manner to a double-lined spectroscopic binary (SB2). Secondly, the $\frac{1}{2}P_2$-period term makes it possible to determine the relative (or mutual) inclination of the two orbits, i.e. the spatial configuration of the triplet. 

Taking into account the above considerations, the ETV analysis was carried out as follows. First, a general linear least-squares method was applied to search for the best fit in the following form:
\begin{equation}
f(E)=c_0+c_1E+c_2E^2+\sum_{j=1}^2\left(a_j\sin j\omega E+b_j\cos j\omega E\right),
\end{equation}
where the frequency was taken from the DFT analysis, and was held fixed. Note that its physical meaning is $\omega=2\pi\frac{P_{\rmn{e}1}}{P_{\rmn{e}2}}$, where $P_{\rmn{e}1}$ and $P_{\rmn{e}2}$ stand for the eclipsing periods of the close and wide binaries. These quantities, strictly speaking, are neither equal to the anomalistic periods $P_1$ and $P_2$ (which appear in the amplitudes of the dynamical terms) [e.~g. for $\gamma$ systematic velocity $P_{\rmn{e}i}=P_i\left(1+\frac{\gamma}{c}\right)$], nor necessarily constant, especially when $c_2\neq0$. Nevertheless, for our purposes, these differences are not significant. 

We carried out two fitting procedures: one for the complete data series, and another only for short-cadence $Q7-Q9$ data. Instead of estimating and using individual measurement errors for each data points, we applied a simple weighting scheme. Namely, weights $\sigma_i=0\fd0005$ and $\sigma_i=0\fd001$, estimated from the eclipse time determination procedure, were chosen for short-cadence and long-cadence minima, respectively. After a preliminary fit, points above the $3\sigma$ limit were removed, and the procedure was reiterated. We list our results from the two data sets in Table~\ref{tab: ETV-lsq}, while the corresponding fitted curves are shown in Fig.~\ref{Fig:O-Clinear}. We also show the phased graph in Fig.~\ref{fig:ETVphase}. The polynomial terms (i.e., $\sum c_iE^i$) were subtracted from this latter curve. In Table~\ref{tab: ETV-lsq}, along with the direct output of the least-squares fits, the derived physical and geometrical quantities, and their standard errors are also tabulated. 

Before analysing the individual Fourier-contributions, we should stress, however, that there is a discrepancy of about 0.05 days between the wide-orbit's period obtained here from the LITE solution and the one determined from the deep eclipses directly (see later in Sect.~\ref{subsect:widesystem}). This is quite significant, as during the measured 17 cycle-long interval it would result in a shift of about $0.85$ days in the occurrence of the eclipse events. Our light curve solution (Sect.~\ref{sect:lcanalysis}) clearly shows that the correct period is the one obtained from the deep minima times in Sect.~\ref{subsect:widesystem}, and not the present one. The origin of this discrepancy is unclear. It might be caused by the observations of shallow minima being absent around the extrema of the LITE-orbit. A firm resolution will require further investigations on a longer time interval. Fortunately, this period difference is too small to influence the analysis of the Fourier terms described below.

\begin{figure}
\includegraphics[width=84mm]{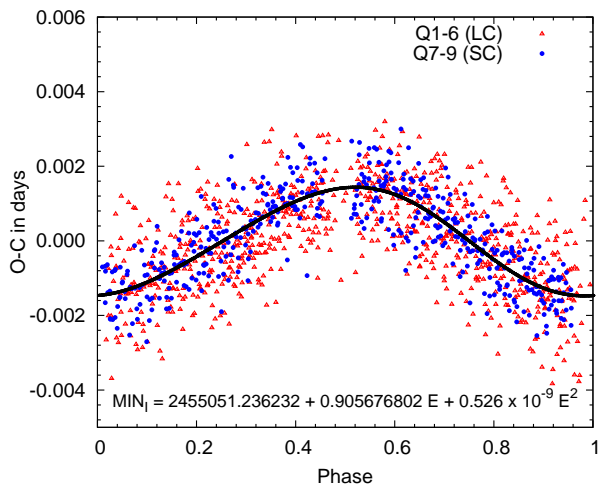}
 \caption{The phased ETV curve together with the best linear lsq-fit for the $Q1-Q9$ data. Note that the quadratic term has been subtracted.}
 \label{fig:ETVphase}
\end{figure}

\subsubsection[]{Short-period variations: light-time effect}

Considering the light-time contribution first, its most important output is the physical size of the light-time orbit of component $B$ (at least as a function of inclination $i_2$). Together with the semi-major axis of component $A$'s orbit (obtained from radial velocity measurements), this yields the physical masses of the wide binary (i.e., the mass of the giant component and the total mass of the close binary). Note that, as one can see in Table~\ref{tab: ETV-lsq}, the ratio $m_A/m_{AB}$ has a significantly lower standard error than the masses individually and, furthermore, it does not depend on the inclination $i_2$. Nevertheless, there is clearly a significant discrepancy between the mass ratios and the masses derived from the two solutions. The mass ratio depends strongly on the amplitude of the LITE term. However, the mass of the giant component resulting from the pure, better-quality short-cadence data accurately confirms the value derived from previous results and astrophysical estimations of \citet{der11}. Consequently, in the followings we adopt this second ($Q7-Q9$ SC-data only) solution.

The second parameter coming from the LITE term, the phase information, is less useful, but we may use it for an indirect checking of the accuracy of our solution. This value estimates a primary eclipsing mid-minimum (i.e. mid-transit of the small binary in front of the giant component) at BJD $55045.21\pm0\fd16$. By the use of the direct ETV-determined ephemeris of the wide binary (see Sect.~\ref{subsect:widesystem}) we measure phase $\phi=0\fp998$ for this event, i.~e., the $\phi=0$ phase occurred at BJD $55045.28762$, which is clearly within the formal error. 

\subsubsection[]{Short-period variations: dynamical effects}

Now we turn to the dynamical term. The corresponding Fourier coefficients ($a_2$, $b_2$) are almost two orders of magnitude smaller than those of the LITE terms, and they are close to the standard errors. Consequently, the following results should be considered with great caution. From the amplitude we get $\sin^2i_\rmn{m}\approx0.05$, which is large enough to marginally verify the omission of the nodal contribution, but not large enough to give a numerically trustable output. From this result we obtain two different values for the relative inclination. However, as will be shown in the Discussion, we can rule out the retrograde orientation photometrically. Therefore, the corresponding angles are calculated only for prograde relative orbits. By combining the mutual inclination, the phase term ($u_\rmn{m2}$) and the visible inclination ($i_2$) -- the latter being known from the light curve solution -- we can calculate the complete 3D orbit of the triple system. In Table~\ref{tab: ETV-lsq} we also give the difference of the longitudes of the nodes ($\Delta\Omega$) on the sky, as well as the visible inclination $i_1$ of the close system. Since $i_1$ is also known from the light curve solution, this result might help to resolve the $\Omega$ ambiguity, and also serves as an accuracy check for our solution. 

Both solutions seem to indicate a significant $\left(13\degr-15\degr\right)$ misalingnment between the two orbital planes. If this fact were real, a precession of the two orbital planes would occur around the invariable plane of the triple system. It can be shown \citep[see~e.~g.][]{soderhjelm75,borkovitsetal07}, that the orbital inlination of the close binary would then vary cyclically with an amplitude of $28\degr-30\degr$ on a time-scale of $13-14$ years. Furthermore, the fact that the phase term $u_\rmn{m2}$ is close to $90\degr$ or $270\degr$ (i.~e. the observable inclinations ($i_1$ and $i_2$) have very similar numerical values) shows that this hypothetical effect would produce the fastest $i_1$ variations at the present epoch. This means that during the $Q1-Q9$ observational interval we should have observed more than $10\degr$ variation in the visible inclination ($i_1$) of the close pair. This variation would have resulted in significant changes in the eclipse depths of the shallow minima. However, according to our analysis (next Section) there is no sign of any eclipse-depth variations in the close system, and so we have to exclude this possibility. Consequently, the presence of the first harmonic in the DFT-spectrum cannot be explained by the non-coplanarity of the orbits.

Having ruled out both the eccentricity of the orbit(s) and the noncoplanarity of the orbital planes, we examined further possibilities by considering the effects of higher-order dynamical terms. Although all the dynamical terms considered e.~g. by \citet{borkovitsetal03,borkovitsetal11} and \citet{agoletal05} disappear for coplanar and circular orbits, this happens only within the frame of the applied approximation. The octuple and higher-order terms of the perturbation function cause non-vanishing contributions even in this case, as it was shown e.~g. by \citet{soderhjelm84,fordetal00}. In order to check the magnitude of such forces, we integrated the motion numerically and calculated the simulated times of minima. In our integration both the Newtonian point-mass and the non-dissipative tidal terms were included. The applied numeric integrator was described in \citet{borkovitsetal04}. An analysis of the DFT spectrum of this higher-order, numerically-generated (and evenly sampled) ETV curve revealed the presence of the first few harmonics of the orbital periods at a 90\% significance level. As the amplitudes of these peaks are lower by approximately two magnitudes than that of the questionable first harmonic in the observed curve, we can conclude that  these higher-order effects are also insufficient to explain the structure of the Fourier space. Therefore, we cannot currently give any plausible dynamically originated explanation for the $P_2/2$-period term in the ETV. 

\begin{table}
 \caption{Fitted and derived parameters (and their formal errors in the last digits) from the general linear least-squares fit to the ETV curve.}
 \label{tab: ETV-lsq}
 \begin{tabular}{@{}lcc}
  \hline
  Parameter & $Q1-Q9$ & $Q7-Q9$ \\
  \hline
  $f_0\left(=\frac{P_\mathrm{1e}}{P_\mathrm{2e}}\right)$ & \multicolumn{2}{c}{$0.019897(2)$} \\
  $c_0$ & $-0.000018(51)$ & $0.004275(1206)$ \\
  $c_1$ & $-0.0000002(3)$ & $-0.0000125(40)$ \\
  $c_2$ & $0.5(4)\times10^{-9}$ & $10.8(33)\times10^{-9}$ \\
  $a_1$ & $0.001040(28)$ & $0.001128(34)$ \\
  $b_1$ & $-0.001004(27)$ & $-0.001028(33)$ \\
  $a_2$ & $-0.000001(29)$ & $0.000006(35)$ \\
  $b_2$ & $0.000089(26)$ & $0.000067(31)$ \\
  \hline
  $T_\rmn{Bab-primin}$ [BJD]& $55051.236232(51)$ & $55051.240526(1206)$ \\
  $P_\rmn{1e}$ [day] & $0.9056768(3)$ & $0.9056645(40)$ \\
  $\Delta P_\rmn{1e}$ [day/cycle] & $1.1(8)\times10^{-9}$ & $21.6(66)\times10^{-9}$ \\
  \hline
  $P_\rmn{2e}$ [day] & $45.518(4)$ & $45.517(5)$ \\
  $a_\rmn{B}\sin i_2$ [R$_\odot$]& $54(1)$ & $57(1)$ \\
  $\left(u_\rmn{AB}\right)_0$ [$\degr$] & $-44(1)$ & $-42(1)$\\
  $T_\rmn{AB-primin}$ [BJD] & $55045.4(1)$ & $55045.2(2)$ \\
  \hline
  $a_\rmn{A}\sin i_2^a$ [R$_\odot$]& \multicolumn{2}{c}{33.43(5)} \\
  $m_\rmn{A}/m_\rmn{AB}$ & $0.617(5)$ & $0.629(5)$ \\
  $m_\rmn{AB}\sin^3i_2$ [M$_\odot$]  & $4.30(15)$ & $4.76(20)$ \\
  $m_\rmn{A}\sin^3i_2$ [M$_\odot$]  & $2.65(10)$ & $3.00(13)$ \\
  \hline
  $m_\rmn{A}/m_\rmn{AB}\sin^2i_\rmn{m}$ & $0.042(12)$ & $0.031(14)$ \\ 
  $i_\rmn{m}^b$ [$\degr$]& $15(2)$ & $13(3)$ \\
  $u_\rmn{m2}$ [$\degr$]& $91(9)$ or $271(9)$ & $95(15)$ or $275(15)$ \\
  \hline
  $i_2^c$ [$\degr$] & \multicolumn{2}{c}{87.7} \\
  $i_1^{b,d}$ [$\degr$] & $88(2)$ or $88(2)$ & $87(3)$ or $89(3)$ \\
  $\Delta\Omega^{b,c}$ [$\degr$]& $15(2)$ or $-15(2)$ & $13(3)$ or $-13(3)$\\
  \hline
 \end{tabular}

$^a$: taken from \citet{der11}; \\
$^b$: $180\degr-i_\rmn{m}, 180\degr-i_1, 180\degr-\Delta\Omega$ give equivalent solutions; \\
$^c$: fixed from the light curve solution; \\
$^d$: The second values are valid for $u_\rmn{m2}+180\degr$.

\end{table}

\subsubsection{Secular variations}
\label{subsubsect:secularvariation}

As mentioned above, the ETV curve shows weak evidence for continuous orbital period changes with a contant rate during the whole observational interval. In order to investigate this feature, we consider the $Q1-Q9$ dataset with longer time coverage, instead of the previously used $Q7-Q9$ SC data. The quadratic ephemeris, calculated from this solution, for the shallow minima is
\begin{equation}
MIN_I=2\,455\,051.23623(5)+0.9056768(3)E+0.5(4)\times10^{-9}E^2,
\label{eq:smallO-Cquad}
\end{equation} 
from which the rate of the constant period change is found to be
\begin{equation}
\frac{\Delta P}{P}\sim\dot P=2\frac{c_2}{c_1^2}\sim0.038\mathrm{~s/yr}.
\label{Eq:DeltaP/P}
\end{equation}

The origin of this variation is not clear. As we mentioned, any orbital precession can be ruled out due to the almost exact coplanarity. Due to the detached system geometry, none mass loss, mass exchange or magnetic cycles can be considered, as a reason. Gravitational effects induced by an additional, more distant and faint companion, could be responsible. Moreover, some interaction (e.g. tidal, magnetic or other) with the giant component might also be the source of this phenomenon. Further observations and investigations are needed to clarify the origin of the secular variations. 

\subsection[]{The wide system}
\label{subsect:widesystem}

For the deep minima the following linear ephemeris was found by a linear least-squares fit:
\begin{equation}
MIN_I \mathrm{~[BJD]~}=2\,455\,499.9962(4) + 45\fd4711(2)\times E.
\label{Eq:greatO-C}
\end{equation}
%
%
Due to the coverage of 17 orbital cycles only, and a large scatter of about $0.03$ days, no periodic or secular trend can be identified in the ETV curve. The relatively large scatter may arise from the irregular, intrinsic variations of the chromospherically active giant component. As it was shown by \citet{kalimerisetal02}, starspots can alter the measured mid-minimum times by $\sim0.01$ days. Evidence for starspots (and even of eclipses of spotted regions) will be given in the Discussion. Therefore, we conclude that during the 2.1 year-long observed time interval, the period of the outer orbit remained constant. 

\section[]{Light curve analysis}
\label{sect:lcanalysis}

\begin{figure*}
\includegraphics[width=55mm]{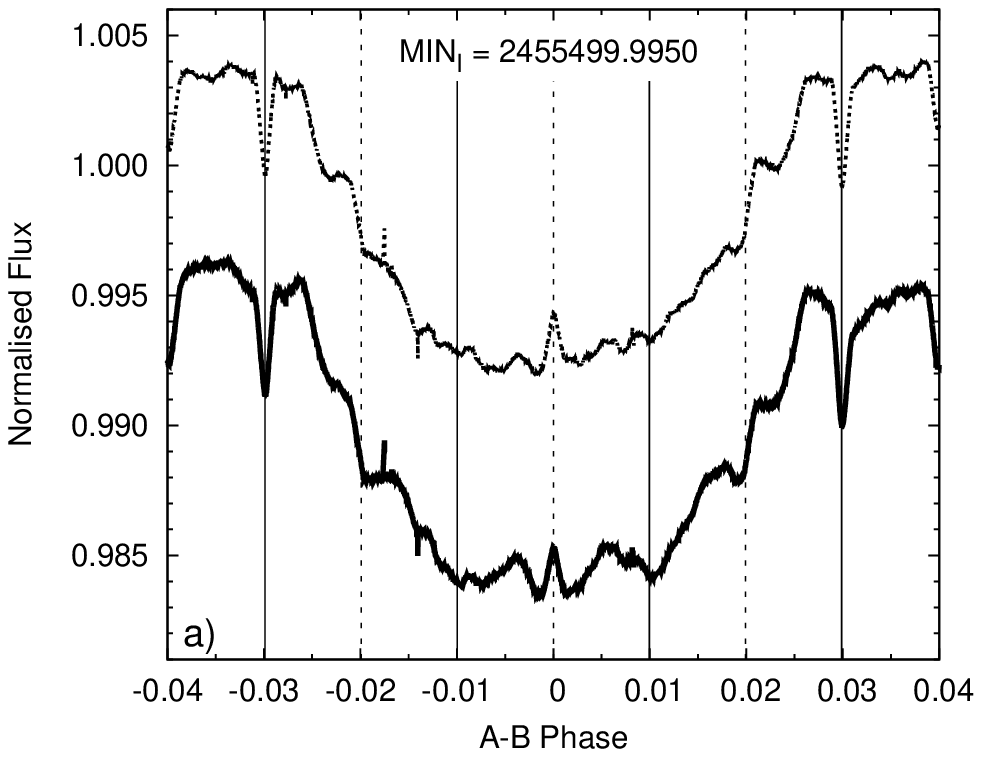}\includegraphics[width=55mm]{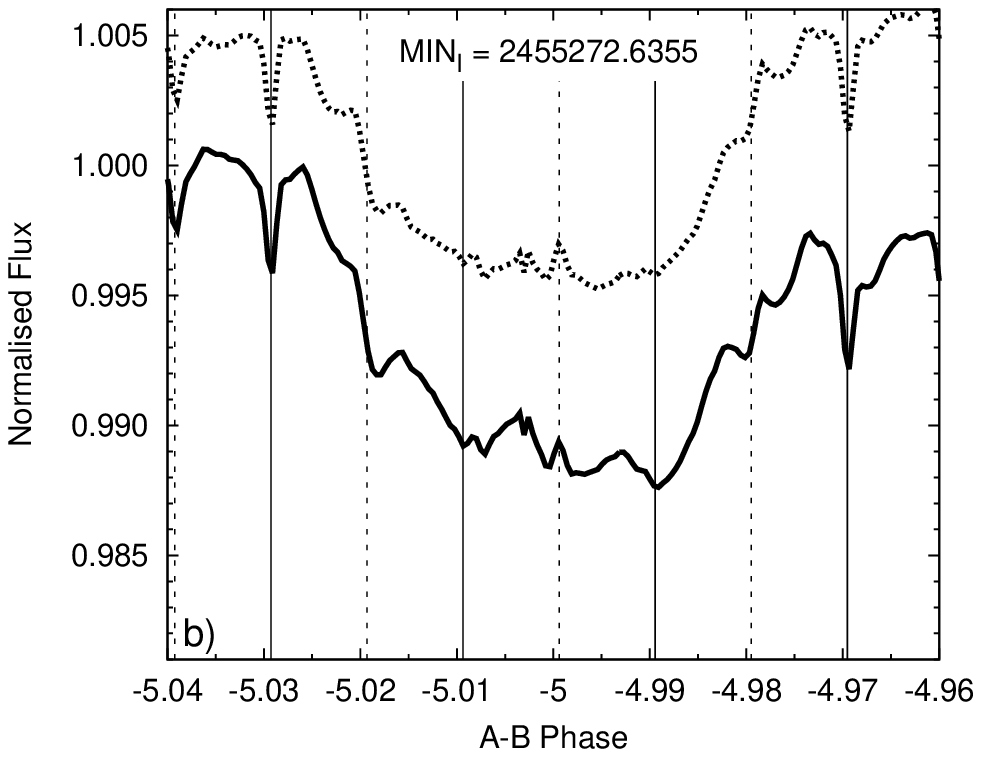}\includegraphics[width=55mm]{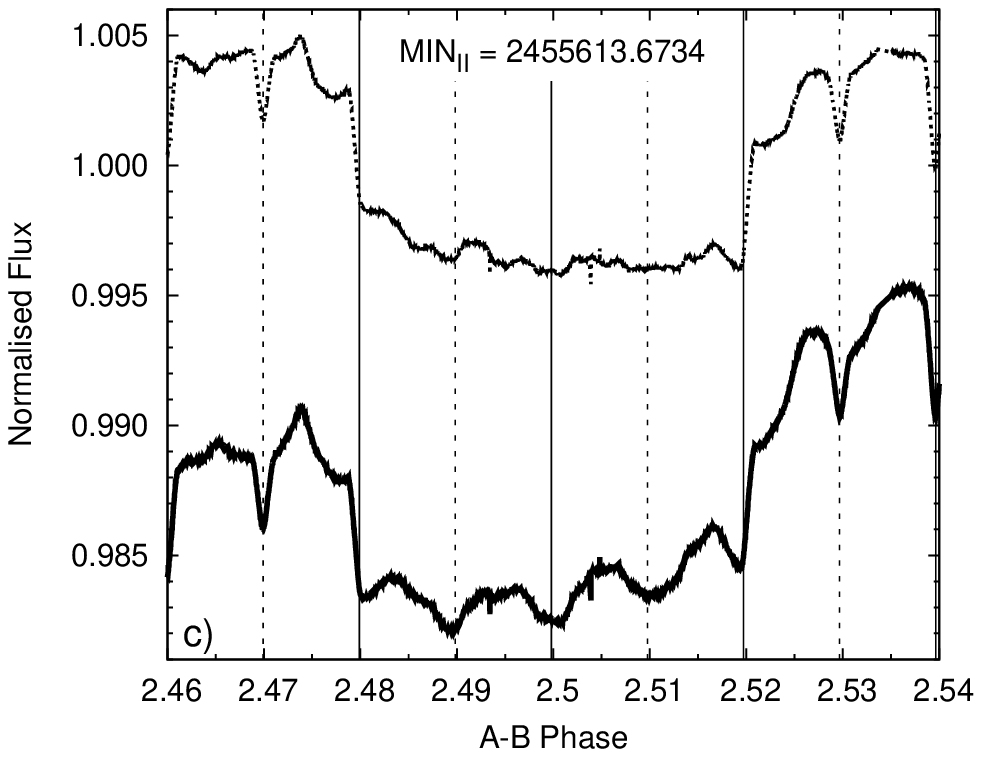}
\includegraphics[width=55mm]{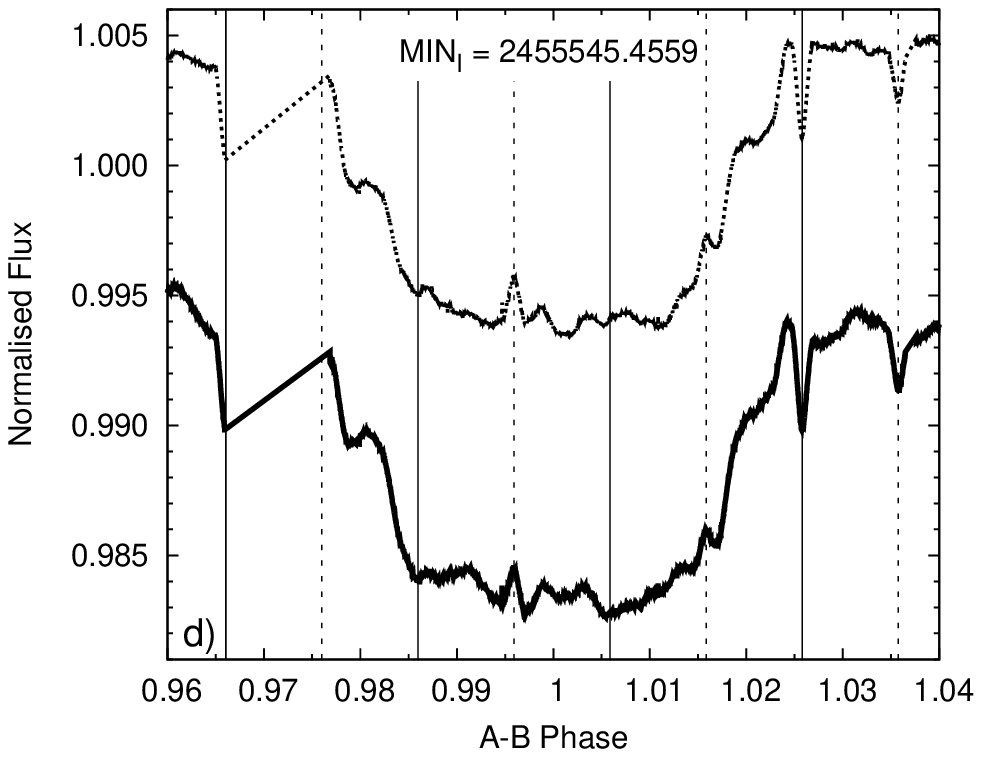}\includegraphics[width=55mm]{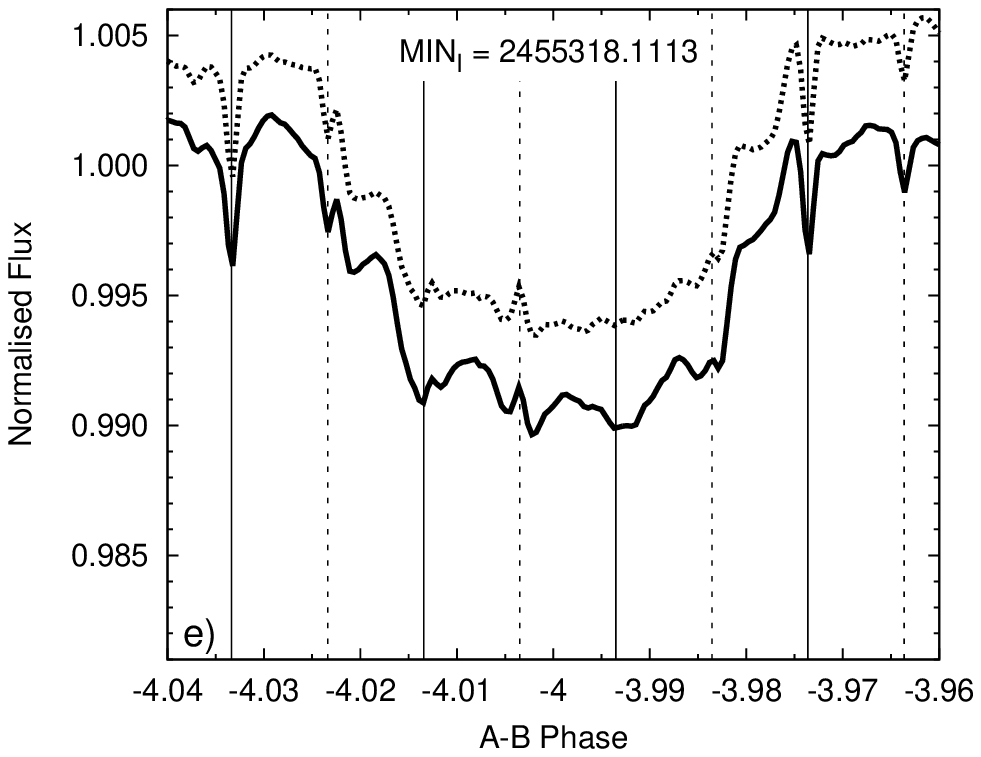}\includegraphics[width=55mm]{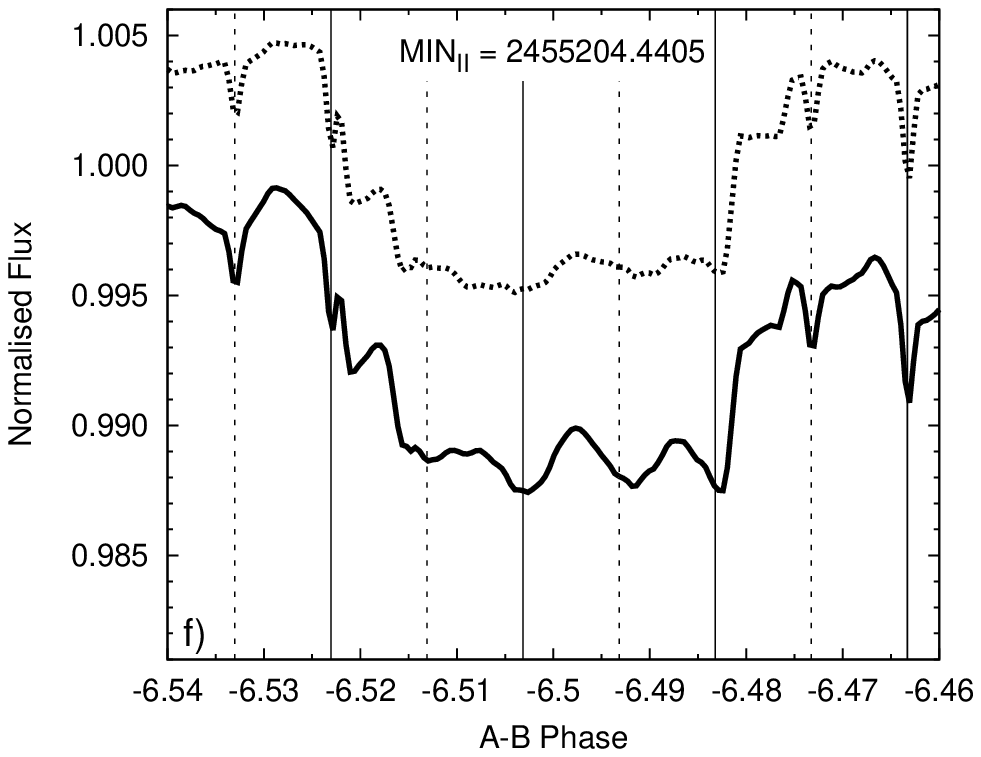}
\includegraphics[width=55mm]{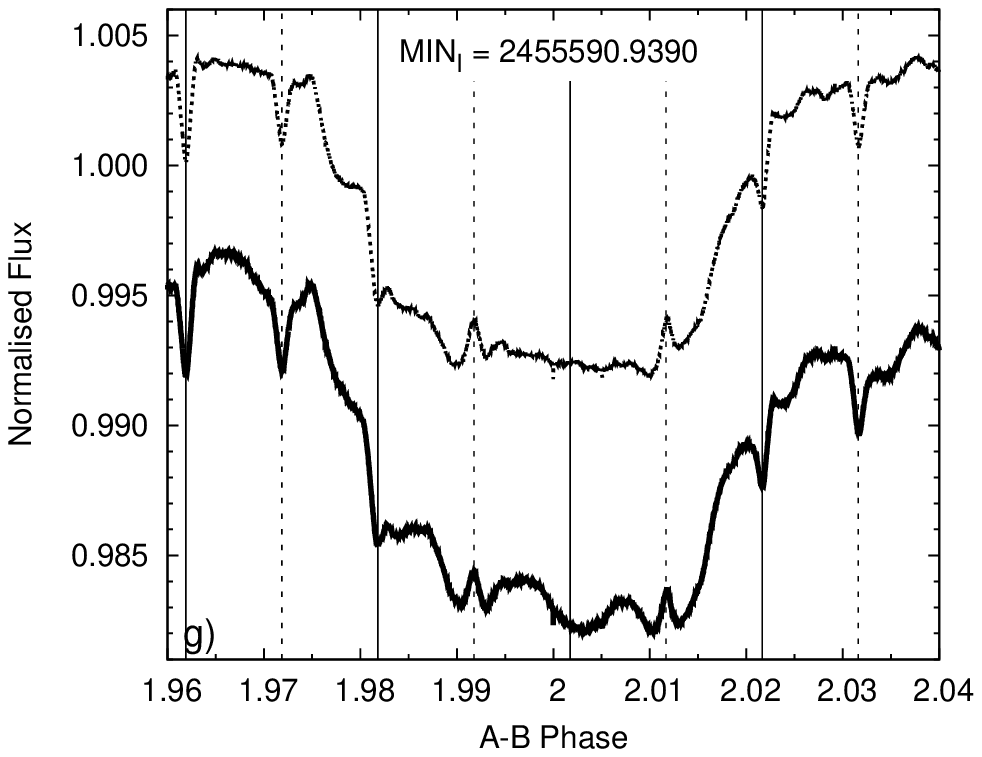}\includegraphics[width=55mm]{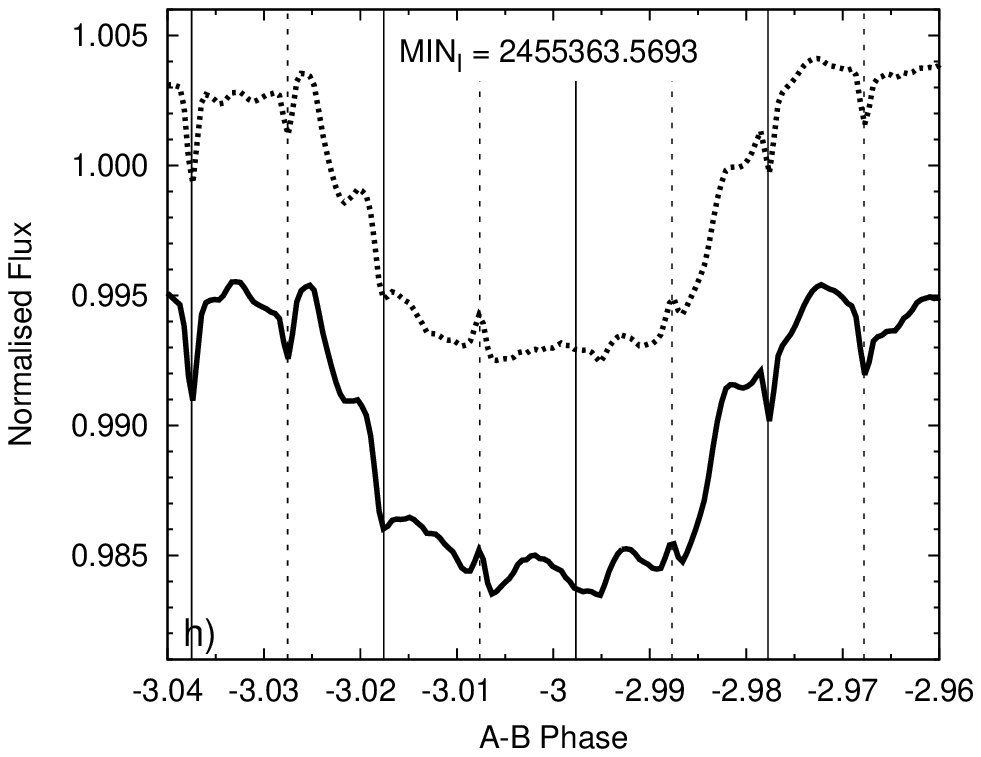}\includegraphics[width=55mm]{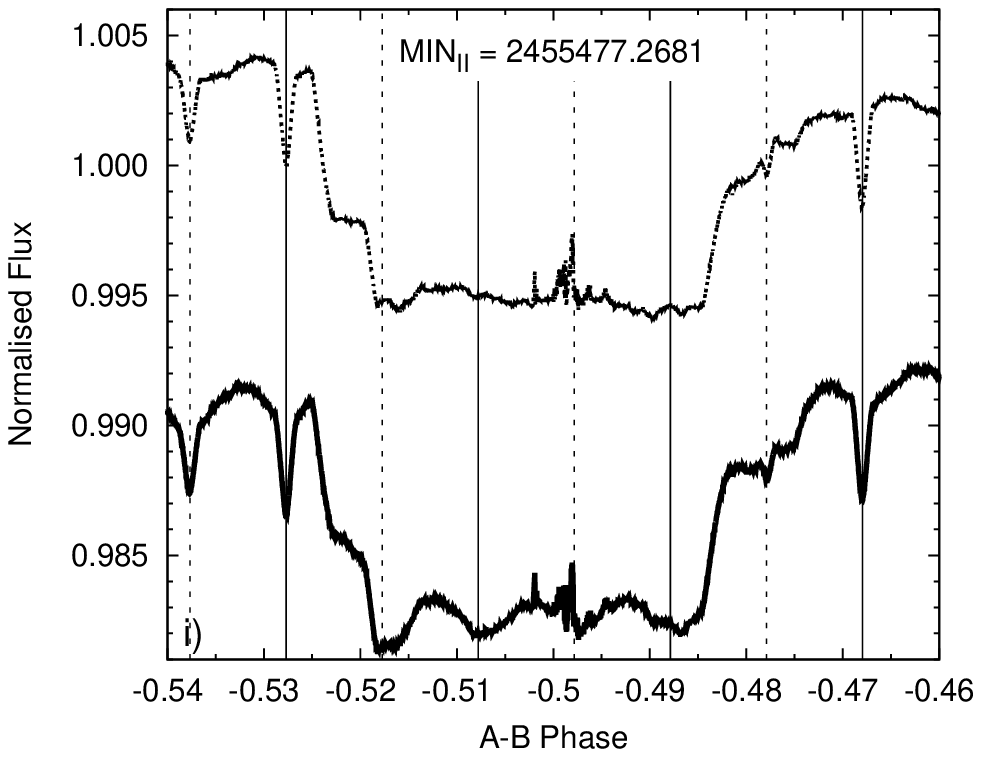}
\caption{Examples for three of the five ``families'' of the outer eclipses. Solid curves are the raw (uncorrected) flux curves, while dotted ones are corrected for the intrinsic variations. Solid and dashed vertical lines denote the small primary and secondary mid-minima, respectively. Note the flatness of the bottom of the primary minimum-curves in panel $b$ (especially with respect to its counterpart in panel $a$), which might be the consequence of a transit in front of a spotted region.}
\label{fig:eclipsecycles}
\end{figure*}

\subsection{Light curve characteristics}
\label{subsect:lcchar}

\begin{figure}
\includegraphics[width=84mm]{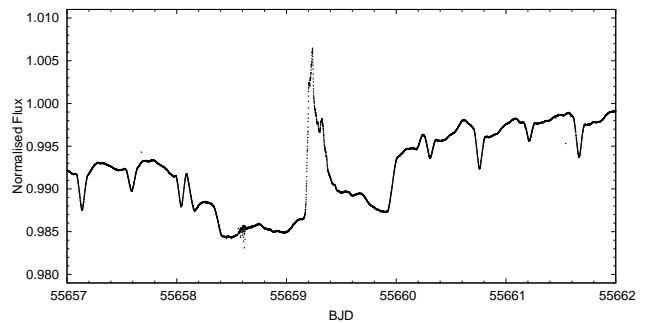}
 \caption{A possible flare event at BJD 55659 (in Q9) within a secondary deep minimum. The location and the amplitude of the eruption demonstrate clearly, that if it is a real flare event, it must have occurred on the giant component.}
 \label{Fig:q9flare}
\end{figure}

The light curve of HD~181086 has at least five different components:
\begin{itemize}
\item[(i)-(ii)] The eclipsing features of both the close inner ($Ba-Bb$), and the wide outer ($A-B$) binary subsystems. This category includes not only the eclipses themselves, but also other effects coming from the close binarity, i.~e., the ellipsoidal variations arising mainly from the tidally distorted shape of the giant component $A$. As we will show below, relativistic Doppler-beaming also produces a contribution. The reflection effect occurs in the close binary, but is negligible for the wide system \citep[c.~f.][]{zuckeretal07}. The characteristic time-scales of these variations are equal to the observed eclipsing periods $P_1$, $P_2$ of the two subsystems. Note that the period ratio is almost exactly $P_1:P_2=5:251$, hence, in every fifth revolution on the wide orbit, the shallow eclipses occur at approximately the same orbital phases of the wide system. Since the shape and the duration of the deep eclipses are remarkably altered by the varying positions of the close binary members, this resonance naturally defines five different deep eclipse patterns (or eclipse families, which are analoguous to the Saros cycles). Furthermore, considering two consecutive deep primary eclipses of a given ``family'' (which occur at cycle numbers $E=n$ and $E=n+5$, respectively), the intervening deep secondary eclipse of the same ``family'' (located at $E=n+2.5$) has a similar egress and ingress pattern, but with a $0.5$ close-orbital phase shift, i.~e. with an interchange between the shallow primary and secondary minima. In Fig.~\ref{fig:eclipsecycles} we plotted some typical members (both primary and secondary) of three of the five ``families''.
\item[(iii)] The strictly periodic and regular light curve variations are strongly altered and distorted by irregular or semi-regular brightness changes with more or less similar amplitudes. This feature may come from the intrinsic variations of the giant primary, and suggests that this star is a chromospherically active object. Some evidence for large spots can be seen in the different depths and shapes of primary deep minima (compare Fig.~\ref{fig:eclipsecycles}$a$ and Fig.~\ref{fig:eclipsecycles}$b$): when the close binary transits across a darker region, the minimum is shallower. The irregular variation seems to be continuous, showing certain quasi-periodicities on a 1--2 month time-scale, and could have some connection with the orbital and/or rotational periods of the giant component.
\item[(iv)] There are further, small amplitude oscillations in the light curve with the half of the sinodic period of the close system with respect to the giant, which strongly indicates a tidal origin.
\item[(v)] Finally, flare events were also observed during some of the observational runs. If these transients have their origins in HD~181086 then, at least in one case, we can be sure that it comes from the giant component, since the flare event at BJD 2\,455\,659 (in Q9) occurred during the secondary minimum of the wide system, i.~e., when the close pair was totally occulted (Fig.~\ref{Fig:q9flare}).
\end{itemize}

In the present analyis, we mainly focus on the eclipsing features [$(i)-(ii)$] of the light curve. As mentioned above, the presence of mutual eclipses in both subsystems makes it possible (at least theoretically) to infer some additional, otherwise unobtainable, physical and geometrical parameters from the light curve solution. For example, both the fine structure and the variable length of the ingress and egress phases of the deep minima reveal information on the mutual inclination of the two subsystems in such a way that even the usual $i$, $180\degr-i$ ambiguity can be resolved, i.~e. we can decide whether the revolutions of the two subsystems are prograde or retrograde relative to each other. Furthermore, the combination of the shallow and deep eclipses gives an independent solution for the photometric mass-ratio in both the close and in the wide systems. (In Appendix~\ref{AppA}, some examples are given for mining the extra information coded into the mutual eclipse geometry.)

\begin{figure*}
\includegraphics[width=84mm]{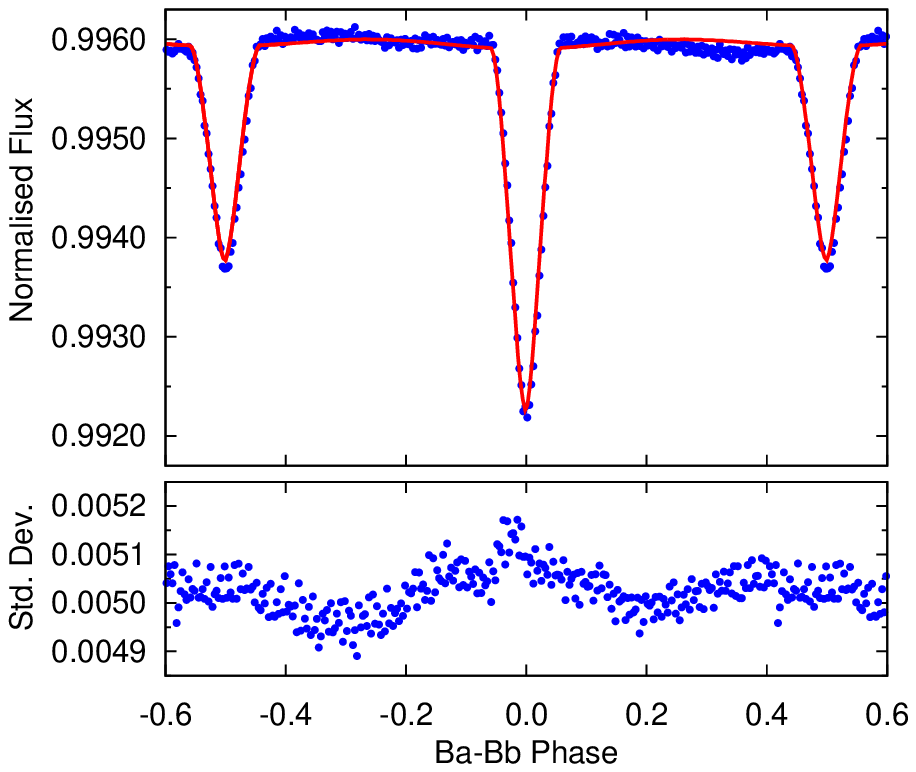}\includegraphics[width=84mm]{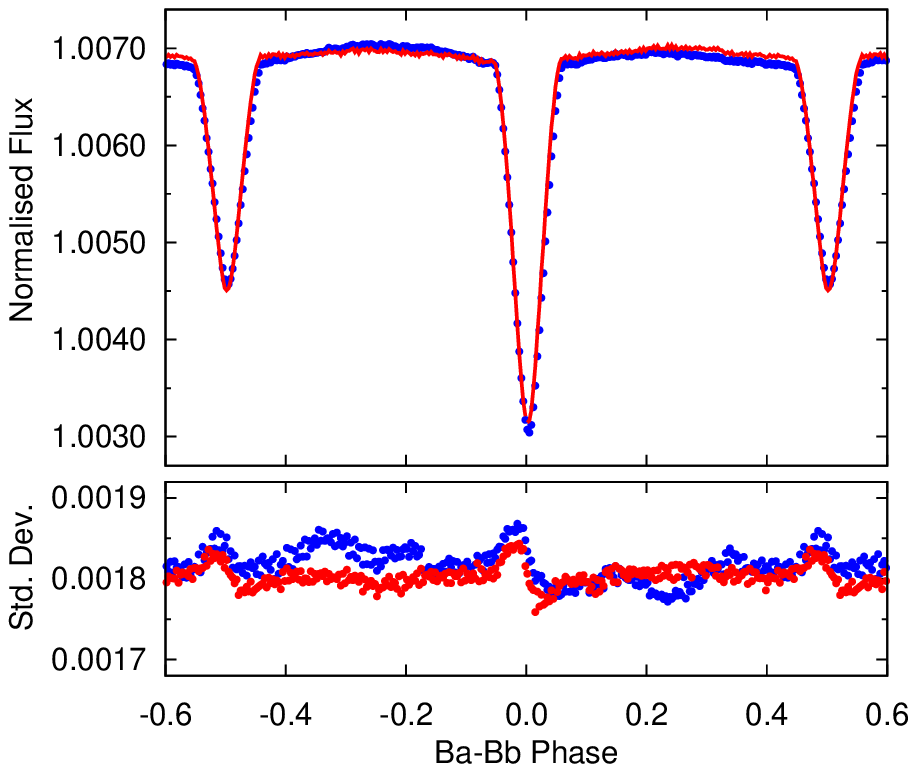}
 \caption{{\it Left panel:} The binned, averaged light curve of the $Ba-Bb$ close binary for the $Q7-Q9$ SC data (upper blue circles) with a typical fit yielded by PHOEBE (red line), and the standard deviations of the binned data with respect the average value of each individual cells (down). {\it Right panel:} The binned, averaged light curve of the $Ba-Bb$ close binary for the detrended $Q7^*-Q9^*$ SC data (upper blue circles) with a similarly processed typical solution curve yielded by our new synthetic code (red line), and the standard deviations of the binned data with respect the average value of each individual cells (down) for both the detrended observed data (blue), and the solution one (red). Note that the bottom curves do not represent the residuals of the upper solution curves.}
 \label{Fig:Bablcav}
\end{figure*}

\subsection{Method of the analysis}
\label{subsect:method}

In order to carry out this analysis, as a first step we had to separate the different kinds of variations in the light curve. While the removal of the transients (or flares) was straightforward, and the small-amplitude tidally generated oscillations do not modify significantly the eclipsing structure, the subtraction of the long-term intrinsic variations was a difficult problem. We resorted to a step-by-step iterative process, in some steps very similar to a filtering in Fourier space.

First, we obtained the averaged light curve of the close, $Ba-Bb$ binary. Since one {\it Kepler} quarter covers $\sim100$ cycles, we expect that those brightness variations which are independent of the close binary's orbital revolution would average out. We therefore binned and averaged the out-of-deep-eclipses parts of our light curves according to the eclipsing phase of the close binary. We applied this process for six different datasets: the three short-cadence data-series ($Q7$, $Q8$, $Q9$) were taken individually, and also together, the long cadence $Q1-Q6$ data together, and, finally, we converted the short cadence data into long cadence ones, and averaged the whole $Q1-Q9$-long LC dataset into an additional light curve. We tried different binning numbers, and found 300 as an optimal solution, providing sufficient time-resolution and still containing enough data points in each cell for an effective averaging. (We have also corrected the phase values for LITE, although, since the cell size was approximately equal to the full amplitude of the ETV [see the previous section], it had only a minor effect on the accuracy.) Then we obtained a light curve solution with the PHOEBE code \citep{prsazwitter05}. Most of the initial parameters were adopted from \citet{der11}. The effect of the giant component at this stage was considered simply (and crudely) as a constant third light. The initial values of this latter quantity were taken from the depth of the deep secondary eclipses (where only the giant component is visible). In the left panel of Fig.~\ref{Fig:Bablcav} we plot the $Q7-Q9$ short-cadence average, together with its PHOEBE solution curve. 

\begin{figure}
\includegraphics[width=84mm]{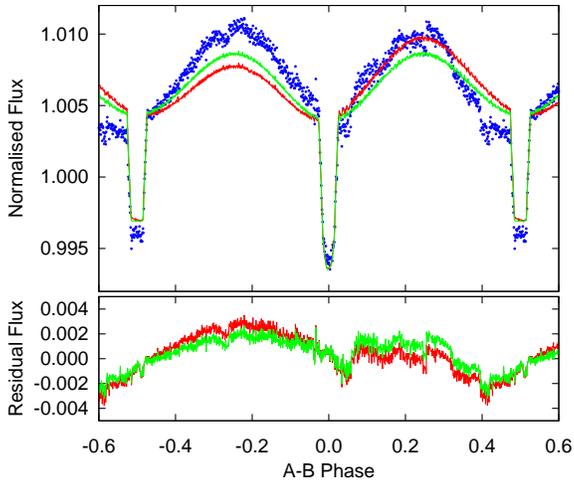}
 \caption{The binned, averaged light curve of the $AB$ outer binary for the total $Q1-Q9$ long cadence dataset (blue circles, and the synthetic eclipsing light curve averaged on the same way with and without Doppler-beaming (red and green, respectively).}
 \label{Fig:ABlcav}
\end{figure}

We also averaged the wide binary's light curve in a similar manner. In this case we divided one orbital revolution into 1000 bins (see Fig.~\ref{Fig:ABlcav}). Note that the whole $Q1-Q9$ time interval spans only $\sim17$ orbital cycles, and  there are also some gaps in the data. Therefore, we cannot expect a well-averaged light curve even for the full dataset. Furthermore, such an averaging smooths out the shoulders in the ingress and egress phases of the outer minima, which contain the most important geometric information. 

In order to recover this information, we calculated a preliminary net eclipsing and elliptical light curve for the whole triple system. For this we developed a new light curve synthesis code, which calculates the motions, gravitational interactions and mutual eclipses of the three stars simultaneously. The main characteristics of our code are described in Appendix~\ref{AppB}.

For the computation of the synthetic curve, most of the input parameters were taken from \citet{der11}, refining their values with our results from the ETV analysis and the close binary's PHOEBE light curve solution. After some very minor trial-and-error fine tunings we found a seemingly satisfactory fit. In Fig.~\ref{Fig:ABlcav} we show two versions of this synthetic curve (subjected to the same averaging process), one including the beaming effect, and the other without. We see that the curve which includes Doppler-beaming (in the order of 1 ppt) gives a better fit. Despite its preliminary stage, the fit is quite satisfactory from the first contact of the deep primary minimum to the next quadrature. The discrepancy in the other portions is probably due to the inefficiency of the averaging. An averaged residual curve is also shown in Fig.~\ref{Fig:ABlcav}.

\begin{figure*}
\includegraphics[width=56mm]{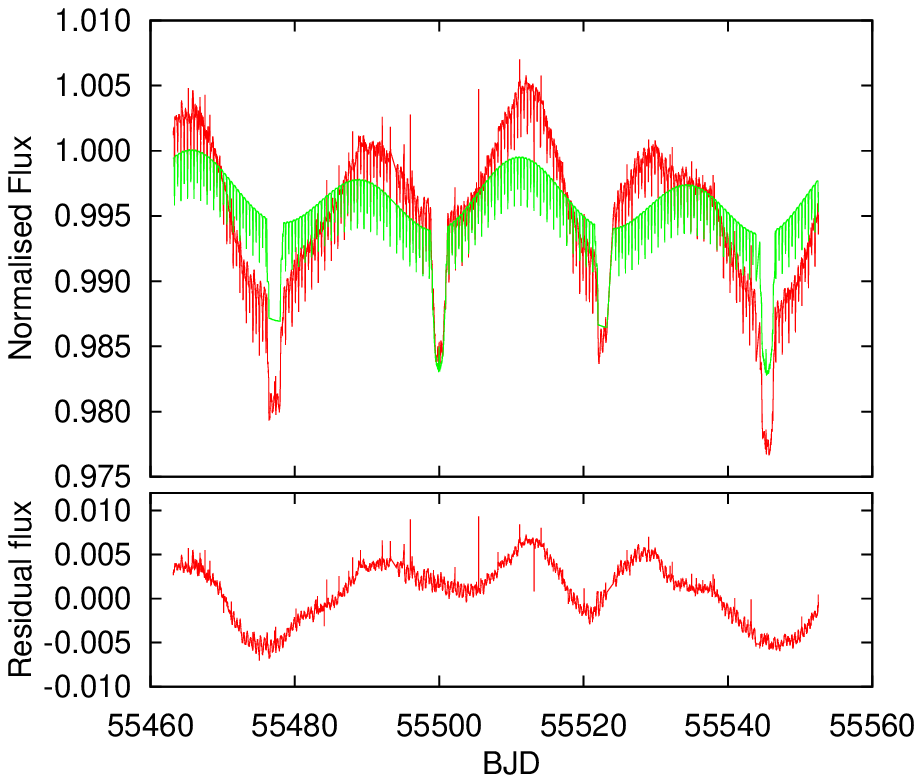}\includegraphics[width=56mm]{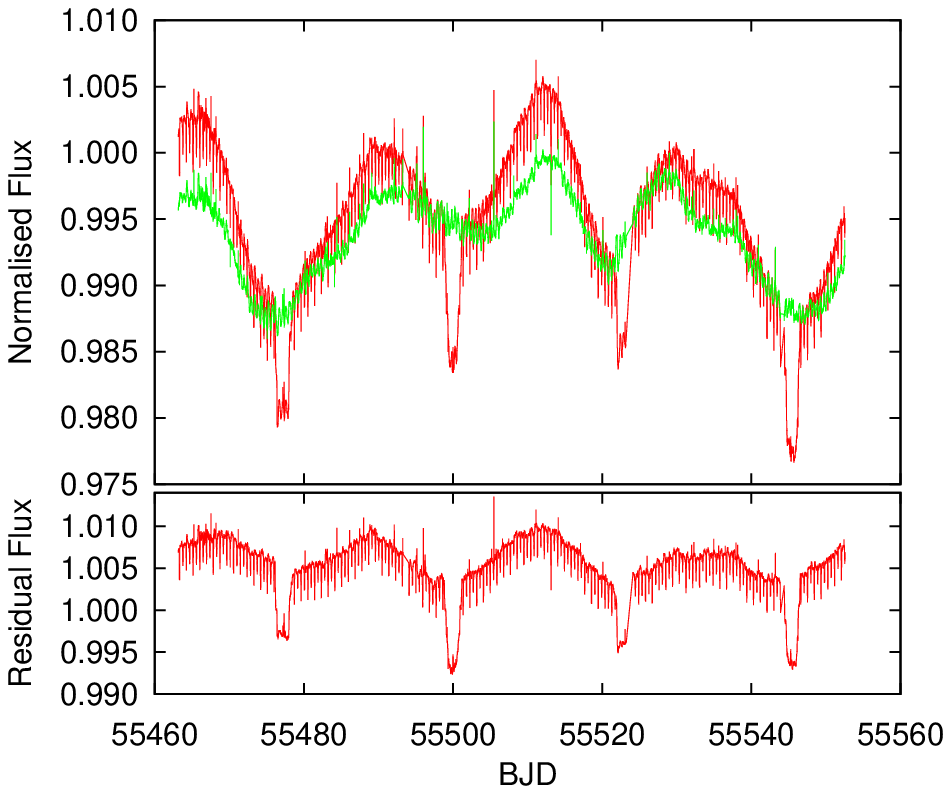}\includegraphics[width=56mm]{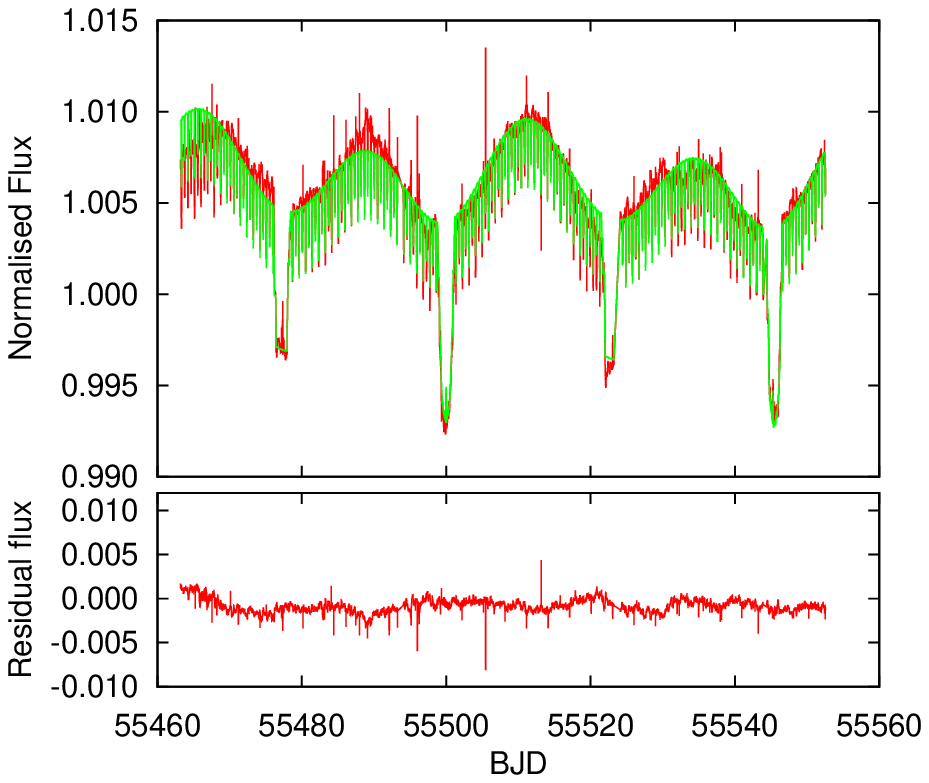}
 \caption{The process of the removal of the intrinsic light curve variations from the raw data for the $Q7$ SC observations. {\it Left panel:} The subtraction of a preliminary synthesized eclipsing light curve (green) from the original $Q7$ data (upper red curve) results a residual curve of the irregular variations (lower red curve). {\it Middle panel:} After a DFT-search of the significant frequencies in the residual curves, the intrinsic variations are represented by the corresponding Fourier polynomial (green), and this latter curve was subtracted from the original data (upper red). The detrended $Q7^*$ data are plotted in the middle lower panel with red color. {\it Right panel:} The final light curve solution (green) was fitted to this $Q7^*$ dataset (upper red). The residual curve can be seen in the bottom panel.}
 \label{Fig:syntheticlc}
\end{figure*}

As a next step, we subtracted this synthetic light curve solution from the raw data. This process was carried out individually for each quarterly dataset. The raw $Q7$ SC-data, the synthetic light curve, and the residual are plotted in the left panel of Fig.~\ref{Fig:syntheticlc}.

\begin{figure*}
\includegraphics[width=56mm]{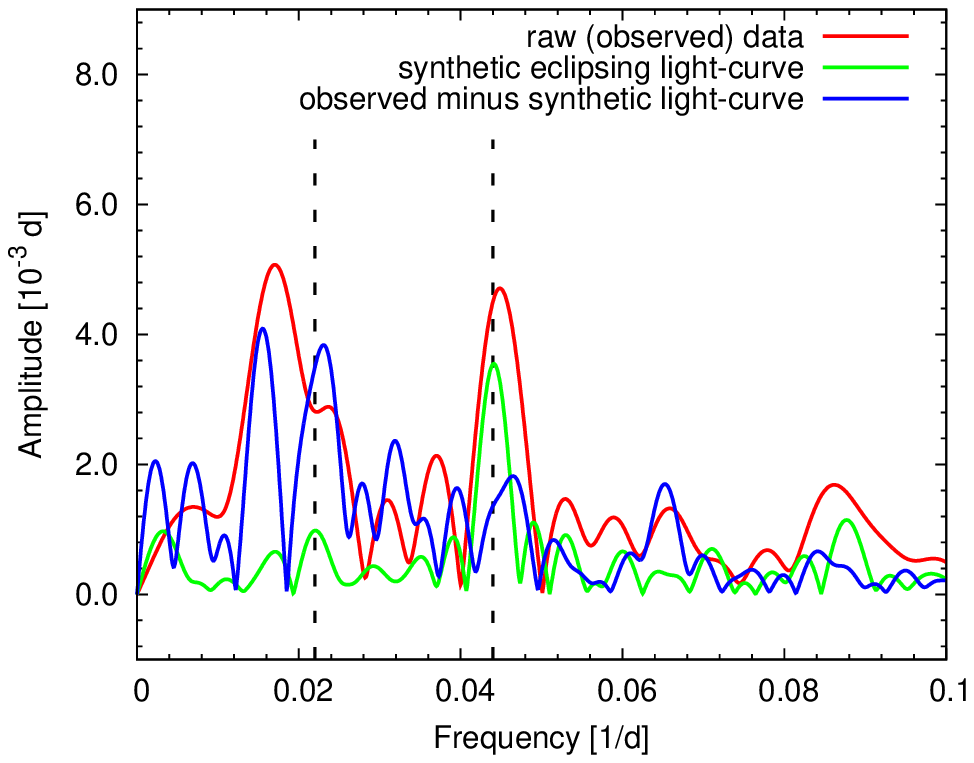}\includegraphics[width=56mm]{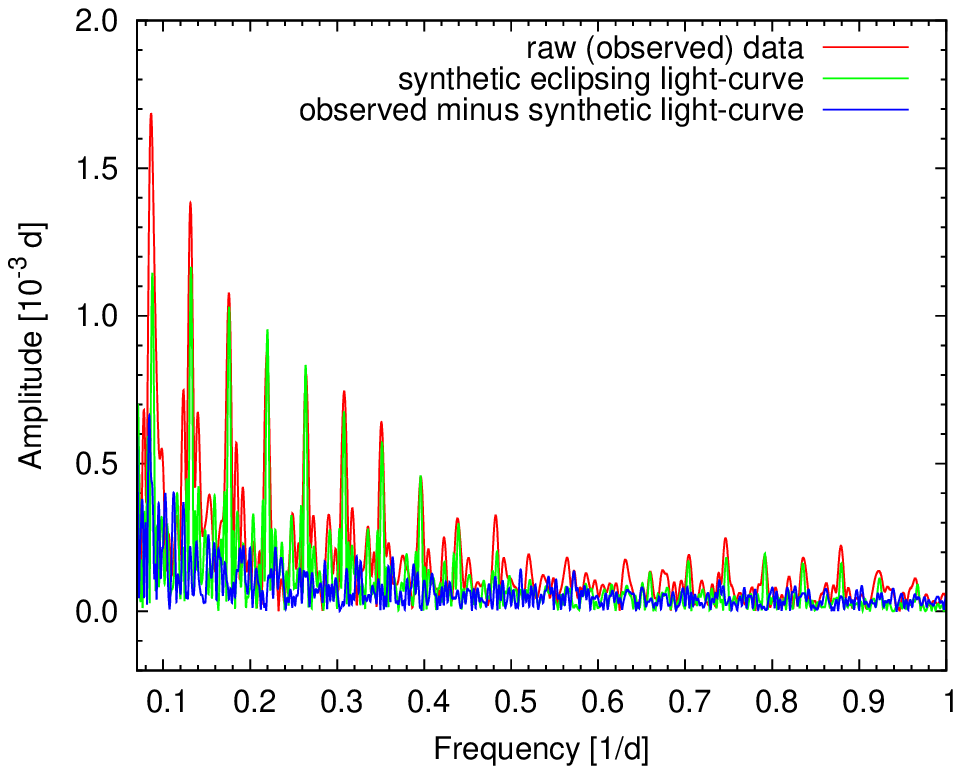}\includegraphics[width=56mm]{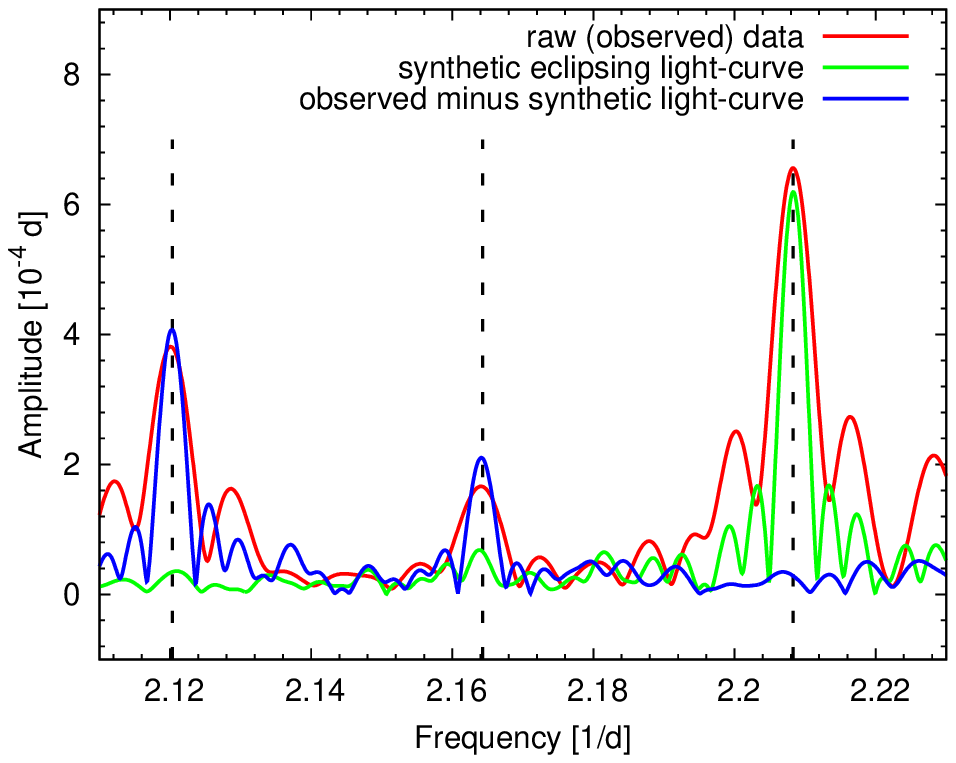}
 \caption{Three different frequency-regions of the DFT spectra obtained for three different datasets of $Q7-Q9$ SC data. Red curves denote the spectrum of the raw data, green ones represent that of the synthetic light curve solution, while blue spectra refer to the residual light curve. Dashed lines in the left panel show the frequencies belonging to the eclipsing period, and its first harmonic. Note the different scales along the $Y$-axes. See text for details.}
 \label{Fig:lc-dft}
\end{figure*}

A discrete Fourier analysis was carried out for the residual curves. This was applied for different datasets. First, in order to get the longest possible homogenous dataset, we made the DFT of the full $Q1-Q9$ LC dataset. We also made DFTs separately for $Q1-Q6$ LC data, and $Q7-Q9$ SC data. We found that the different datasets produced very similar spectra, and consequently, similar significant frequencies. Using the most prominent 10-15 frequencies, we fitted sinusoidal curves to the residual light curves. We found the best solutions, when we fitted two consecutive quarter-data together. Finally, these Fourier polynomials were subtracted from the original observational data. As a final result, we obtained such a detrended `observational' dataset, which was dominated by the eclipsing nature of the triple system. This set was used for further analysis. The step-by-step process for the $Q7$ SC-data is shown in the panels of Fig.~\ref{Fig:syntheticlc}, while three segments of the $Q7-Q9$-SC DFT spectrum are plotted in Fig.~\ref{Fig:lc-dft}. The right panel of Fig.~\ref{Fig:Bablcav}, showing the close binary's averaged light curve for the detrended $Q7^*-Q9^*$ data, illustrates the effectiveness of this procedure. (The bottom right panel of the Figure also contains an indirect evidence for the lack of short-term variations in the inclination $i_1$: a change in the eclipse depth would imply an increase of the point-to-point scatter during the eclipses, which is not seen to occur.)

In the next stage we made a grid-search analysis with our code on the detrended $Q7^*$ LC-dataset. We chose this quarter because of its relatively regular, less-distorted shape. The fitted parameters were as follows: the two mass-ratios $q_{1,2}$, the (fractional) stellar radii $R_{\rmn{A,Ba,Bb}}$, temperatures of the close binary members $T_{\rmn{Ba,Bb}}$, one of the three stellar luminosities in Kepler-band (the other two were calculated), the two orbital periods $P_{1,2}$, two epochs $T_{0-\rmn{1,2}}$, two observable inclinations $i_{1,2}$, and the relative longitude of the node of the two orbits on the sky $\Delta\Omega$, while other parameters were kept as fix ones. Logarithmic limb-darkening formulae were applied (equivalent with $ld=2$ constraint of the WD and PHOEBE code), with coefficients taken directly from PHOEBE code. The $k_j$ internal structure constants were taken from the tables of \citet{claretgimenez92}. 

In order to estimate the accuracy and reliability of the obtained parameters, we repeated our procedure for the other quarters. This enabled us to estimate the influence of the residual distorted, spotted features of the pre-processed light curves on the solutions. All the fixed and fitted parameters, as well as their estimated errors, and some derived quantities are listed in Table~\ref{tab: syntheticfit}.

\begin{figure*}
\includegraphics[width=60mm]{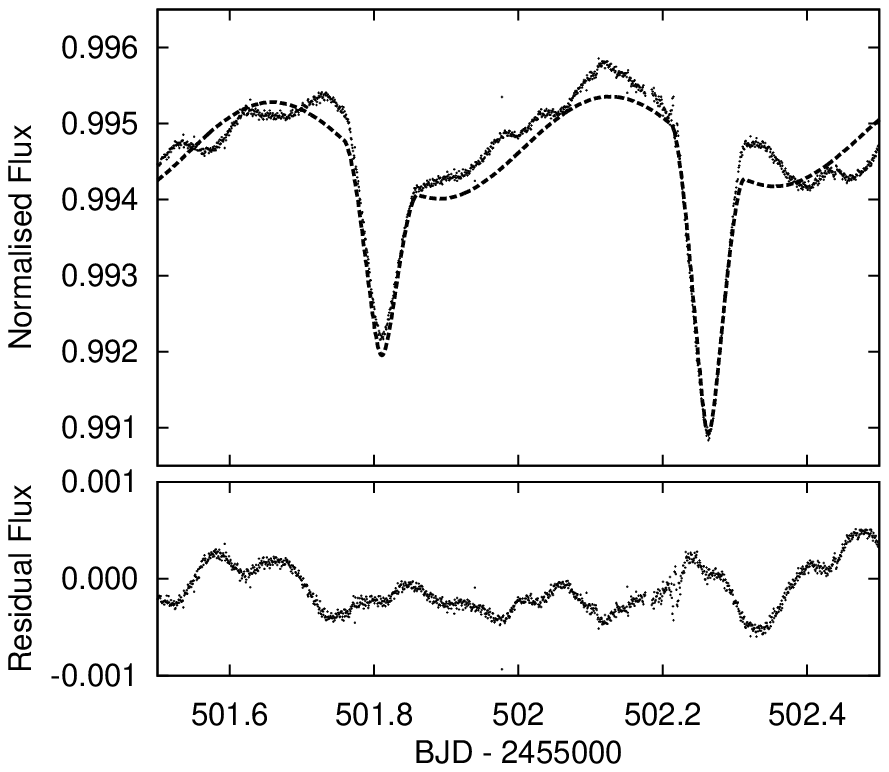}\includegraphics[width=60mm]{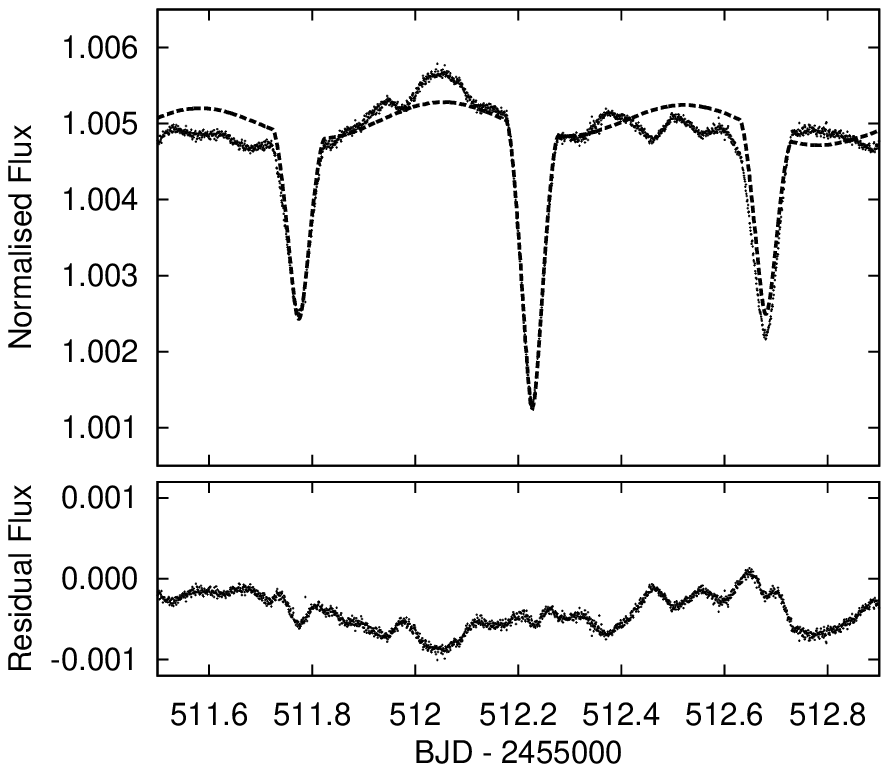}\includegraphics[width=60mm]{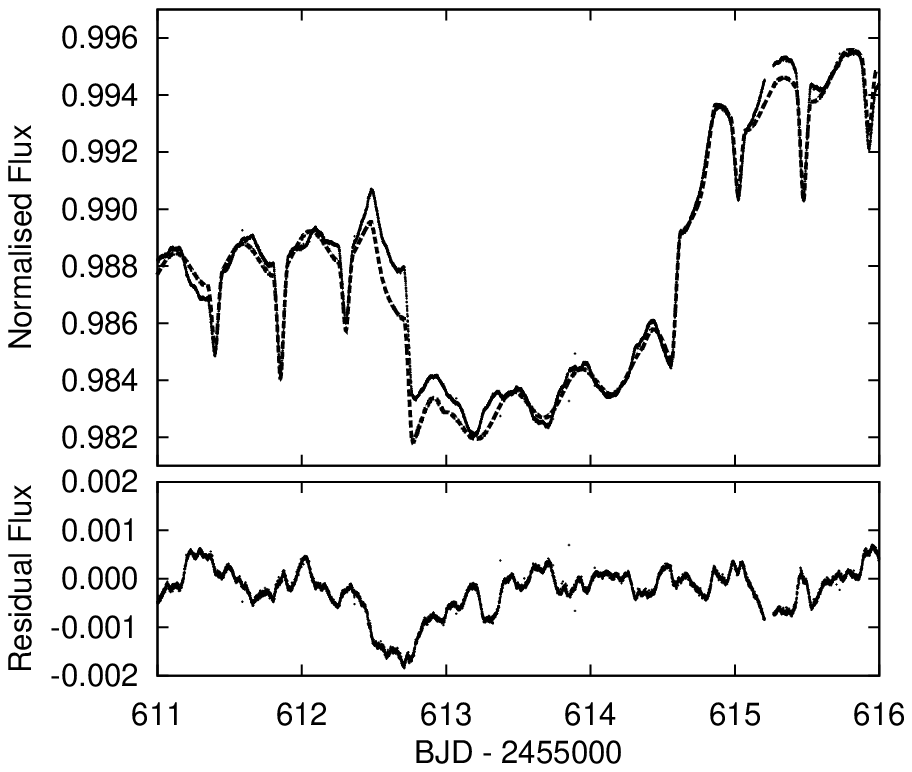}
 \caption{Three zoom-ins into the $Q7$ and $Q8$ light curves with solutions, and residuals. {\it Upper panels:} Points show the original $Q7$ and $Q8$ data, while dashed lines represent the sum of the solution for the detrended $Q7^*$ and $Q8^*$ data, and the Fourier-modelled intrinsic variations. {\it Bottom panels:} The residual curves. The left and middle panels show small parts of the $Q7$ data just after a large primary minimum, and in the following quadrature phase, respectively, while the right panel represents the surroundings of a large secondary minimum from the more distorted quarter $Q8$.}
 \label{Fig:lcveglegzoomin}
\end{figure*}

Our final solution for $Q7$ data is plotted in the panels of Fig.~\ref{Fig:lcveglegzoomin} for some characteristic parts of the curve. 


\begin{table}
 \caption{Stellar and orbital parameters derived from the combined ETV and synthetic light curve analysis. (The numbers in parantheses are the estimated errors in the last digits.)}
 \label{tab: syntheticfit}
 \begin{tabular}{@{}l|ll|l}
  \hline
\multicolumn{4}{c}{orbital parameters} \\
\hline
   & \multicolumn{3}{c}{subsystem} \\
   & \multicolumn{2}{c}{Ba--Bb} & A--B \\
  \hline
  $P$ [d] & \multicolumn{2}{c}{$0.9056768(2)$} & $45.4711(2)$ \\
  $T_\mathrm{MIN I}$ [BJD] & \multicolumn{2}{c}{$2455051.23623(5)$} & $2455499.9962(4)$ \\
  $a$ [R$_\odot$] & \multicolumn{2}{c}{$4.777(39)$} & $90.31(72)$\\
  $e$ & \multicolumn{2}{c}{$0.0$} & $0.0$ \\
  $\omega$ & \multicolumn{2}{c}{$-$} & $-$ \\ 
  $i$ [deg] & \multicolumn{2}{c}{$86.7(14)$} & $87.5(2)$ \\
  $\Delta\Omega$ [deg] & \multicolumn{3}{c}{$0.0(5)$} \\
  $i_\mathrm{m}$ [deg] & \multicolumn{3}{c}{$0.8(14)$} \\
  \hline
  $q$ & \multicolumn{2}{c}{$0.95(3)$} & $0.595(5)$ \\
  $L_\mathrm{sec}/L_\mathrm{TOT}$ & \multicolumn{2}{c}{$0.3468$} & $0.0078$\\
  \hline  
\multicolumn{4}{c}{stellar parameters} \\
\hline
   & Ba & Bb &  A \\
  \hline
 \multicolumn{4}{c}{fitted and/or derived parameters} \\
 \hline
 \multicolumn{4}{c}{relative quantities} \\
  \hline
 $r_\rmn{pole}$  & $0.1798$ & $0.1664$ & $0.1376$ \\
 $r_\rmn{side}$  & $0.1808$ & $0.1672$ & $0.1379$ \\
 $r_\rmn{point}$ & $0.1826$ & $0.1687$ & $0.1382$ \\
 $r_\rmn{back}$  & $0.1822$ & $0.1684$ & $0.1382$ \\
 \hline
 \multicolumn{4}{c}{absolute quantities} \\
  \hline 
 $m$ [M$_\odot$] & $0.915(34)$ & $0.870(43)$ & $3.0(1)$ \\
 $R$ [R$_\odot$] & $0.865(10)$ & $0.800(20)$ & $12.46(15)$ \\
 $T_\mathrm{eff}$ [K]& $5100(100)$ & $4675(100)$ & $5100(100)$ \\
 $L_\mathrm{bol}$ [L$_\odot$] & $0.447(37)$ & $0.270(27)$ & $92.812(7615)$ \\
 $\log g$ [dex] & $4.53$ & $4.58$ & $2.73$\\
 \hline
 \multicolumn{4}{c}{fixed quantities} \\
 \hline
 $k_2$         & $0.020$   & $0.020$   & $0.033$   \\
 $\beta$       & $0.32$    & $0.32$    & $0.32$    \\
 $A$           & $0.5$     & $0.5$     & $0.5$     \\
 $x_\rmn{bol}$ & $0.71476$ & $0.71476$ & $0.71159$ \\
 $y_\rmn{bol}$ & $0.13026$ & $0.13026$ & $0.12561$ \\
 $x_\rmn{K}$   & $0.70835$ & $0.70835$ & $0.70074$ \\
 $y_\rmn{K}$   & $0.16354$ & $0.16354$ & $0.16609$ \\
 \hline
 \end{tabular}


\end{table}

\section[]{Discussion and Conclusions}

We have determined a new set of physical parameters for all three components in the system. Our results have roughly an order of magnitude lower random errors than was achievable after the discovery by \citet{der11}. Furthermore, we were able to exploit the unique geometry to infer new parameters that were previously beyond reach.

For the previously determined parameters, we find excellent agreement with the new values. For example, the primary's radius, combining the Hipparcos parallax with CHARA/PAVO onterferommetry, was measured by \citet{der11} to be $R_\mathrm{A}=12.4\pm1.3\rmn{R}_\odot$. Now we have determined $R_\mathrm{A}=12.46\pm0.15\rmn{R}_\odot$ by combining the stellar masses from the ETV study with the simultaneous light curve analysis. Similarly, the ETV analysis plus the SB1 radial velocity measurements yielded a primary mass of $m_\rmn{A}=3.0\pm0.1\rmn{M}_\odot$, which agrees with the estimated mass from evolutionary tracks in \citet{der11}. All in all, the derived physical parameters draw a consistent picture of the system, proving that despite the difficulties in the light curve modelling, our method yields robust results.

A preliminary comparison with models from the BASTI \citep{pietrinfernietal04} and Dartmouth \citep{dotteretal08} databases shows that the fundamental properties for all three components are consistent with solar-metallicity isochrones with ages $\sim 300-500$\,Myr, although the dwarf radii appear to be significantly larger than expected. More detailed comparison using the near model-independent properties presented here will allow powerful tests of stellar evolutionary theory, such as tidal effects on the mass-radius relation for low-mass stars in close-in binary systems \citep[see, e.~g.,][]{krausetal11}.


One important question in relation to the giant primary is its evolutionary stage, being located in a part of the H-R diagram where $H$-shell-burning stars ascending the first red giant branch overlap closely with $He$-core-burning giants (in other words, there is an age uncertainty that cannot be resolved from the evolutionary tracks alone). Dynamical considerations can help here, too, via comparing the orbital configurations with theoretical tidal circularization time-scales. According to Eq.~(7) in \citet{verbuntphinney95}, which was based on the works of \citet{zahn77,zahn89}, a binary with the same parameters as HD~181068 A and B (=Ba+Bb) is expected to be circularized under a period limit of $P_\mathrm{circ}\sim15$~days for $H$-shell burning primary. With the observed $P_2\sim45$ days and the perfectly circular orbit, theory implies indirectly that the primary must be older, so that in the $He$-core burning phase. The question, however, is more complicated because of the binary nature of the secondary. This causes additional complications by the tidal oscillations that are expected to affect the convective envelope of the primary. It is not known if the tidal damping is effective enough to shorten significantly the circularization time.

\begin{figure*}
\includegraphics[width=84mm]{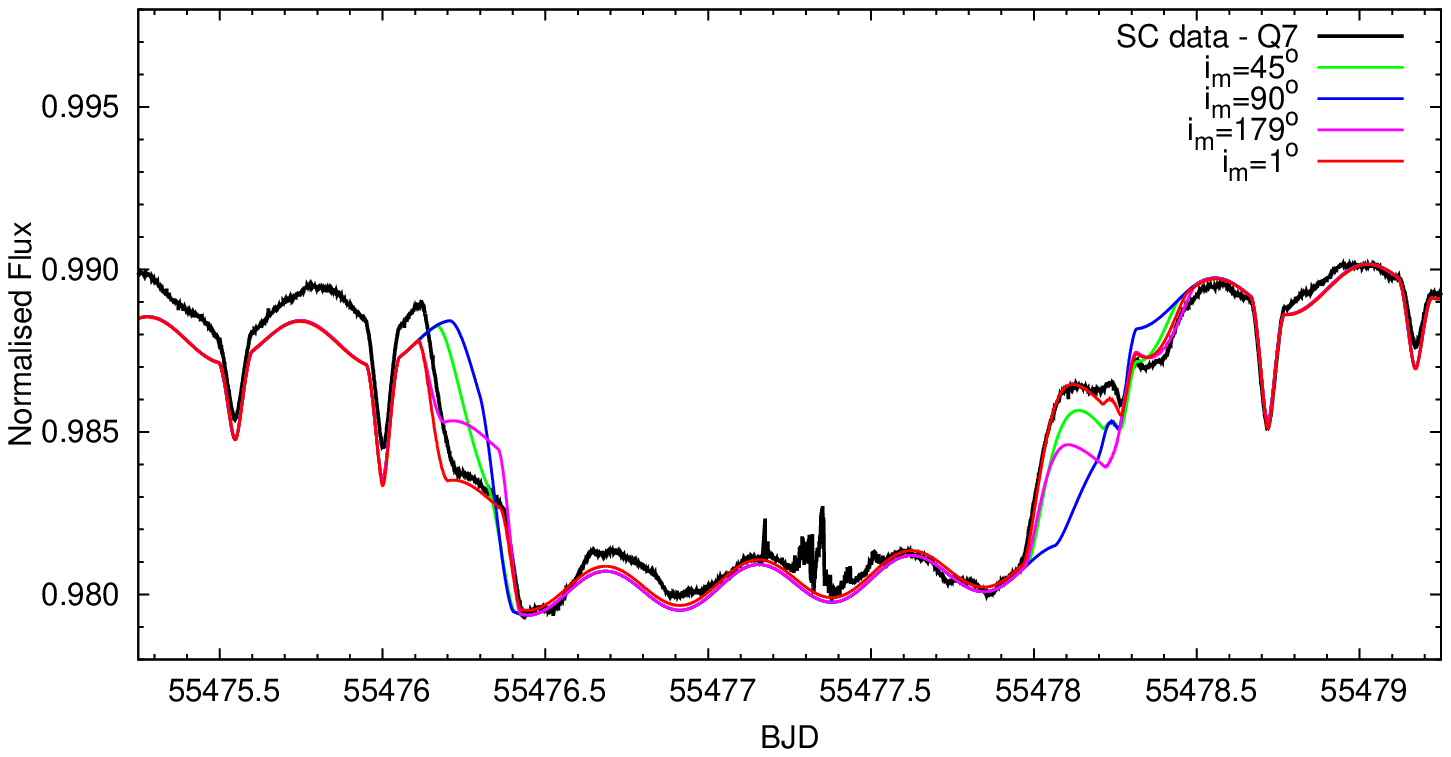}\includegraphics[width=84mm]{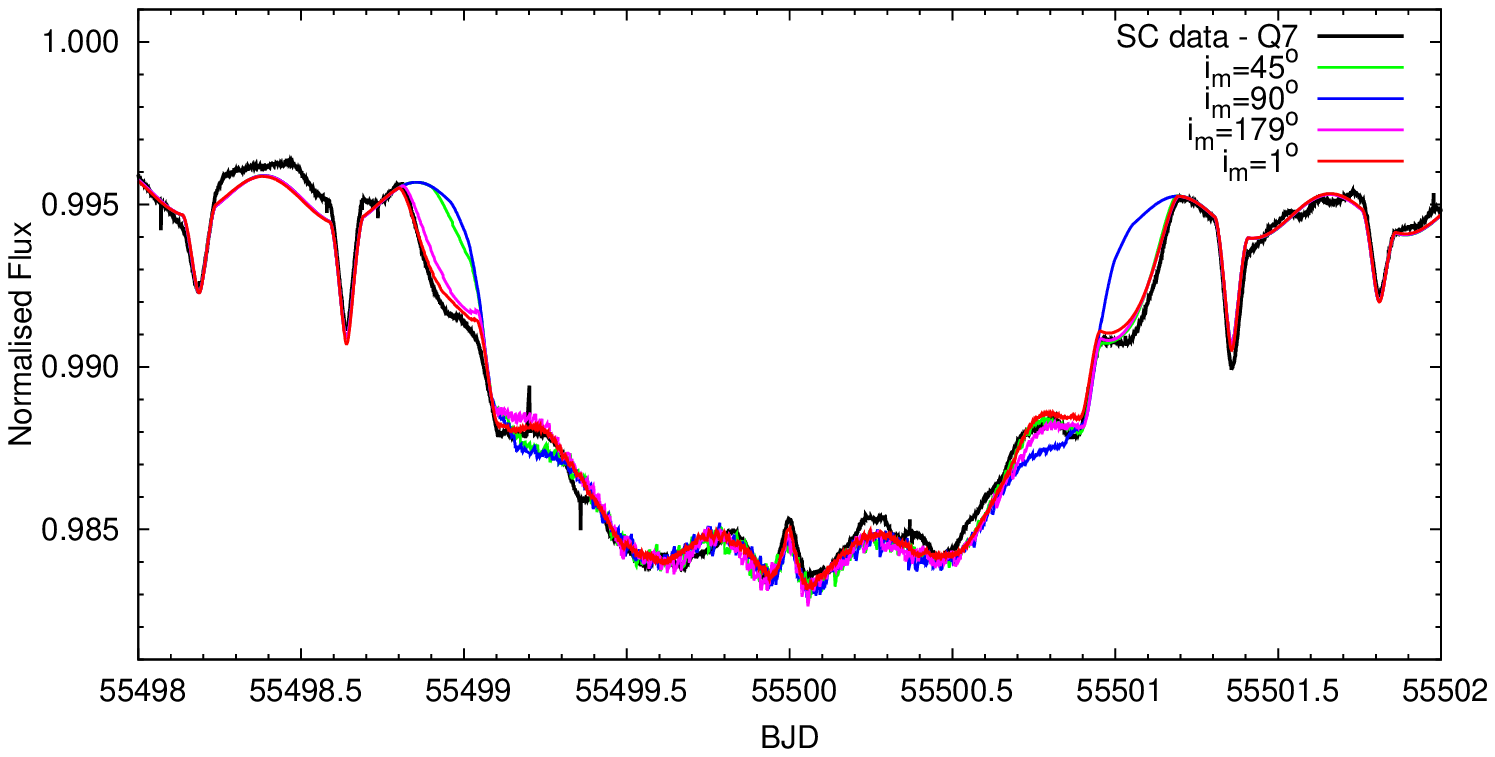}
 \caption{Synthetic light curve for an outer secondary (left), and a primary (right) minimum,  calculated with different mutual inclinations. See text for details.}
 \label{fig:mutualincl}
\end{figure*}

Considering the other orbital parameters, our solution for the orbital inclination of the close binary ($i_1=86\fdg7\pm1\fdg4$) has a relatively large uncertainty. This is not surprising because, being a partially eclipsing pair, the observable inclination is very sensitive to any additional third light, i.~e. in the present case, for the light of the giant primary, and particularly, for the continuous variation of this extra amount of light. The inclination of the wide system was found to be ($i_2=87\fdg5\pm0\fdg2$). From these two values alone, in the absence of any other information, we would be able to say nothing about the mutual inclination of the system. However, the simultaneous light curve fit of this triple eclipsing system provides a direct and powerful method for determining this quantity. As mentioned above, this comes from both the fine structure and the timings of the ingress and egress phases of the deep eclipses. This is illustrated in Fig.~\ref{fig:mutualincl}, where we plotted the first two deep eclipses of the $Q7$ SC light curve. The only difference between the different colored curves is the $\Delta\Omega$ parameter, and consequently, the mutual inclination. While the out-of-deep-eclipse parts, and the totality-of-eclipse periods of the light curves are identical, the ingress/egress fine structures, and the moments of the contacts differ significantly, and this makes the $\Delta\Omega$ adjustable parameter (and so the mutual inclination) to a well-determined quantity. Furthermore, even the $i_\rmn{m}=1\degr$ curve is definitely separable from its retrograde counterpart $i_\rmn{m}=179\degr$. (Note, however, that this separation is only possible when the masses or the radii are different in the close binary.) A combination of the obtained $\Delta\Omega$ parameter with the two observable inclinations results in a mutual inclination of $i_\mathrm{m}=0\fdg8\pm1\fdg4$, which suggests an exact coplanarity. This is in accordance with the lack of the eclipse-depth variation of the shallow eclipses.  

Finally, we briefly comment on the other features of the light curves. First we consider the irregular, or semiregular brightness changes, which likely originate from chromospheric activity. Evidence of the presence of spotted regions on the giant's surface was shown in the previous section (see e.~g. Fig.~\ref{fig:eclipsecycles}). Further characteristics can be deduced from the comparative investigation of the DFT spectra of the raw observed light curve, the synthetic and the residual ones (Fig.~\ref{Fig:lc-dft}). What can be seen well even at the first glance is that in the low frequency domain (left panel), the spectrum of the observed data remarkably departs from that of the synthetic data. While in the synthetic eclipsing, ellipsoidal data the dominant frequency corresponds to the half eclipsing period of the wide system, the highest peak of the original data is located about the eclipsing period itself. Furthermore, this latter peak is clearly a double one, whose two peaks are already well separated in the spectrum of the residual light curve (i.~e. after the removal of the eclipsing and ellipsoidal features). In our interpretation these two peaks might have a rotational origin. The good correspondance of this pair of peaks with the orbital period proves the synchronised rotation of the primary, while its splitting might give an evidence of differential rotation \citep[see~e.~g.][]{olahetal03}. Note, that the spectroscopically obtained $v_\rmn{rot}\sin i=14\rmn{~kms}^{-1}$ \citep{der11} for $R_\rmn{A}=12.5\rmn{~R}_\odot$, and $\sin i_2=87\fdg4$ result in $P_\rmn{rot}=45\fd474$, which is also in very strong correspondance with this result.
   
%

Considering the high-frequency end of the DFT spectra (right panel of Fig.~\ref{Fig:lc-dft}), three distinct peaks can be identified at $f_1=2.20829\rmn{~d}^{-1}$, $f_2=2.16431\rmn{~d}^{-1}$ and $f_3=2.12032\rmn{~d}^{-1}$, from which three, the first is exactly the half of the eclipsing period of the close binary, while the other two are $f_2=f_1-f_0$, and $f_3=f_1-2f_0$, where $f_0=0.04398\rmn{~d}^{-1}$ corresponds to the half of the eclipsing period of the wide system. Such a way, the tidal origin of this small amplitude oscillation on the surface of the giant primary is out of question. These oscillatory features will be investigated in details in a forthcoming paper.



\section*{Acknowledgements}

This project has been supported by the Hungarian OTKA Grants K76816, K83790 and MB08C 81013, ESA PECS C98090, the ``Lend\"ulet-2009'' Young Researchers Program of the Hungarian Academy of Sciences and the European Community's Seventh Framework Programme (FP7/2007-2013) under grant agreement no. 269194. AD and RSz has been supported by the J\'anos Bolyai Research Scholarship of the Hungarian Academy of Sciences. AD was supported by the Hungarian E\"otv\"os fellowship. Funding for this Discovery Mission is provided by NASA's Science Mission Directorate. The Kepler Team and the Kepler Guest Observer Office are recognized for helping to make the mission and these data possible.
TB thanks Professor R. E. Wilson, Drs. K. Ol\'ah and Sz. Csizmadia for the valuable discussions on the questions of light curve modelling.
AD thanks Dr. A. Simon for the technical assistance.

\appendix

\section[]{Determination of system parameters from the mutual eclipses}
\label{AppA}

\begin{figure*}
\includegraphics[width=84mm]{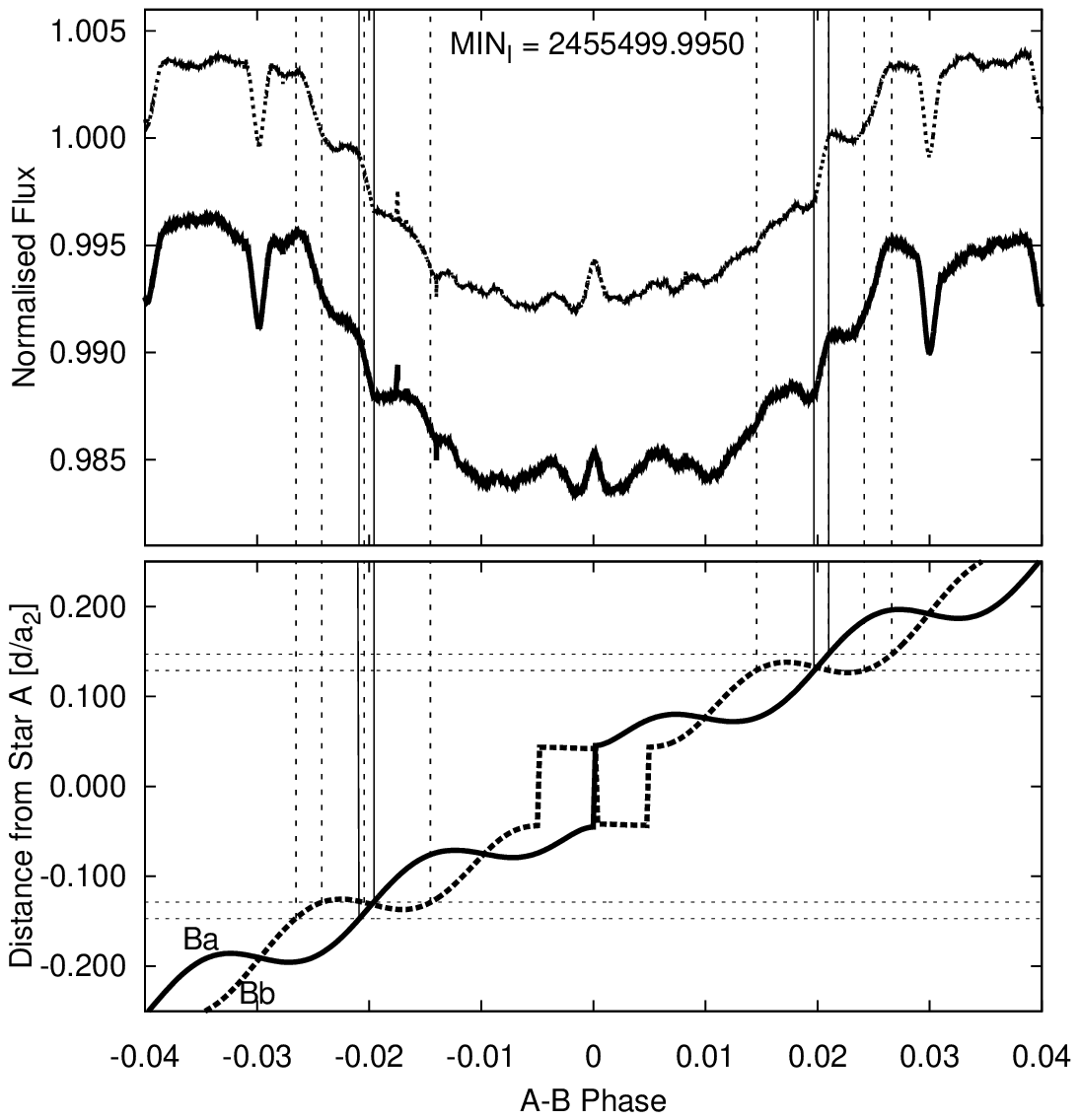}\includegraphics[width=84mm]{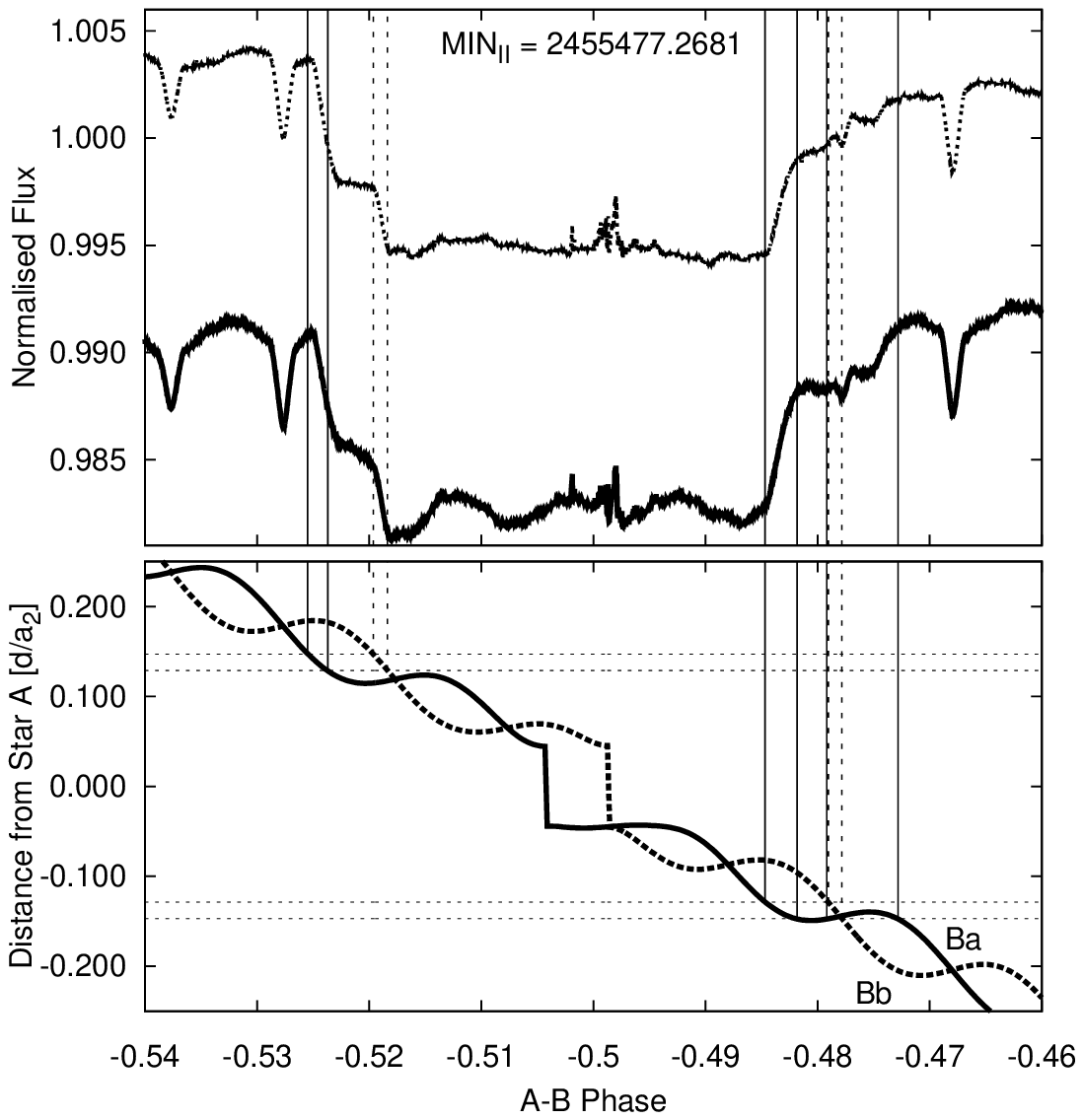}
\caption{The ``walzer'' of the close pair in front of (left), and behind (right) the giant primary, projected on the sky. The upper panels are identical with Figs.~\ref{fig:eclipsecycles}$i$ and $a$, respectively. The dashed horizontal lines denote the $R_\rmn{A}\pm\overline{R_\rmn{Ba,b}}$ distances from the center of mass of the giant, i.~e. the outer and inner contact places. Vertical lines connect the moments of the different contacts with the corresponding light curve points.}
 \label{fig:tanc}
\end{figure*}

In this Appendix we show examples of how several system parameters can be determined from the geometry of the large, mutual eclipses. Strictly speaking, the most important condition for the validity of the following calculations is not the mutuality of the shallow and the deep eclipses themselves, but rather the fact that due to the hierarchical configuration of the triple system, the deep eclipses contain some mixtures of individual eclipses of the two members of the close pair with respect to the more distant giant component, which produce small changes in the deep eclipse configurations from eclipse to eclipse, and even between the ingress and egress phases of the same event. Our algorithm is a natural extension of the well-known methodology of determination of the relative radii of the stars (with respect to their separation, and as a function of their orbital inclination) purely from the eclipse geometry, commonly used from the very beginning of eclipsing binary studies.

The usual method in binaries with spherical components is well known: the sky-oriented distance of the stellar disc centers is $R_1+R_2$ at the first and last contacts (i.~e., at the start of ingress and at the egress phases), and if the eclipses are total (either transit or occultation), the same distance is $R_1-R_2$ at the second and third contacts (i.~e., at the end of the ingress and at the start of the egress phases). Then expressing the projected distances with the orbital elements and time, and measuring the eclipse durations (both the one from the first to the last contacts, and the totality length from the second to the third contacts), the individual fractional radii of the stars can be determined. 

For our triple star configuration, the egress and ingress phases of the deep eclipses show a complex pattern. The two dwarf members of the close binary may enter in front of or behind the giant's disk individually, or even simultaneously (see Fig.~\ref{fig:tanc}). Additionally, during an entry the stars' velocities, directions and distances (both physical and projected) relative to the giant component change continuously, producing variable length and shape in the egress and ingress patterns. Anyhow, no matter how complex an egress or ingress pattern is in itself, every eclipse event contains one and only one first, second, third, and fourth contacts. And furthermore, assuming that a given contact is not strongly altered by a just ongoing shallow eclipse event, we can simply and unambiguously decide which member of the close binary takes part in the given contact event. For example, in case of prograde revolution, the very first contact of a primary transit is produced by the eclipser of the last shallow eclipse event, i.~e., if the last event was a small secondary minimum, then the very first contact of the large primary transit is produced by the primary of the close pair.

\begin{figure}
\includegraphics[width=84mm]{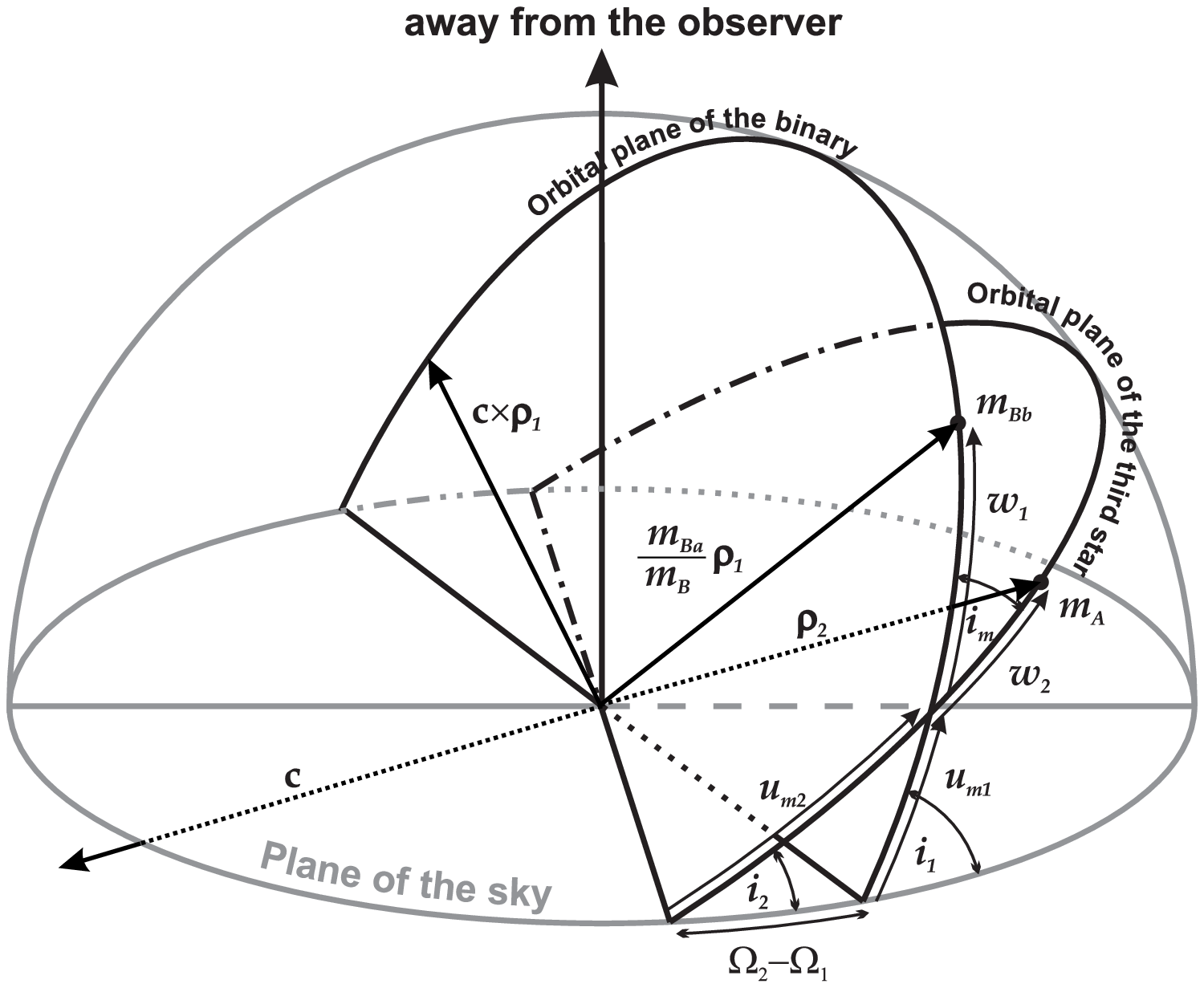}
 \caption{An illustration of the angles and other quantities used in the text.}
 \label{fig:krsz}
\end{figure}

Let us consider the projected distances at the disk centres in the moments of the contacts. In the present situation the projected distance between the eclipser and the eclipsed stars no longer will be the projected radius vector of a Keplerian relative orbit, but comes from the superposition of two Keplerian orbits: the absolute orbit of the close binary members around their center of mass (CM), and the relative orbit of this CM around the giant component. The most convenient and practical description of the present scenario uses Jacobian vectors. The first Jacobian vector ($\vec{\rho}_1$) is directed from $m_\rmn{Ba}$ to $m_\rmn{Bb}$, i.~e., it is the radius vector of the close binary's relative orbit, while the second one ($\vec{\rho}_2$) originates from the CM of the close pair, and ends in $m_\rmn{A}$, i.~e. it is the radius-vector in the wide pair (see Fig.~\ref{fig:krsz}). With these notations, the position vectors connecting the three stars mutually are
\begin{eqnarray}
\vec{d}_\rmn{BaBb}&=&\vec{\rho}_1, \\
\vec{d}_\rmn{BaA}&=&\vec{\rho}_2+\frac{q_1}{1+q_1}\vec{\rho}_1, \\
\vec{d}_\rmn{BbA}&=&\vec{\rho}_2-\frac{1}{1+q_1}\vec{\rho}_1,
\end{eqnarray}
where, as before, $q_1$ denotes the mass ratio of the close pair. In the usual astrometric frame of reference the right-handed $x$ and $y$ coordinate axes lie in the plane of the sky, while the $z$ axis points outwards from the observer. In the astrometric convention, $x$ points to the celestial north pole. In case of photometry, however, both the eclipsing light curve, and the radial velocity are invariant with respect to any rotation in the plane of the sky, and so, in the absence of any additional information on the spatial orientation of the intersection of the orbital plane and the sky (i.~e., $\Omega$), we are free to use any orientation for the $x$ axis. In the context of modelling eclipsing binaries, the coordinate equations take their simplest form if one of the axes in the plane of the sky coincides with the nodal line. In this case, the other axis gives the direction of the projected orbital angular momentum vector. In the present case, however, we cannot use this latter formal simplicity, because of the differing orbital planes of the close and the wide orbits.

It is well known from the textbooks of celestial mechanics and/or astrometry that in such a frame of reference the cartesian coordinates of a Keplerian orbit can be written as
\begin{eqnarray}
x&=&r[\cos(v+\omega)\cos\Omega-\sin(v+\omega)\sin\Omega\cos i], \\
y&=&r[\cos(v+\omega)\sin\Omega+\sin(v+\omega)\cos\Omega\cos i], \\
z&=&r\sin(v+\omega)\sin i,
\end{eqnarray}
What is important for us is the projected distances onto the plane of the sky, instead of the spatial ones. In vectorial forms e.~g.
\begin{equation}
d^{xy}_{\rmn{BaA}}=\sqrt{\left[\vec{\rho_2}+\frac{q_1}{1+q_1}\vec{\rho_1}-\left(\vec{\rho_2}\cdot\vec{z}+\frac{q_1}{1+q_1}\vec{\rho_1}\cdot\vec{z}\right)\vec{z}\right]^2},
\end{equation}
or, with orbital elements,
\begin{eqnarray}
d^{xy}_{\rmn{BaBb}}&=&\rho_1\sqrt{1-\sin^2i_1\sin^2u_1}, \\
d^{xy}_{\rmn{BaA}}&=&\rho_2\left[1-\sin^2i_2\sin^2u_2\right. \nonumber \\
&&+2\frac{q_1}{1+q_1}\frac{\rho_1}{\rho_2}\left(\lambda-\sin i_1\sin u_1\sin i_2\sin u_2\right) \nonumber \\
&&\left.+\left(\frac{q_1}{1+q_1}\frac{\rho_1}{\rho_2}\right)^2\left(1-\sin^2i_1\sin^2u_1\right)\right]^{1/2}, \label{Eq:dxyBaA} \\ 
d^{xy}_{\rmn{BbA}}&=&\rho_2\left[1-\sin^2i_2\sin^2u_2\right. \nonumber \\
&&-2\frac{1}{1+q_1}\frac{\rho_1}{\rho_2}\left(\lambda-\sin i_1\sin u_1\sin i_2\sin u_2\right) \nonumber \\
&&\left.+\left(\frac{1}{1+q_1}\frac{\rho_1}{\rho_2}\right)^2\left(1-\sin^2i_1\sin^2u_1\right)\right]^{1/2},
\label{Eq:dxyBbA}
\end{eqnarray}
where $u_i=v_i+\omega_i$ gives the true longitude of the given object measured from the node and furthermore, 
\begin{equation}
\lambda=\cos w_1\cos w_2+\sin w_1\sin w_2\cos i_\mathrm{m}
\end{equation}
is the direction cosine between vectors $\vec{\rho}_1$ and $\vec{\rho}_2$, in which expression $w_i=u_i-u_{\rmn{m}i}$ denotes the true longitude measured from the intersection of the two orbital planes, while $u_{\rmn{m}i}$ is a nodal longitude-like quantity, namely the angular distance of the intersection of the given orbital plane from the sky (cf.~Fig.~\ref{fig:krsz}). It can also be seen in this figure that the three inclinations form angles of that spherical triangle, the sides of which are the three node-like arcs $\Delta\Omega$, $u_\rmn{m1}$ and $u_\rmn{m2}$. Consequently, the two observable inclinations ($i_1$, $i_2$) and the difference of the nodes of the close and wide orbits ($\Delta\Omega=\Omega_2-\Omega_1$) unambiguously determine the remaining quantities (i.~e., $i_\rmn{m}$ and $u_\rmn{m}$-s) with the copious identifications of the spherical triangles, from which some of the most useful ones in the present context are as follows:
\begin{eqnarray}
\cos i_\rmn{m}&=&\cos i_1\cos i_2+\sin i_1\sin i_2\cos\Delta\Omega, \\
\sin i_\rmn{m}\cos u_{\rmn{m}2}&=&-\cos i_1\sin i_2+\sin i_1\cos i_2\cos\Delta\Omega, \\
\sin i_\rmn{m}\sin u_{\rmn{m}2}&=&\sin i_1\sin\Delta\Omega, \\
\cos i_1&=&\cos i_2\cos i_\rmn{m}-\sin i_2\sin i_\rmn{m}\cos u_{\rmn{m2}}, \\
\cos u_\rmn{m1}&=&\cos u_\rmn{m2}\cos\Delta\Omega+\sin u_\rmn{m2}\sin\Delta\Omega\cos i_2. \nonumber \\
\end{eqnarray} 

At this point we note that in case of coplanarity Eqs.~(\ref{Eq:dxyBaA},\ref{Eq:dxyBbA}) become more simple, not only due to $\sin i_1=\sin i_2$, but also because then the direction cosine $\lambda$ is simply
\begin{equation}
\lambda=\cos(u_2-u_1).
\end{equation}

In what follows, we assume that both the inner and the outer orbits are circular, as it happens to be in HD~181068. In this case $\rho_{1,2}\equiv a_{1,2}$, which means a substantial simplification in our treatment. At this point we are in the position to give the functional dependencies of the stellar sizes from the orbital elements. Namely, for any contact of the deep eclipses
\begin{equation}
R_{A}\pm R_{Ba,Bb}=a_2f_{Ba,Bb}(i_1,i_2,\Delta\Omega,q_1,a_1/a_2;u_1,u_2),
\label{eq:RAB}
\end{equation}
where the plus and minus signs hold for outer and inner contacts, respectively. Moreover, for the partial shallow eclipses for the outer contacts we can also write
\begin{equation}
R_{Ba}+R_{Bb}=a_1f_{Bab}(i_1;u_1).
\label{eq:RBab}
\end{equation}
In these equations the independent (time-like) variables are hidden in the $u_i$-s, while the other parameters are constants. The $u_i$ longitudes are very closely related to the eclipsing phases. It is well known \citep[see~e.~g.~][]{gimenezgarcia83} that for circular orbits in the moment of a mid-minimum $u=90\degr$ or $u=270\degr$. Considering the shallow minima, in the present situation, as $\vec{\rho}_1$ points toward the secondary of the close pair, i.~e., $u_1$ refers to the relative orbit of the secondary, or $Bb$-component, therefore $u_1=90\degr$ for the secondary minimum, and $u_1=270\degr$ for the primary one. So the connection of $u_1$ with the eclipsing phase of the close pair is very simple, $u_1=360\degr\times\phi_1-90\degr$, or converting it to time directly, $u_1(t)=\frac{2\pi}{P_1}(t-T_0)-\frac{\pi}{2}$, where $T_0$ is a mid-primary minimum time. 

The calculation of $u_2$ requires a bit of extra care. First, we have to keep in mind that $\vec{\rho}_2$ is oriented from component $B$ (i.~e. the centre of mass of the close pair) towards component $A$, and so in the present formulae $u_2$ refers to the relative orbit of the main, giant component $A$ around the smaller and fainter component $B$, i.~e., the secondary of the wide system. Consequently, in this case $u_2=90\degr$ formally means not the secondary but the primary deep minima. Furthermore, the determination of a reference mid-minimum time is not so simple. Because the positions of the close binary members on their orbits are different at the beginning and at the end of a deep eclipse, the mid-minimum does not occur exactly at half-time between the first and the last contacts (similarly to the eccentric case). Nevertheless, by the use of an averaged light curve (like the one in Fig.~\ref{Fig:ABlcav}), or of a radial velocity curve, or from the average of several approximately determined mid-minima times we can obtain a satisfactory reference mid-minimum moment, and then $u_2(t)$ can be calculated in the same way as $u_1$.

In the next step we use the fundamental difference between a traditional simple eclipsing binary and our triple system. In a single eclipsing binary all the eclipses are similar, i.~e., all contacts occur at the same orbital phase, longitude ($u$) and, furthermore, for circular case the eclipses are geometrically symmetric in time around their midpoint. Due to the latter, for the first and fourth (last) contacts $\sin^2u_\rmn{I}=\sin^2u_\rmn{IV}$, and a similar relation can be written for the inner contacts. As a consequence, we have only one equation for $\frac{R_1+R_2}{a}$ and one for $\frac{R_1-R_2}{a}$, and so, some extra way is needed to resolve the inclination dependence. In opposition, in case of our deep eclipses, the configuration of the first and last contacts, and the second and third contacts as well, generally vary from eclipse to eclipse, and they are even different within the same event. Consequently, we can get separate sets of equations (\ref{eq:RAB}) for different $u_1$ and $u_2$, from which the unknown parameters can in principle be determined by numerical methods.

Furthermore, once $a_1/a_2$ is known, $q_2$ can also be calculated easily by the use of Kepler~III:
\begin{equation}
\frac{q_2}{1+q_2}=\left(\frac{a_1}{a_2}\right)^3\left(\frac{P_2}{P_1}\right)^2.
\label{Eq:q2}
\end{equation}
The individual relative radii of the three stars can also be derived from the combination of Eqs.~(\ref{eq:RAB}), even without the use of Eq.~(\ref{eq:RBab}). On the other hand, these individaul radii can be determined also in the case when both the deep and the shallow minima are partial.

Several difficulties arise, however, during the practical application of this method. For example, in our case of HD~181068, the light curve distortions of non-eclipsing origin cause difficulties in the accurate determination of locations (and so times) of the contacts. Furthermore, our stars are not exactly spherical, and finally, due to the $5:251$ mean-motion resonance, we can get only a limited number of different eclipse configurations. On the other hand, there are some additional results that may serve as auxiliary sources of information. For example, Eq.~(\ref{eq:RBab}) could be used as an additional equation for $i_1$, or even $a_1/a_2$. Radial velocity and/or ETV results provide further constraints or equations. 

In the followings we give a practical example for HD~181068. To do this, we use our simulated light curve solution. By this trick we avoid the practical problem of correct and accurate identification of contact times, since our purpose is simply to demonstrate the theoretical effectiveness of this method. Furthermore, our finding about the orbital coplanarity makes our formulae as simple as possible, therefore the whole calculation can be performed analytically.

\begin{table}
 \caption{Contact times for outer secondary eclipse events}
 \label{tab: contacttimes}
 \begin{tabular}{@{}llllll}
  \hline
No & contact & star & MBJD & $u_1$ & $u_2$ \\
\hline
-0.5 & I & Ba & 55476.1096 & 313\fdg75441 & 260\fdg89609\\
     &II & Bb & 55476.4245 & 78\fdg92486 & 263\fdg38919 \\
     &III& Ba & 55477.9677 & 332\fdg33560 & 275\fdg60691 \\
     &IV & Ba & 55478.4722 & 172\fdg87067 & 279\fdg60111 \\
 \hline
 0.5 & I & Ba & 55521.5177 & 3\fdg14288 & 260\fdg39810 \\
     &II & Ba & 55522.0279 & 205\fdg94365 & 264\fdg43742 \\  
 \hline
 1.5 &III& Bb & 55568.9434 & 134\fdg51263 & 275\fdg87372  \\
 \hline
 2.5 & I & Bb & 55612.4733 & 157\fdg33030 & 260\fdg50577 \\
     &II & Bb & 55612.9903 & 2\fdg83403 & 264\fdg59893  \\
     &III& Bb & 55614.3571 & 186\fdg12706 & 275\fdg42007 \\
 \hline
 3.5 & II& Bb & 55658.3965 & 51\fdg46726 & 264\fdg08590 \\
     &III& Ba & 55659.9241 & 298\fdg67710 & 276\fdg18012 \\
     &IV & Bb & 55660.2422 & 65\fdg11954 & 278\fdg69856 \\
 \hline
 4.5 & I & Ba & 55703.4516 & 320\fdg54080 & 260\fdg79316 \\
     &II & Bb & 55703.7629 & 84\fdg28028 & 263\fdg25777 \\
     &III& Ba & 55705.3125 & 340\fdg23499 & 275\fdg52616 \\
     &IV & Ba & 55705.8234 & 183\fdg31398 & 279\fdg57102\\
 \hline
 \end{tabular}
 
 \end{table}
 
In Table~\ref{tab: contacttimes} we list the available contact moments for the large secondary eclipses occuring within the $Q7-Q9$ SC data. We used only secondary occultations to ignore the additional uncertainty arising from the limb-darkening during the primary transits. According to the last column (i.~e., $u_2$), one can see that its value may vary by a few tenths of degree for both the inner and outer contacts. Although this may look like a small variation, note, however, that a typical shift of $0\fdg5$ in $u_2$ translates to a change in the occurrence of the corresponding event of $\delta t\sim1.5$ hours. 

According to the above table, we have $3$ different outer and $3$ different inner contact moments for star $Ba$, while for $Bb$ these numbers are $2$ and $5$, respectively. We say different, as events $E=-0.5$ and $E=4.5$ belong to the same eclipse-family, and consequently, the corresponding moments are so similar to each other that we counted them only once. However, even in this case, we can write more equations than what is necessary. 

As an example, we show a concrete calculation. Using the first and last contacts of the $E=-0.5$  and the first contact of $E=0.5$ events, substituting the corresponding $u$ values into Eq.~(\ref{Eq:dxyBaA}), and subtrating their squares from each other we get three different equations, from which we need only two. As such,
\begin{eqnarray}
\sin^2i\left(\alpha_2+\beta_2x+\gamma_2x^2\right)+\delta_2x&=&0, \\
\sin^2i\left(\alpha_3+\beta_3x+\gamma_3x^2\right)+\delta_3x&=&0,
\end{eqnarray}
where
\begin{equation}
x=\frac{a_\rmn{Ba}}{a_2},
\end{equation}
and
\begin{eqnarray}
\alpha_{2,3}&=&(\sin^2u_2)_1-(\sin^2u_2)_{2,3}, \\
\beta_{2,3}&=&2[(\sin u_1)_1(\sin u_2)_1-(\sin u_1)_i(\sin u_2)_{2,3}], \\
\gamma_{2,3}&=&(\sin^2u_1)_1-(\sin^2u_1)_{2,3}, \\
\delta_{2,3}&=&2[\cos(u_1-u_2)_{2,3}-\cos(u_1-u_2)_1,
\end{eqnarray}
respectively. Eliminating the $\delta$-terms and using the fact that $\sin^2i\neq0$, we get a second order equation in $x$, i.~e. the orbital ratio. Obtaining $x$, the inclination can simply be calculated from any of the two equations. A similar treatment can be applied to the other component $Bb$. In this case, as we had only two outer contact times, we used the first two inner contact moments for the second equation. (Note that, for $Bb$, the signs of $\beta$ and $\delta$ coefficients should be changed!) When the two ratios $a_\rmn{Ba}/a_2$ and $a_\rmn{Bb}{a_2}$ are known, both $q_1$ and $a_1/a_2$ can be immediately calculated, and then $q_2$ follows too. Finally, the fractional radii of the three stars are also easily detemined. Remaining at the present sample, we first determine $R_\rmn{A}$ and $R_\rmn{Bb}$ from the two (one outer and one inner) contact equations of component $Bb$, and then $R_\rmn{Ba}$ can be calculated from any of the outer contact equations for $Ba$-star, without using any inner contact moments. (On the other hand, we can also calculate inner contact moments for component $Ba$, of course.) There is, however, another possibility to determine all these quantities without using inner contact times. This is because we can also use the outer contact-equation of the shallow eclipses. Thus the theoretically minimal data needed are the moments of: one outer contact for a shallow eclipse, three outer contacts for deep eclipses of one of the components, and two outer contacts for the other component. This means that we do not need any inner contact times (so the method works for partial eclipses, too), the usually better measurable outer contact times are sufficient. In Table~\ref{Tab: analres} we give our results. For comparison we also give the corresponding parameters of the synthetic light curve.

Finally, combining these results with the LITE solution, which returns $a_\rmn{B}\sin i$ in physical units (see Sect.~2), we can also calculate all the masses and stellar and orbital sizes in physical dimensions. So, we can conclude that in the case of triply eclipsing hierarchical systems (i.~e., where all the three objects eclipse each other at least partially, but not necessarily simultaneously) a high-precision single-band photometry of the eclipses is at least in principle efficient for determining all the above presented quantities in physical units. 

(The above calculations were made for coplanar and circular orbits. In the non-coplanar case our equations can be numerically solved in a similar way. The eccentric case is more complicated, but the asymmetry of the eclipses both in length and in phase gives all the required information too, so the difficulty is only practical.)

\begin{table}
 \caption{Analytic results from eclipse geometry and the original values}
 \label{Tab: analres}
 \begin{tabular}{@{}llllll}
  \hline
parameter & contacts used & calculated & original  \\
\hline
$\frac{a_\rmn{Ba}}{a_2}$ &(I,IV)$_{-0.5}$,I$_{0.5}$ & $0.023$ & $0.026$\\
$i$ & (I,IV)$_{-0.5}$,I$_{0.5}$ & $83\fdg7$ & $86\fdg7$/$87\fdg5$ \\
$\frac{a_\rmn{Bb}}{a_2}$ &II$_{-0.5}$,III$_{1.5}$,I$_{2.5}$,IV$_{3.5}$ & $0.027$ & $0.029$\\
$i$ &II$_{-0.5}$,III$_{1.5}$,I$_{2.5}$,IV$_{3.5}$ & $87\fdg0$ & $86\fdg7$/$87\fdg5$ \\
\hline
$q_1$ && $0.85$ & $0.95$ \\
$a_1/a_2$ && $0.049$ & $0.053$ \\
$q_2$ && $0.439$ & $0.545$ \\
\hline
$R_\rmn{A}/a_2$ & from $Ba$ & $0.170$ & $0.138$ \\
$R_\rmn{A}/a_2$ & from $Bb$ & $0.139$ & $0.138$ \\
$R_\rmn{Ba}/a_2$ && $0.009$ & $0.010$ \\
$R_\rmn{Bb}/a_2$ && $0.010$ & $0.009$ \\
\hline
 \end{tabular}
 
 \end{table}

\section[]{Light curve synthesis code for hierarchical triple systems}
\label{AppB}

The code is largely based on the well-known Wilson-Devinney program, which is being continuously developed from its first version \citep{wilsondevinney71} up to now \citep{wilson08,wilsonvanhamme09}. \citep[See also][~Chapters VI and VII.]{kallrathmilone09}. The PHOEBE Scientific Reference \citep{prsa06} was also used as a cook book. Some of the subroutines were borrowed directly from the Fortran code of the WD program (converting them from Fortran to C). However, a number of significant alterations were also applied. First, our code calculates the motion and positions of all the three stars, and naturally, the mutual eclipses (i.e. when an eclipse event of the close binary occurs in front of the disk of component $A$, or during the egress or ingress phase of the wide eclipse events). The mutual tidal interaction of the three stars is computed for every moment. In order to do this in a more simple way, instead of the usual two-mass point Roche-model, the stellar surfaces (and the local gravities as well) were calculated from the first order (linear) series expansions of the potential of a moderately distorted spherical body. In this formalism the stellar radius can be written in the following form:
\begin{equation}
r=R\left(1+\sum_{j=2}^4f_j+g_2\right),
\end{equation}
where the amplitudes of the first order tidal distortions caused by star $k$ on star $i$ are
\begin{equation}
f_j^{(i\leftarrow k)}=\left(1+2k_j^{(i)}\right)\frac{m_k}{m_i}\left(\frac{R_i}{\rho_{ik}}\right)^{j+1}P_j\left(\lambda''_{ik}\right),
\end{equation}
while the amplitude of the rotational distortion of star $i$ is
\begin{equation}
g_2^{(i)}=-\frac{\omega_i^2R_i^3}{3Gm_i}P_2\left(\nu'_i\right).
\end{equation}
In the equations above $R_i$ stands for the undistorted radius, $k_j^{(i)}$ denotes the $j$-th internal structure constant of star $i$, $\rho_{ik}$ the distance of the two stars, $\omega_i$ the rotational angular velocity of the star, while the direction cosines in the arguments of the given Legendre polynomials P$_j$ are the angle between the radius vector of the given surface element and the axis of the tidal bulge (practically the radius vector connects the centre of mass of the two stars) in the tidal terms ($\lambda''$), and the angle between the same surface element and the axis of stellar rotation ($\nu'$). See \citet{kopal78}, Chapter~II for details.

Strictly speaking, for strongly distorted systems this formalism is less accurate than the closed form of the Roche-model, but in the present situation, due to the moderate distortion of the present stars, it is adequate. Furthermore, besides the obvious advantage coming from linearity, it also treats the stars more realistically, as it no longer attributes infinite central mass densities to them.

Another improvement is the inclusion of the relativistic Doppler-beaming effect. The contribution to the total beaming is calculated for each surface cell of the three stars individually, and the radial velocity contribution coming from the stellar rotation is also taken into account, so theoretically the code is able to model the beaming analogous of the Rossiter-McLaughlin effect too. (Regardless, in the present system this is insignificant.) The beaming effect is well illustrated in Fig.~\ref{Fig:ABlcav}. 

Finally, we have also included the light-time effect. It is applied only to the wide subsystem, meaning that the positions of component $A$ and component $B$ were calculated in different moments according to their different distances from the observer, and furthermore, a second time-delay was also calculated in modelling the momentary tidal fields effect on each component.

In its present version, the program has $3\times11$ star-specific global physical parameters (masses, radii, tidal distortion [$k_{2..4}$] parameters, effective temperatures, chemical abundances, gravity darkening, two bolometric limb darkening coefficients, and bolometric albedos), $3\times3+1$ filter-specific parameters (luminosities, two limb darkening coefficients, and fourth light), $2\times6+1$ orbital parameters (the six orbital elements of the two orbits and the systemic radial velocity), and finally, $3\times6$ Eulerian angles and angular velocities describing stellar rotations. There are also several flags which turn on and off various constraints between different variables. Some of them are identical with those used in the WD program, but there are additional ones, due to the specific model. For example, instead of masses (which are undefined in the case of a simple two-body eclipsing light curve), one mass (usually $m_\rmn{A}$), and two mass ratios ($q_1$ and $q_2$) can also be used as input parameters. Similarly, instead of absolute radii, the use of fractional radii is more practical for light curve solution, although if the mass of the tertiary ($m_\rmn{A}$) and the outer mass-ratio ($q_2$) is known and fixed from ETV-solution, and of course, the orbital periods are also fixed, the use of absolute or fractional radii is fully equivalent.

\label{lastpage}


\begin{thebibliography}{55}

\bibitem[\protect\citeauthoryear{Agol et al.}{2005}]{agoletal05} Agol E., Steffen J., Sari R., Clarkson W., 2005, \mnras, 359, 567

\bibitem[\protect\citeauthoryear{Baron et al.}{2012}]{baronetal12} Baron, F. et al., 2012, \apj, 752, 20

\bibitem[\protect\citeauthoryear{Borkovits et al.}{2011}]{borkovitsetal11} Borkovits T., Csizmadia Sz., Forg\'acs-Dajka E., Heged\"us T., 2011, \aap, 528, A53

\bibitem[\protect\citeauthoryear{Borkovits et al.}{2003}]{borkovitsetal03} Borkovits T., \'Erdi B., Forg\'acs-Dajka E., Kov\'acs T., 2003, \aap, 398, 1091

\bibitem[\protect\citeauthoryear{Borkovits et al.}{2004}]{borkovitsetal04} Borkovits T., Forg\'acs-Dajka E., Reg\'aly Zs., 2004, \aap, 426, 951

\bibitem[\protect\citeauthoryear{Borkovits et al.}{2007}]{borkovitsetal07} Borkovits T., Forg\'acs-Dajka E., Reg\'aly Zs., 2007, \aap, 473, 191

\bibitem[\protect\citeauthoryear{Borucki et al.}{2010}]{bor10}
 Borucki et al., 2010, Science, 327, 977

\bibitem[Carter et al.(2011)]{carteretal11} Carter, J.A. et al., 2011, Science, 331, 562 

\bibitem[\protect\citeauthoryear{Carter et al.}{2012}]{carteretal12} Carter, J. A. et al., 2012, Science, 337, 556

\bibitem[\protect\citeauthoryear{Claret \& Gim\'enez}{1992}]{claretgimenez92} Claret, A., Gim\'enez, A., \aaps, 96, 255

\bibitem[Derekas et al.(2011)]{der11} Derekas A. et al., 2011, Science, 332, 216

\bibitem[\protect\citeauthoryear{Dotter et al.}{2008}]{dotteretal08}Dotter, A., Chaboyer, B., Jevremovi{\'c}, D., Kostov, V., Baron, E., Ferguson, J.~W., 2008, \apjs, 178, 89 

\bibitem[\protect\citeauthoryear{Doyle et al.}{2011}]{doyleetal11} Doyle, L. R. et al., 2011, Science, 333, 1602

\bibitem[\protect\citeauthoryear{Fabrycky \& Tremaine}{2007}]{fabryckytremaine07} Fabrycky, D., Tremaine, S., 2007, \apj, 669, 1298

\bibitem[Feiden, Chaboyer, \& Dotter(2011)]{fei11}
Feiden, G. A., Chaboyer, B., Dotter, A., 2011, \apj, 740, 25

\bibitem[\protect\citeauthoryear{Ford et al.}{2000}]{fordetal00} Ford, E. B., Kozinsky, B., Rasio, F. A., 2000, \apj, 535, 385

\bibitem[Gies et~al.(2012)]{gie12}
 Gies, D. R., Williams, S. J., Matson, R. A., Guo, Z., Thomas, S. M., Orosz, J. A., Peters, G. J., 2012, AJ, 143, 137

\bibitem[Gilliland et~al.(2010)]{gil10}
 Gilliland R. L., et al., 2010, PASP, 122, 131
 
\bibitem[\protect\citeauthoryear{Gim\'enez \& Garcia-Pelayo}{1983}]{gimenezgarcia83} Gim\'enez, A., Garcia-Pelayo, J. M., 1983 \apss, 92, 203 
 

\bibitem[Jenkins et~al.(2010a)]{jen10a}
 Jenkins  J.~M., et al., 2010a, \apj, 713, L87

\bibitem[Jenkins et~al.(2010b)]{jen10b}
 Jenkins  J.~M., et al., 2010b, \apj, 713, L120

\bibitem[\protect\citeauthoryear{Kalimeris et~al.}{2002}]{kalimerisetal02} 
 Kalimeris, A., Rovithis-Livaniou, H., Rovithis, P. 2002, \aap, 387, 969
 
\bibitem[\protect\citeauthoryear{Kallrath \& Milone}{2009}]{kallrathmilone09} Kallrath, J., Milone, E. F., 2009, {\it Eclipsing Binary Stars: Modeling and Analysis}, Astronomy and Astrophysics Library, Springer, DOI: 10.1007/978-1-4419-0699-1\_7

 
\bibitem[Koch et al.(2010)]{koc10} Koch  D.~G., et al., 2010, \apj, 713, L79
 
\bibitem[\protect\citeauthoryear{Kopal}{1978}]{kopal78} Kopal, Z., 1978, {\it Dynamics of close binary systems}, Astrophysics and Space Science Library. Vol. 68, Dordrecht, D. Reidel Publ. Co.

\bibitem[\protect\citeauthoryear{Kraus et al.}{2011}]{krausetal11} Kraus, A.~L., Tucker, R.~A., Thompson, M.~I.. Craine, E.~R., Hillenbrand, L.~A., 2011, \apj, 728, 48

\bibitem[Lehmann et al.(2012)]{leh12} Lehmann, H., Zechmeister, M., Dreizler, S., Schuh, S., Kanzler, R., 2012, \aap, 541, 105
 
\bibitem[\protect\citeauthoryear{Lestrade et al.}{1993}]{lestradeetal93} Lestrade, J.-F., Phillips, R. B., Hodges, M. W., Preston, R. A., 1993, \apj, 410, 808

\bibitem[\protect\citeauthoryear{Lissauer et al.}{2011}]{lissaueretal11} Lissauer, J. J. et al., 2011, Nature, 470, 53

\bibitem[\protect\citeauthoryear{Mayer}{1990}]{mayer90} Mayer P., 1990, BAICz, 41, 231 

\bibitem[\protect\citeauthoryear{O'Brien et al.}{2011}]{obrienetal11} O'Brien, D. P. et al., 2011, \apj, 728, 111 

\bibitem[\protect\citeauthoryear{Ol\'ah et al.}{2003}]{olahetal03} Ol\'ah, K., Jurcsik, J., Strassmeier, K. G., 2003, \aap, 410, 685

\bibitem[\protect\citeauthoryear{\"Ozdemir et al.}{2003}]{ozdemiretal03} \"Ozdemir S., Mayer P., Drechsel H., Demircan O., Ak, H., 2003, \aap, 403, 675

\bibitem[\protect\citeauthoryear{P\'al}{2012}]{pal12} P\'al, A., 2012, \mnras, 420, 1630

\bibitem[\protect\citeauthoryear{Peterson et al.}{2011}]{petersonetal11} Peterson, W. M., Mutel, R. L., Lestrade, J.-F., G\"udel, M., Goss, W. M., 2011, \apj, 737, 104

\bibitem[\protect\citeauthoryear{Pietrinferni et al.}{2004}]{pietrinfernietal04} Pietrinferni, A., Cassisi, S., Salaris, M., Castelli, F., 2004, \apj, 612, 168

\bibitem[\protect\citeauthoryear{Piirola}{2010}]{piirola10} Piirola, V., 2010, ASPC, 435, 225

\bibitem[\protect\citeauthoryear{Pop \& Vamo\c s}{2012}]{popvamos12} Pop, A., Vamo\c s, C. 2012, NewAstr, 17, 667

\bibitem[\protect\citeauthoryear{Pr\v sa}{2006}]{prsa06} Pr\v sa, A., 2006, {\it PHOEBE Scientific Reference. PHOEBE version 0.30}, University of Ljubljana, Faculty of Mathematics and Physicy, Dept. of Astrophysics 

\bibitem[\protect\citeauthoryear{Pr\v sa \& Zwitter}{2005}]{prsazwitter05} Pr\v sa, A., Zwitter, T., 2005, \apj, 628, 426

\bibitem[\protect\citeauthoryear{Qian et al.}{2012}]{qianetal12} Qian, S.-B., Liu, L., Zhu, L.-Y., Dai, Z.-B., Fern\'andez Laj\'us, E., Baume, G. L., 2012, \mnras, 422, L24

\bibitem[\protect\citeauthoryear{Ragozzine \& Holman}{2010}]{ragozzineholman10} Ragozzine, D., Holman, M. J., 2010, arXiv:1006.3727

\bibitem[\protect\citeauthoryear{Sanborn \& Zavala}{2012}]{sanbornzavala12} Sanborn, J. J., \& Zavala, R. T., JASS, 29, 63

\bibitem[\protect\citeauthoryear{S\"oderhjelm}{1975}]{soderhjelm75} S\"oderhjelm, S., 1975, \aap, 42, 229

\bibitem[\protect\citeauthoryear{S\"oderhjelm}{1984}]{soderhjelm84} S\"oderhjelm, S., 1984, \aap, 141, 232

\bibitem[Steffen et al.(2011)]{steffenetal11} Steffen J.H. et al., 2011, \mnras, 417, L31


\bibitem[\protect\citeauthoryear{Tokovinin}{2008}]{tokovinin08} Tokovinin, A., 2008, \mnras, 389, 925

\bibitem[\protect\citeauthoryear{Verbunt \& Phinney}{1995}]{verbuntphinney95} Verbunt, F., Phinney, E. S., 1995, \aap, 296, 709


\bibitem[\protect\citeauthoryear{Welsh et al.}{2012}]{welshetal12} Welsh, W. F. et al., 2012, Nature, 481, 475

\bibitem[\protect\citeauthoryear{Wilson}{2008}]{wilson08} Wilson, R. E., 2008, \apj, 672, 575

\bibitem[\protect\citeauthoryear{Wilson \& Devinney}{1971}]{wilsondevinney71} Wilson, R. E., Devinney, E. J., 1971, \apj, 166, 605

\bibitem[\protect\citeauthoryear{Wilson \& Van Hamme}{2009}]{wilsonvanhamme09} Wilson, R. E., Van Hamme, W., 2009, \apj, 699, 118

\bibitem[\protect\citeauthoryear{Zahn}{1977}]{zahn77} Zahn, J.-P., 1977, \aap, 57, 383

\bibitem[\protect\citeauthoryear{Zahn}{1989}]{zahn89} Zahn, J.-P., 1989, \aap, 220, 112 

\bibitem[\protect\citeauthoryear{Zucker et al.}{2007}]{zuckeretal07} Zucker, S., Mazeh, T., Alexander, T., 2007, \apj, 670, 1326

\end{thebibliography}
\end{document}